%% file: ms.tex
\documentclass[a4paper,iop,twocolumn,numberedappendix,appendixfloats]{emulateapj}
\usepackage[breaklinks,colorlinks=true,linkcolor=blue,anchorcolor=blue,citecolor=blue,urlcolor=blue,draft=false]{hyperref}
\usepackage{color}
\usepackage{graphicx}
\usepackage{amsmath} 
\usepackage{amssymb}
\usepackage{mathrsfs}
\usepackage{placeins}
\usepackage{times}
\usepackage{xcolor}
\usepackage{xspace}

\hypersetup{pdfauthor={Keiichi Umetsu},pdftitle={CLASH-WL2D}}
\usepackage{natbib}
 
\newcommand{\simgt}{\lower.5ex\hbox{$\; \buildrel > \over \sim \;$}}
\newcommand{\simlt}{\lower.5ex\hbox{$\; \buildrel < \over \sim \;$}}
\newcommand{\llangle}{\langle\!\langle}
\newcommand{\rrangle}{\rangle\!\rangle}

\newcommand{\Mvir}{M_\mathrm{vir}}

\newcommand{\rvir}{r_\mathrm{vir}}

\newcommand{\Nbin}{N_\mathrm{bin}}
\newcommand{\Npix}{N_\mathrm{pix}}

\newcommand{\kpch}{\mathrm{kpc}\,h^{-1}}
\newcommand{\Mpch}{\mathrm{Mpc}\,h^{-1}}
\newcommand{\Msunh}{M_\odot\,h^{-1}}
\newcommand{\Msun}{M_\odot\,h_{70}^{-1}}

\newcommand{\percent}{\ensuremath{\%}}

\def\bc{\mbox{\boldmath $c$}}
\def\btheta{\mbox{\boldmath $\theta$}} 
\def\bnabla{\mbox{\boldmath $\nabla$}}
\def\bkappa{\mbox{\boldmath $\kappa$}}
\def\bSigma{\mbox{\boldmath $\Sigma$}} 

\def\blambda{\mbox{\boldmath $\lambda$}}

\def\bd{\mbox{\boldmath $d$}}
\def\bs{\mbox{\boldmath $s$}}
\def\bp{\mbox{\boldmath $p$}}
\def\bR{\mbox{\boldmath $R$}}

\def\singlebond{\@makechembond\@ne}
\def\doublebond{\@makechembond\tw@}
\def\triplebond{\@makechembond\thr@@}    
\def\ge{\geqslant}
\def\le{\leqslant}

\def\red{}

\lefthead{Umetsu et al.}
\righthead{Projected Dark and Baryonic Structure of 20 CLASH Clusters}
\usepackage{datetime}
\usepackage{version}    
\input{citation_fix.tex}

\begin{document}
\title{The Projected Dark and Baryonic Ellipsoidal Structure of 20 CLASH
Galaxy Clusters\altaffilmark{*}}
\author{Keiichi Umetsu\altaffilmark{1}}         
\author{Mauro Sereno\altaffilmark{2,3}}         
\author{Sut-Ieng Tam\altaffilmark{4}}           
\author{I-Non Chiu\altaffilmark{1}}             
\author{Zuhui Fan\altaffilmark{5,6}}            
\author{Stefano Ettori\altaffilmark{2,7}}       
\author{Daniel Gruen\altaffilmark{8,9,10}}      
\author{Teppei Okumura\altaffilmark{1}}         
\author{Elinor Medezinski\altaffilmark{11}}     
\author{Megan Donahue\altaffilmark{12}}         
\author{Massimo Meneghetti\altaffilmark{13}}    
\author{Brenda Frye\altaffilmark{14}}           
\author{Anton Koekemoer\altaffilmark{15}}       
\author{Tom Broadhurst\altaffilmark{16,17}}     
\author{Adi Zitrin\altaffilmark{18}}            
\author{Italo Balestra\altaffilmark{19,20}}     
\author{Narciso Ben\'itez\altaffilmark{21}}     %
\author{Yuichi Higuchi\altaffilmark{1}}         
\author{Peter Melchior\altaffilmark{11}}        %
\author{Amata Mercurio\altaffilmark{22}}        
\author{Julian Merten\altaffilmark{23}}         
\author{Alberto Molino\altaffilmark{24}}        
\author{Mario Nonino\altaffilmark{20}}          
\author{Marc Postman\altaffilmark{15}}          
\author{Piero Rosati\altaffilmark{25}}          
\author{Jack Sayers\altaffilmark{26}}           
\author{Stella Seitz\altaffilmark{27,28}}       %

\altaffiltext{*}
 {Based in part on data collected at the Subaru Telescope,
  which is operated by the National Astronomical Society of Japan.}
\email{keiichi@asiaa.sinica.edu.tw}
\altaffiltext{1}
 {Institute of Astronomy and Astrophysics, Academia Sinica,
  P.~O. Box 23-141, Taipei 10617, Taiwan}
\altaffiltext{2}
 {INAF - Osservatorio di Astrofisica e Scienza dello Spazio di Bologna,
 via Piero Gobetti 93/3, I-40129 Bologna, Italy}
\altaffiltext{3}
 {Dipartimento di Fisica e Astronomia, Universit\`a di Bologna, via
 Piero Gobetti 93/2, I-40129 Bologna, Italy}
\altaffiltext{4}{Centre for Extragalactic Astronomy, Department of
Physics, Durham University, Durham DH1 3LE, U.K.}
\altaffiltext{5}{South-Western Institute for Astronomy Research, Yunnan University, Kunming 650500, China}
\altaffiltext{6}{Department of Astronomy, School of Physics, Peking University, Beijing 100871, China}
\altaffiltext{7}{INFN, Sezione di Bologna, viale Berti Pichat 6/2, I-40127 Bologna, Italia}
\altaffiltext{8}{SLAC National Accelerator Laboratory, Menlo Park, CA 94025, USA}
\altaffiltext{9}{KIPAC, Physics Department, Stanford University,
Stanford, CA 94305, USA}
\altaffiltext{10}{Einstein Fellow}
\altaffiltext{11}{Department of Astrophysical Sciences, 4 Ivy Lane, Princeton, NJ 08544, USA}
\altaffiltext{12}{Department of Physics and Astronomy, Michigan State
University, East Lansing, MI 48824, USA} 
\altaffiltext{13}{INAF-Osservatorio Astronomico di Bologna, via Ranzani
1, I-40127 Bologna, Italy}
\altaffiltext{14}{University of Arizona, 933 N. Cherry Ave, Tucson, AZ 85721}
\altaffiltext{15}{Space Telescope Science Institute, 3700 San Martin
Drive, Baltimore, MD 21218, USA}
\altaffiltext{16}{Department of Theoretical Physics, University of Basque Country UPV/EHU, E-48080 Bilbao,
Spain}
\altaffiltext{17}{IKERBASQUE, Basque Foundation for Science, E-48013
Bilbao, Spain}
\altaffiltext{18}{Physics Department, Ben-Gurion University of the Negev,
P.O. Box 653, Be'er-Sheva 8410501, Israel}
\altaffiltext{19}{University Observatory Munich, Scheinerstrasse 1,
D-81679 Munich, Germany}
\altaffiltext{20}{INAF-Osservatorio Astronomico di Trieste, Via Tiepolo
11, I-34131 Trieste, Italy}
\altaffiltext{21}{Instituto de Astrof\`isica de Andaluc\`ia (CSIC),
Glorieta de la Astronom\`ia s/n, Granada, 18008, Spain}
\altaffiltext{22}{INAF-Osservatorio Astronomico di Capodimonte, via Moiariello
16, 80131 Napoli, Italy}
\altaffiltext{23}{Oxford University, Keble Road, Oxford OX1 3RH, United Kingdom}
\altaffiltext{24}{Universidade de S\~ao Paulo, Instituto de Astronomia, Geof\'isica e Ci\^encias Atmosf\'ericas, Rua do Mat\~ao 1226, 05508-090, S\~ao Paulo, Brazil}
\altaffiltext{25}{Universit\`a di Ferrara, Via Saragat 1, I-44122
Ferrara, Italy} 
\altaffiltext{26}{Division of Physics, Math, and Astronomy, California Institute of
Technology, Pasadena, CA 91125}
\altaffiltext{27}{Universit\"ats-Sternwarte, M\"unchen, Scheinerstr. 1,
D-81679 M\"unchen. Germany}
\altaffiltext{28}{Max-Planck-Institut f\"ur extraterrestrische Physik,
Giessenbachstr. 1, D-85748 Garching} 
\begin{abstract}
 We reconstruct the two-dimensional (2D) matter distributions in 20
 high-mass galaxy clusters selected from the CLASH survey by using the
 new approach of performing a joint weak gravitational
 lensing analysis of 2D shear and azimuthally averaged magnification
 measurements. This combination allows for a complete analysis of the
 field, effectively breaking the mass-sheet degeneracy.
 In a Bayesian framework, we simultaneously constrain the mass profile
 and morphology of each individual cluster, assuming an
 elliptical Navarro--Frenk--White halo characterized 
 by the mass, concentration,
 projected axis ratio, and position angle (PA) of
 the projected major axis. We find that spherical mass estimates
 of the clusters from azimuthally averaged weak-lensing
 measurements in previous work are in excellent agreement with our results
 from a full 2D analysis.
 Combining all 20 clusters in our sample, we detect the elliptical shape of
 weak-lensing halos at the $5\sigma$ significance level
 {\red within a scale of}
 $2\Mpch$.  The median projected axis ratio is $0.67\pm 0.07$
 at a virial mass of
 $\Mvir=(15.2\pm 2.8)\times 10^{14}M_\odot$,
 which is in agreement with theoretical predictions
 from recent numerical simulations of the standard collisionless cold
 dark matter model. We also study misalignment statistics of
 the brightest cluster galaxy, X-ray, thermal Sunyaev--Zel'dovich
 effect, and strong-lensing morphologies with respect to the
 weak-lensing signal. Among the three baryonic tracers studied here, we
 find that the X-ray morphology is best aligned with the weak-lensing
 mass distribution, with a median misalignment
 angle of $|\Delta\mathrm{PA}|=21^\circ\pm 7^\circ$.
 We also conduct a stacked quadrupole shear analysis of the 20
 clusters assuming that the X-ray major axis is aligned
 with that of the projected mass distribution. This yields a consistent
 axis ratio of $0.67\pm 0.10$, suggesting again a tight alignment
 between the intracluster gas and dark matter.
\end{abstract}   
 


\section{Introduction}
\label{sec:intro}

Galaxy clusters, as the most massive objects formed in the universe,
represent fundamental probes of cosmology. Clusters contain rich
information about the initial conditions for structure formation,
the emergence of large-scale structure over cosmic time, and the
properties of dark matter (DM). 
In the standard picture of hierarchical structure formation, determining
the abundance of rare massive clusters above a given mass provides a
powerful test of growth of structure \citep[e.g.,][]{Bahcall+Fan1998}
because cluster-scale halos populate the exponential tail of the cosmic
mass function.  
Statistical properties of clusters can thus be used to yield unique
constraints on cosmological parameters and 
models of cosmic structure formation
\citep{Allen+2004,Vikhlinin+2009CCC3,Mantz2010}, 
 complementing standard cosmological probes, such as
cosmic microwave background (CMB) anisotropy, large-scale
galaxy clustering, distant supernova, and cosmic shear observations.  



The most critical ingredient for cluster-based tests of structure
formation is the distribution and amount of DM in cluster halos.
In this context, the standard $\Lambda$ cold dark matter ($\Lambda$CDM)
model and its variants, such as self-interacting DM
\citep[SIDM;][]{Spergel+Steinhardt2000}
and wave DM \citep[$\psi$DM;][]{Schive2014psiDM},
provide a range of observationally testable predictions.
$N$-body simulations in the standard $\Lambda$CDM model reveal that
clusters form through  
successive mergers of groups and smaller clusters, as well as through
the smooth accretion of matter from the surrounding filamentary structure
\citep{Colberg+2000,Shaw+2006},
leading to a highly anisotropic geometry in which infall and merging
of matter tend to take place along preferential directions. The process
results in the emergence of the filamentary 
cosmic web, as observed in galaxy redshift surveys
\citep{Colless+2011,Tegmark+2004,Geller+2011}.
Cluster halos are formed in overdense regions where the filaments 
intersect.
The shape of halos is generally triaxial
with a preference for prolate shapes
\citep{2002ApJ...574..538J,Shaw+2006},
reflecting the collisionless nature of DM
\citep{Ostriker+Steinhardt2003}.
On average, older halos are more relaxed and are thus more spherical.
Since more massive halos form later on average, 
clusters are thus expected to be more elongated than
less massive systems \citep{Despali2014}.
Accretion of matter from the surrounding large-scale environment also
plays a key role in determining the shape and orientation of cluster
halos. 
The halo orientation tends to be in the preferential infall direction  
of the subhalos and hence aligned along the surrounding filaments 
\citep{Shaw+2006}. The shape and orientation of galaxy clusters thus
provide an independent test of models of structure formation.


Gravitational lensing offers a direct probe of the cluster mass
distribution through observations of  
weak shear lensing
\citep[e.g.,][]{WtG1,Gruen2014,Hoekstra2015CCCP,Melchior2017des,Sereno2017psz2lens,Medezinski2018planck}, 
weak magnification lensing
\citep[e.g.,][]{Hildebrandt+2011,Umetsu+2011,Coupon+2013,Chiu2016magbias,Tudorica2017sparcs},
strong gravitational lensing
\citep[e.g.,][]{2005ApJ...621...53B,Zitrin+2013M0416,Jauzac2015a2744,Cerny2017relics,Diego2018a370},
and the combination of these effects
\citep[e.g.,][]{Umetsu+2011stack,Umetsu2015A1689,Umetsu2016clash,Coe+2012A2261,Medezinski+2013}.
The critical advantage of gravitational lensing is its unique ability to
map the mass distribution independently of
assumptions about their physical or dynamical state. Cluster lensing
thus provides a direct and powerful way to test predictions of halo
density structure dominated by DM.  

Cluster lensing observations have established that the projected total
mass distribution within individual and ensemble-averaged clusters can
be well described by sharply steepening density profiles 
\citep{Umetsu+2011stack,Umetsu2014clash,Umetsu2016clash,Newman+2013a,Okabe+2013,Okabe+Smith2016}, 
with a near-universal shape \citep{Niikura2015,Umetsu+Diemer2017}, as
predicted for halos dominated by collisionless DM in quasi-gravitational equilibrium
\citep[e.g.,][]{1996ApJ...462..563N,1997ApJ...490..493N,Hjorth+2010DARKexp,DARKexp2}.
Subsequent cluster lensing studies targeting lensing-unbiased samples
\citep{Merten2015clash,Umetsu2016clash,Du2015,Okabe+Smith2016,Cibirka2017}
have shown that the degree of halo concentration derived for these clusters
agrees well with theoretical models that are  
calibrated for recent $\Lambda$CDM cosmologies with a relatively high
normalization \citep{Bhatt+2013,Dutton+Maccio2014,Meneghetti2014clash,Diemer+Kravtsov2015}.
These results are in support of the standard explanation for DM as
effectively collisionless and non-relativistic on  sub-Mpc scales and
beyond, with an excellent match between cluster lensing data and
$\Lambda$CDM predictions. 

The CLUster Multi-Probes in Three Dimensions
\citep[CLUMP-3D;][]{Sereno2017clump3d} program aims to study
intrinsic three-dimensional (3D) properties of high-mass galaxy
clusters and to test models of cluster formation.
By exploiting rich data sets ranging from the X-ray, through optical, to
radio wavelengths, we can constrain the 3D geometry and
internal structure of individual clusters, together with the equilibrium
status of the intracluster gas residing in cluster DM halos
\citep[e.g.,][]{Morandi2012,Sereno2013glszx}.
In the approach developed by \citet{Sereno2013glszx}, we exploit the  
combination of X-ray and thermal Sunyaev--Zel'dovich effect (SZE)
observations to constrain the line-of-sight elongation of the
intracluster gas in a parametric triaxial framework.
Employing minimal geometric assumptions about
the matter and gas distributions, we then couple the constraints from 
gravitational lensing, X-ray, and SZE data sets in a Bayesian inference
framework \citep{Sereno2013glszx,Sereno2017clump3d,Umetsu2015A1689}.  
This multi-probe method allows constraints on the intrinsic
shape and orientation of the matter and gas distributions to be improved
without assuming hydrostatic equilibrium.

As part of the CLUMP-3D program, we present in this paper a
two-dimensional (2D) weak-lensing analysis of wide-field shear and 
magnification data for a sample of 20 high-mass clusters,
for which high-quality multiwavelength data sets are available from the
CLASH survey \citep{Postman+2012CLASH} and dedicated follow-up programs
\citep{Donahue2014clash,Umetsu2014clash,Umetsu2016clash,Rosati2014VLT,Merten2015clash,Zitrin2015clash,Czakon2015}.   
In this work, we analyze the ground-based weak-lensing data of 
\citet{Umetsu2014clash} 
obtained from deep multiband imaging taken primarily with the
Suprime-Cam on the Subaru Telescope \citep[$34\arcmin\times 
27\arcmin$;][]{2002PASJ...54..833M}.
For our southernmost cluster (RX~J2248$-$4431), which is
not observable from Subaru, we analyze data obtained with the Wide-Field
Imager (WFI) at the ESO 2.2\,m MPG/ESO telescope at La Silla
\citep{Gruen+2013RXJ2248,Umetsu2014clash}.

The primary aim of this paper is to perform an unbiased mass
reconstruction in our 20 cluster fields, from which to simultaneously
constrain the structure and morphology of the cluster mass
distribution, both individually and statistically.
This allows us to compare the position angles of cluster major axes 
determined from our wide-field weak-lensing analysis to those of
baryonic tracers and central lensing maps inferred by
\citet{Donahue2015clash,Donahue2016clash}.
This work has two companion papers: 
the triaxial modeling and ensemble characterization of the 20 CLASH
clusters by
\citet{Chiu2018clump3d}
from a joint analysis
of weak- and strong-lensing data sets and
the multi-probe triaxial modeling of 16 CLASH X-ray-selected clusters by
\citet{Sereno2018clump3d}
from a joint analysis of
weak-lensing, strong-lensing, X-ray, and SZE data sets.

This paper is organized as follows. In Section \ref{sec:basics}, we
describe the basic theory of weak gravitational lensing by galaxy
clusters. After summarizing the properties of the cluster sample and the
observational data, we outline in Section \ref{sec:wl} the formalism and 
procedure for reconstructing the cluster mass distribution from a 2D
weak-lensing analysis of wide-field shear and magnification data. In
Section \ref{sec:results}, we present the results of 2D mass modeling of
weak-lensing maps for our sample.
Section \ref{sec:discussion} is devoted to the discussion of 
the results.
Finally, a summary is given in Section \ref{sec:summary}.

Throughout this paper, we assume a spatially flat $\Lambda$CDM cosmology  
with
$\Omega_\mathrm{m}=0.27$,
$\Omega_\Lambda=0.73$,
and a Hubble constant of 
$H_0 = 100h$\,km\,s$^{-1}$\,Mpc$^{-1}=70h_{70}$\,km\,s$^{-1}$\,Mpc$^{-1}$
with $h=0.7h_{70}=0.7$.
We denote the mean matter density of the universe as
$\rho_\mathrm{m}$ and the critical density of the universe as $\rho_\mathrm{c}$. 
We adopt the standard notation
$M_{\Delta_\mathrm{c}}$ or $M_{\Delta_\mathrm{m}}$ 
to denote the mass enclosed within a sphere of radius
$r_{\Delta_\mathrm{c}}$ or $r_{\Delta_\mathrm{m}}$,
within which the mean overdensity equals 
$\Delta_\mathrm{c} \times \rho_\mathrm{c}(z)$ 
or
$\Delta_\mathrm{m} \times \rho_\mathrm{m}(z)$
at a particular redshift $z$, such that
$M_{\Delta\mathrm{c}}=(4\pi/3)\Delta_\mathrm{c}\rho_\mathrm{c}(z)r_{\Delta\mathrm{c}}^3$
and
$M_{\Delta\mathrm{m}}=(4\pi/3)\Delta_\mathrm{m}\rho_\mathrm{m}(z)r_{\Delta\mathrm{m}}^3$.
We compute the virial mass and radius, $\Mvir$ and $\rvir$,
using an expression for $\Delta_\mathrm{vir}(z)$ based on the
spherical collapse model \citep[Appendix A of][]{1998PASJ...50....1K}.
For a given overdensity $\Delta$, the concentration parameter is
defined as $c_\Delta=r_\Delta/r_\mathrm{s}$.
All quoted errors are $1\sigma$ confidence limits unless otherwise
stated. 

\section{Basics of Cluster Gravitational Lensing}
\label{sec:basics}

\subsection{Shear and Magnification}
\label{subsec:lensing}

The effect of weak gravitational lensing on background sources is
characterized by the convergence, $\kappa$, and 
the shear with spin 2 rotational symmetry,
$\gamma=|\gamma|e^{2i\phi}$
\citep[e.g.,][]{2001PhR...340..291B,Umetsu2010Fermi}. 
The convergence causes an isotropic magnification due
to lensing and is defined as the surface mass density
$\Sigma$ in units of the critical surface density for lensing,
$\kappa=\Sigma/\Sigma_\mathrm{c}$,
where
\begin{equation}
\Sigma_\mathrm{c} = \frac{c^2 D_\mathrm{s}}{4\pi G
 D_\mathrm{l}D_\mathrm{ls}}
 \equiv \frac{c^2}{4\pi G D_\mathrm{l}}\beta^{-1}
\end{equation}
with
$c$ the speed of light,
$G$ the gravitational constant,
and
$D_\mathrm{l}$, $D_\mathrm{s}$, and $D_\mathrm{ls}$ 
the observer--lens, observer--source, and lens--source angular diameter
distances, respectively.
The dimensionless factor $\beta(z, z_\mathrm{l}) = D_\mathrm{ls}/D_\mathrm{s}$
describes the geometric lensing strength as a function of source redshift $z$ and
lens redshift $z_\mathrm{l}$, where $\beta(z,z_\mathrm{l})=0$ for
unlensed objects with $z\le z_\mathrm{l}$. Hence, the shear and
convergence depend on the source and lens redshifts ($z,z_\mathrm{l}$),
as well as on the image position $\btheta$.

 
The gravitational shear $\gamma$  is directly observable from
image ellipticities of background galaxies in the weak regime,
$\kappa\ll 1$.  
The shear and convergence fields are related by \citep{1993ApJ...404..441K}
\begin{equation}
\label{eq:kappa2gamma}
\gamma(\btheta) =
\int\!d^2\theta'\,D(\btheta-\btheta')\kappa(\btheta')
\end{equation}
with $D(\btheta)$ the complex kernel
$D(\btheta)=(\theta_2^2-\theta_1^2-2i\theta_1\theta_2)/(\pi|\btheta|^4)$.
In general, the observable quantity for weak lensing
is not $\gamma$ but the complex reduced shear,
\begin{equation}
\label{eq:redshear}
g =\frac{\gamma}{1-\kappa},
\end{equation}
which remains invariant under the global transformation
$\kappa(\btheta)\to \lambda \kappa(\btheta) + 1-\lambda$
and 
$\gamma(\btheta)\to \lambda \gamma(\btheta)$
with an arbitrary constant $\lambda\ne 0$ (for a fixed
source redshift $z$). This is known as the
mass-sheet  degeneracy  \citep{Schneider+Seitz1995}.
In principle, this degeneracy can be broken or alleviated, for example,
by measuring the magnification factor $\mu$ in the subcritical regime,
\begin{equation}
\label{eq:mu}
\mu = \frac{1}{(1-\kappa)^2-|\gamma|^2}
\equiv \frac{1}{\Delta_\mu},
\end{equation}
which transforms as $\mu(\btheta)\to \lambda^{-2}\mu(\btheta)$.
For simplicity of notation, we often use the inverse magnification
$\Delta_\mu=\mu^{-1}$ rather than the magnification.

\subsection{Source Redshift Distribution}
\label{subsec:Nz}

For statistical weak-lensing measurements, we consider a population of
source galaxies characterized by their mean redshift distribution
function, $\overline{N}(z)$. 
In general, we use different size, magnitude, color, and quality cuts
in background selection for measuring shear and magnification,
which results in different $\overline{N}(z)$.
In contrast to the former analysis, the
latter does not require background sources to be spatially resolved,
while it does require a stringent flux limit against incompleteness
effects \citep{Umetsu2014clash,Chiu2016magbias}. 
The source-averaged lensing depth for a given population ($X=g,\mu$) is
\begin{equation}
\label{eq:depth}
\langle\beta\rangle_X =\left[
\int_0^\infty\!dz\, \overline{N}_X(z) \beta(z)\right]
\left[
\int_0^\infty\!dz\, \overline{N}_X(z)
\right]^{-1}.
\end{equation} 

Let us introduce the relative lensing strength of a given source population
as $\langle W\rangle_X   = \langle\beta\rangle_X  / \beta_\infty$ 
with $\beta_\infty\equiv \beta(z\to \infty, z_\mathrm{l})$
defined relative to a fiducial source in the far background \citep{2001PhR...340..291B}.
The associated critical density is
$\Sigma_{\mathrm{c},\infty}(z_\mathrm{l})=c^2/(4\pi G D_\mathrm{l})\beta_{\infty}^{-1}$.
Hereafter, we use the far-background fields
$\kappa_\infty(\btheta)$ and 
$\gamma_\infty(\btheta)$
to describe the projected mass distribution.

\subsection{Pixelized Mass Distribution}
\label{subsec:massmodel}

We pixelize the convergence field,
$\kappa_\infty(\btheta)=\Sigma_{\mathrm{c},\infty}^{-1}\Sigma(\btheta)$,
into a regular grid of pixels,
and describe $\kappa_\infty(\btheta)$ by a linear combination of basis
functions $B(\btheta-\btheta')$,
\begin{equation}
\label{eq:basis}
\kappa_\infty(\btheta) 
=\Sigma_{\mathrm{c},\infty}^{-1}
\sum_{n=1}^{N_\mathrm{pix}}
 B(\btheta-\btheta_n)\, \Sigma_n.
\end{equation}
To avoid the loss of information due to oversmoothing,
we take the basis function to be the Dirac delta function,
 $B(\btheta-\btheta')=(\Delta\theta)^2\delta^2_\mathrm{D}(\btheta-\btheta')$,
 with $\Delta\theta$ a constant grid spacing \citep{Umetsu2015A1689}.
Our model (signal) is specified by a vector of 
parameters containing cell-averaged surface mass densities
\citep{Umetsu2015A1689},
\begin{equation}
 \bs = \left\{\Sigma_n\right\}_{n=1}^{N_\mathrm{pix}}.
\end{equation}
The complex shear field is expressed as 
\begin{equation}
\label{eq:shear2m}
\gamma_\infty(\btheta)= \Sigma_{\mathrm{c},\infty}^{-1}\sum_{n=1}^{\Npix} 
{\cal D}(\btheta-\btheta_n)\,\Sigma_n
\end{equation}
with ${\cal D}\equiv D\otimes B = \pi^{-1} (\Delta\theta)^2D$ an
effective complex kernel (see Equation (\ref{eq:kappa2gamma})).
Hence, both $\kappa_\infty(\btheta)$ and $\gamma_\infty(\btheta)$ can be
written as linear combinations of mass coefficients.

It is important to note that,
because of the choice of the basis function,
an unbiased extraction of the mass coefficients
$\{\Sigma_n\}_{n=1}^{N_\mathrm{pix}}$
is possible by performing a spatial integral of $\kappa_\infty(\btheta)$
over a certain area. Such operations include 
smoothing (Figure \ref{fig:M0329}),
azimuthal averaging for a mass profile measurement \citep{Umetsu2015A1689},
and fitting with smooth parametric functions (Section
\ref{sec:results}).

\section{Weak-lensing Data and Methodology}
\label{sec:wl}

The present work is a full 2D generalization of the 
weak-lensing study by \citet{Umetsu2014clash},  
who conducted a one-dimensional (1D) weak-lensing analysis of
azimuthally averaged shear and magnification measurements for a sample 
of 20 CLASH clusters (Section \ref{subsec:data}).
A practical limitation of a shear-only analysis is the inherent mass-sheet
degeneracy  \citep{Falco1985,Gorenstein1988,Schneider+Seitz1995,Seitz+Schneider1997,Bradac2004}.
One can substantially alleviate
this degeneracy by using the complementary combination of shear and 
magnification
\citep{Schneider+2000,UB2008,Rozo+Schmidt2010,Umetsu2013}.
Measuring the two complementary effects also allows the internal
consistency of weak-lensing measurements to be tested
\citep{Umetsu2014clash}.
Besides, in the context of the CLUMP-3D program, obtaining accurate mass
maps has the important advantage of being able to identify local mass
structures and to directly compare them with multiwavelength observations.

In this study, we use the ground-based weak-lensing data obtained by the     
CLASH collaboration (Section \ref{subsec:back}).
Our shear catalogs (Section \ref{subsec:shear}) as well as
azimuthally averaged magnification profiles (Section \ref{subsec:magbias})
have already been published in \citet{Umetsu2014clash}.
Data products from the CLASH survey,
including the reduced Subaru images, weight maps, and multiband
photometric catalogs, are publicly available at the Mikulski Archive for
Space Telescopes
(MAST)\footnote{\href{https://archive.stsci.edu/prepds/clash/}{https://archive.stsci.edu/prepds/clash/}}. 
Details of the image reduction, photometry, background source
selection, and the creation of our
weak-lensing shear catalogs are given in \citet[][see their Section
4]{Umetsu2014clash}. More details on weak-lensing systematics are
presented in Section 3 of \citet{Umetsu2016clash}.
Thus, we provide here only a summary of the procedures.
Section \ref{subsec:diff} summarizes the major differences between our 
analysis and those of \citet{Umetsu2014clash} and \citet{Umetsu2016clash}.

\subsection{Cluster Sample}
\label{subsec:data}

\input{table1.tex}

The cluster sample studied in the CLUMP-3D program
\citep{Sereno2017clump3d,Sereno2018clump3d,Chiu2018clump3d}
stems from the wide-field weak-lensing analysis of \citet{Umetsu2014clash}.
The sample comprises two subsamples, one with 16 X-ray-selected
clusters and another with four high-magnification clusters (Table \ref{tab:sample}).
Both subsamples were 
taken from the CLASH survey \citep{Postman+2012CLASH},
a 524 orbit {\em Hubble Space Telescope} ({\em HST})
Multi-Cycle  Treasury program targeting 25 high-mass clusters.

Here, 20 CLASH clusters were selected to have X-ray temperatures
above 5\,keV and to have a high degree of regularity in their X-ray
morphology.
Specifically, these clusters show well-defined central surface
brightness peaks and nearly concentric isophotes in {\em Chandra} X-ray
images \citep{Postman+2012CLASH}.
The CLASH X-ray criteria ensure well-defined cluster centers, reducing
the effects of cluster miscentering (see below).
\citet{Meneghetti2014clash} characterized intrinsic and observational
properties of the CLASH X-ray-selected subsample by analyzing simulated
halos chosen to match these individual clusters in terms of the X-ray
morphological regularity. 
Their simulations suggest that this
subsample is largely free of orientation bias
and dominantly composed of relaxed clusters
($\sim 70\percent$), but it also contains a non-negligible fraction
($\sim 30\percent$) of unrelaxed clusters.
Another subset of five clusters were selected for their high-lensing
magnification. These clusters often turn out to be dynamically
disturbed, complex merging systems 
\citep[e.g.,][]{Zitrin+2013M0416,Medezinski+2013,Balestra2016m0416,Jauzac2017m0717}.
A complete definition of the CLASH sample is given in
\citet{Postman+2012CLASH}.

\citet{Donahue2016clash} presented uniformly estimated X-ray
morphological statistics for the full sample of 25 CLASH clusters using
{\em Chandra} X-ray observations.
Comparing the X-ray morphological properties
between the two CLASH subsamples, they found that the X-ray-selected
subsample is slightly rounder (typical axis ratio of $\sim 0.9$ versus
$\sim 0.8$ measured within an aperture radius of $350\,\kpch$), more
centrally concentrated, and has smaller centroid shifts than the
lensing-selected subsample. In order to understand how typical CLASH
clusters are relative to a complete set of simulated clusters of similar
mass, \citet{Donahue2016clash} also compared high-mass halos from
nonradiative simulations \citep{Sembolini2013music2,Meneghetti2014clash}
with the CLASH {\em Chandra} observations. They found that, overall,
both X-ray- and lensing-selected CLASH clusters are rounder than the
simulated clusters in terms of the X-ray axis ratio.

Following \citet{Umetsu2014clash,Umetsu2016clash}, we use the location 
of the brightest cluster galaxy (BCG) as the cluster center (Table \ref{tab:sample}).
On average, the sample exhibits a small positional offset between the
BCG and X-ray peak, characterized by an rms offset of
$\sigma_\mathrm{off}\simeq 30\,\kpch$
\citep[][]{Umetsu2014clash}.
For the X-ray-selected subsample, the
offset is even smaller,
$\sigma_\mathrm{off}\simeq 11\,\kpch$ \citep{Umetsu2014clash}, because
of the CLASH selection function.
This level of centering offset is sufficiently small compared 
to the range of cluster radii of interest (e.g.,
 $r_\mathrm{2500c}, r_\mathrm{500c}, r_\mathrm{200c}, r_\mathrm{200m}$),
 as well as to the effective resolution of our mapmaking
 (see Section \ref{subsec:maps}).
Accordingly, smoothing from the miscentering effects
\citep[see][]{Johnston+2007a,Umetsu+2011stack} is not expected to
significantly impact our mass, concentration, and shape measurements for
our cluster sample.

 Since the clusters in our sample are highly massive
 \citep[$M_\mathrm{200c}\simeq 14\times 10^{14}\Msun$;][]{Umetsu2016clash}, they 
 can strongly lens background galaxies into multiple images or giant
 arcs in their central region. On the basis of deep multiband
 {\em  HST} images, \citet{Zitrin2015clash} identified many secure sets
 of multiple-image systems in all CLASH clusters except RX~J1532.9+3021
 \citep[see][]{Zitrin2015clash}, our least massive cluster with
 $M_\mathrm{200c}\sim 6\times 10^{14}\Msun$
 \citep[Tables 2 and 3 of][]{Umetsu2016clash}. 
 For our sample, we find
 a median effective Einstein radius\footnote{The effective Einstein
 radius is defined as $\theta_\mathrm{Ein}=\sqrt{A_\mathrm{c}/\pi}$ with
 $A_\mathrm{c}$ the area enclosed within the critical curves.} of
 $\simeq 22\arcsec$ \citep{Umetsu2016clash} for a source redshift of
 $z=2$. We refer to our companion papers
 \citep{Chiu2018clump3d,Sereno2018clump3d}
 for joint analyses of central strong-lensing and 2D weak-lensing 
 constraints. 

\subsection{Photometry and Background Selection}
 \label{subsec:back}

A secure background selection is critical for a cluster weak-lensing
analysis, so that unlensed foreground and member galaxies do not dilute
the lensing signal \citep{BTU+05,Medezinski+2010,Medezinski2017src,Okabe+2013}.
\citet{Umetsu2014clash} employed the color-color (CC) selection method
of \citet{Medezinski+2010} to identify background galaxy populations,
typically using the Subaru/Suprime-Cam $BR_\mathrm{C}z'$
photometry where available
\citep[for a summary, see Table 1 of][]{Umetsu2014clash},
which spans the full optical wavelength range.
The photometric zero points were precisely calibrated to an accuracy of
$\sim 0.01$\,mag, using the
{\em HST} photometry of cluster elliptical galaxies and with a set of
galaxies having spectroscopic redshifts from the CLASH-VLT large
spectroscopic program with VIMOS
\citep[e.g.,][]{Biviano2013,Rosati2014VLT,Annunziatella2015,Grillo2015,Balestra2016m0416},
which obtained thousands of spectroscopic redshifts for cluster 
members and intervening galaxies along the line of sight,
including lensed background galaxies.

The CC-selection method has been calibrated with evolutionary color
tracks of galaxies
\citep{Kotulla2009,Medezinski+2010,Medezinski+2011,Medezinski2017src,Hanami+2012}
and with well-calibrated photometric-redshift (photo-$z$) catalogs such
as COSMOS \citep{Ilbert+2009COSMOS,Laigle2016cosmos}.  
For details of our CC selection,
we refer the reader to Section 4.4 of \citet{Umetsu2014clash} and
Section 3.2 of \citet{Umetsu2016clash}.

For shear measurements, \citet{Umetsu2014clash} combined two distinct
background populations that encompass the red and blue branches of
field galaxies in the CC plane, having redshift distributions
peaked around $z\sim 1$ and $z\sim 2$, respectively
\citep{Medezinski+2010,Medezinski+2011}.

\citet{Umetsu2014clash} used flux-limited samples of red background
galaxies at $z\sim 1$ for magnification bias measurements.   
Faint magnitude cuts were applied in the reddest band
to avoid incompleteness near the detection limit.
Our CC selection avoids incompleteness at the
faint end in the bluer bands
\citep{Hildebrandt2015} because we have
correspondingly deeper photometry in the bluer bands. 
As discussed in \citet{Umetsu2016clash}, this is by design to detect
faint red galaxies as proposed by \citet{Broadhurst1995}.

The mean lensing depths $\langle \beta\rangle$ and
$\langle\beta^2\rangle$ of the respective background samples were 
estimated using the photo-$z$'s of individual galaxies determined with the
BPZ code \citep{Benitez2000,Benitez+2004} 
from our point-spread-function (PSF) corrected photometry typically in five
Suprime-Cam bands \citep[see Table 1 of][]{Umetsu2014clash}.    
An excellent statistical agreement was obtained between the
$\langle\beta\rangle$ estimates from the BPZ measurements in the CLASH
fields and those from the COSMOS photo-$z$ catalog,
with a median offset of $0.27\percent$ and a field-to-field
scatter of $5.0\percent$ \citep{Umetsu2014clash}.

\subsection{Two-dimensional Weak-lensing Shear Analysis}
\label{subsec:shear}

\subsubsection{Reduced Shear Field}
\label{subsubsec:shear}


We use the 2D reduced shear field averaged on a grid as the primary
constraint from our wide-field weak-lensing observations. 
From shape measurements of background galaxies,
the source-averaged reduced shear $g_n=g(\btheta_n)$ is measured 
on a regular Cartesian grid of pixels ($n=1,2,..., \Npix$) as
\begin{equation}
\label{eq:bin_shear} 
g_n
=
\left[
\displaystyle\sum_k
S(\btheta_{(k)},\btheta_n)
w_{(k)}g_{(k)}
\right]
\left[
\displaystyle\sum_{k} 
S(\btheta_{(k)},\btheta_n)w_{(k)}
\right]^{-1} 
\end{equation}
where $S(\btheta,\btheta')$ is a spatial window function,
$g_{(k)}$ is an estimate of $g(\btheta)$ for the $k$th galaxy at
$\btheta_{(k)}$,  
and 
$w_{(k)}$ is its statistical weight,
$w_{(k)} = 1/(\sigma^2_{g(k)}+\alpha^2_g)$,
with $\sigma^2_{g(k)}$ the error variance of $g_{(k)}$.
Following \citet{Umetsu2014clash}, we choose $\alpha_g=0.4$,
a typical value of the mean rms
$(\overline{\sigma_g^2})^{1/2}$ found in Subaru observations
\citep[][]{Umetsu+2009,Oguri2010LoCuSS,Okabe+Smith2016}. 

The theoretical expectation (denoted by a hat symbol) for the estimator
(\ref{eq:bin_shear}) is approximated by \citep{Seitz+Schneider1997,Umetsu2015A1689}
\begin{equation}
\label{eq:g_ave}
\widehat{g}(\btheta_n) \approx \frac{\langle W\rangle_g
 \gamma_\infty(\btheta_n)}{1-f_{W,g} \langle W \rangle_g \kappa_\infty(\btheta_n)},
\end{equation}
where $\langle W\rangle_g$ is the source-averaged relative lensing
strength (Section \ref{subsec:Nz}), and
$f_{W,g}=\langle W^2\rangle_g/\langle W\rangle_g^2$ is a dimensionless
correction factor of the order unity.
The error variance  $\sigma_{g,n}^2$ for $g_n$ is expressed as
\begin{equation}
 \begin{aligned}
\label{eq:bin_shearvar}
\sigma^2_{g,n}=
 \frac{ \sum_k S^2(\btheta_{(k)},\btheta_n)w_{(k)}^2 \sigma^2_{g(k)} }
{ \left[
 {\sum_k S(\btheta_{(k)},\btheta_n)w_{(k)}}
  \right]^{2}}.
  \end{aligned}
\end{equation}
We adopt the top-hat window
of radius $\theta_\mathrm{f}$ \citep{Merten+2009,Umetsu2015A1689},
$S(\btheta,\btheta')=H(\theta_\mathrm{f}-|\btheta-\btheta'|)$, with
$H(x)$ the Heaviside function defined such that
$H(x)=1$ if $x\ge 0$ and $H(x)=0$ otherwise.
The shape-noise covariance matrix for $g_{\alpha,n}=g_\alpha(\btheta_n)$ 
is then given as \citep{Oguri2010LoCuSS}\footnote{In
\citet{Oguri2010LoCuSS}, $\sigma_g$ denotes the per-component dispersion
due to shape noise.} 
\begin{equation}
 \label{eq:Cshape}
 \left(C_g^\mathrm{shape}\right)_{\alpha\beta,mn}
 = 
  \frac{1}{2}\delta_{\alpha\beta} \sigma_{g,m} \sigma_{g,n}
  \xi_{H}(|\btheta_m-\btheta_n|),
\end{equation}
where
the indices $\alpha$ and $\beta$ run over the two components of the
reduced shear ($\alpha,\beta=1,2$),
$\delta_{\alpha\beta}$ denotes the Kronecker delta, and
$\xi_H(x)$ is the autocorrelation of a pillbox of 
radius 
$\theta_\mathrm{f}$ \citep{1999ApJ...514...12W,Park+2003,Umetsu2015A1689}, 
\begin{equation}
\label{eq:xi_H}
 \xi_{H}(x)=\frac{2}{\pi}\left[\cos^{-1}\left(\frac{x}{2\theta_{\rm
  f}}\right)-\left(\frac{x}{2\theta_{\rm
  f}}\right)\sqrt{1-\left(\frac{x}{2\theta_\mathrm{f}}\right)^2}\right]
\end{equation}
for $|x|\le 2\theta_\mathrm{f}$
and $\xi_{H}(x)=0$ for $|x|>2\theta_\mathrm{f}$.

\subsubsection{CLASH Weak-lensing Shear Data}
\label{subsubsec:shearmeas}

In this study, we directly measure 2D reduced shear maps in 20 CLASH
cluster fields using the wide-field shear catalogs obtained by the CLASH 
collaboration \citep{Umetsu2014clash}. 
The shear measurement pipeline of \citet{Umetsu2014clash} is based on
the IMCAT package \citep[][hereafter KSB]{1995ApJ...449..460K} with modifications 
incorporating the improvements developed by \citet{Umetsu+2010CL0024}. 
 
Briefly summarizing, the key feature in our analysis pipeline is that
only those galaxies detected with sufficiently high significance,
$\nu_g>30$, are used to model the isotropic PSF correction as a function
of object size and magnitude.  
Here, $\nu_g$ is the peak detection significance given by the IMCAT
peak-finding algorithm {\sc hfindpeaks}. A very similar procedure was
employed by the LoCuSS collaboration in their weak-lensing study of 50
clusters based on Subaru/Suprime-Cam data \citep{Okabe+Smith2016}. 
Another key feature is that we select those galaxies isolated in
 projection for the shape measurement, reducing the impact of crowding 
 and blending \citep[for details, see][]{Umetsu2014clash}.
After the close-pair rejection, objects with low detection
 significance $\nu_g<10$ were excluded from our analysis.
 All galaxies with usable shape measurements are matched with 
 those in our CC-selected samples of background galaxies
 (Section \ref{subsec:back}), ensuring that each galaxy is detected in
 both the reddest CC-selection band and the shape measurement band.
 Applying conservative selection criteria (Section \ref{subsec:back}),
 \citet{Umetsu2014clash} find 
 a typical surface number density of $n_\mathrm{g}\simeq
 12$\,galaxies\,arcmin$^{-2}$ for their weak-lensing-matched background
 catalogs \citep[][their Table 3]{Umetsu2014clash}.  

In Appendix \ref{appendix:test}, we show the results of our shape 
measurement test based on simulated Subaru/Suprime-Cam images
provided by M. Oguri.
From this test, we find that the reduced shear signal can be recovered
up to $|g_\alpha|\simeq 0.3$ 
with $m_\alpha \simeq -0.05$ of the multiplicative calibration
bias and $|c_\alpha|< 10^{-3}$ of the residual shear offset
(for details, see Appendix \ref{appendix:test}),
where the observed and true values of the reduced shear are related by
\citep[][]{2006MNRAS.368.1323H,2007MNRAS.376...13M}
\begin{equation}
 \label{eq:gcal}
 g_\alpha^\mathrm{obs}=(1+m_\alpha)g_{\alpha}^\mathrm{true}+c_\alpha.
\end{equation}
\citet{Umetsu2014clash} included for each galaxy a shear
calibration factor with $m_1=m_2=-0.05$ to account for residual
calibration. 
The degree of multiplicative bias $m_\alpha$ depends on the seeing
conditions and the PSF properties (Figure \ref{fig:test2}), so that 
the variation with the PSF properties limits the shear calibration
accuracy to $\delta m_\alpha\sim 0.05$ \citep[][Section 3.3]{Umetsu+2012}.
We note that the same simulation data set was used by the LoCuSS
collaboration \citep{Okabe+Smith2016} to test their shape measurement
pipeline, and \citet{Okabe+Smith2016} found a similar level
of shear calibration bias ($m_\alpha\sim -0.03$) from their mock
observations.


For all cluster fields in our sample, the estimated values for
$\langle\beta\rangle_g$ and $f_{W,g}$ are 
summarized in Table 3 of \citet{Umetsu2014clash}. The calibration
uncertainty in $\langle \beta\rangle_g$ is marginalized over in our
joint analysis of shear and magnification data (Section \ref{subsec:massrec}).

\subsection{Weak-lensing Magnification Analysis}
\label{subsec:magbias}

\subsubsection{Flux Magnification Bias}

Lensing magnification can influence the observed surface
number density of background sources, amplifying their apparent 
fluxes and expanding the area of sky
\citep{Hildebrandt+2011,Umetsu+2011,Morrison+2012,DES2016wlmag,Tudorica2017sparcs}.
The former effect
increases the number of sources detectable above the limiting
flux, whereas the latter reduces the effective observing area in
the source plane, reducing the number of sources per solid angle. 
The net effect is known as magnification bias
\citep{1995ApJ...438...49B} and depends on the slope 
of the intrinsic source luminosity function.
Since a given flux limit corresponds to different luminosities
at different source redshifts, number counts of distinctly
different source populations probe different 
regimes of magnification bias \citep{Umetsu2013}.

Deep multiband photometry can be used to sample the faint
end of the luminosity function of red quiescent galaxies lying at
$z\sim 1$ \citep[e.g.,][]{Ilbert2010}.
For such a source population,
the effect of magnification bias is dominated by the
geometric area distortion, because there are relatively few fainter
galaxies that can be magnified into the flux-limited sample. 
This effect results in a net depletion of source
counts, a phenomenon known as negative magnification bias or
weak-lensing depletion
\citep[e.g.,][]{Broadhurst1995,1998ApJ...501..539T,BTU+05,UB2008,Umetsu+2011,Umetsu+2012,Umetsu2014clash,Umetsu2015A1689,Ford2012,Coe+2012A2261,Medezinski+2013,Radovich2015planck,Ziparo2016locuss,Wong2017}. 
A practical advantage of this technique,  at the expense of very
deep multicolor imaging, is that the effect is
not sensitive to the exact form of the source luminosity
function \citep{Umetsu2014clash}.

In the weak-lensing regime,
the shift in magnitude $\delta m=2.5\log_{10}\mu$ due to magnification
is small compared to the range in which the slope of the luminosity function
varies. The number counts can then be approximated by a power law at the 
limiting magnitude $m_\mathrm{lim}$. The expectation value (denoted by a
hat symbol) for the lensed counts at source redshift $z$ is then
expressed as \citep{1995ApJ...438...49B}
\begin{equation} 
 \label{eq:magbias}
\widehat{N}_\mu(\btheta,z| <m_\mathrm{lim}) = \overline{N}_\mu(z|<m_\mathrm{lim})
   \,\Delta_\mu(\btheta,z)^{1-2.5s}
\end{equation}  
with $\overline{N}_\mu(z|<m_\mathrm{lim})$ the unlensed mean counts per cell
and $s$ the logarithmic count slope evaluated at $m=m_\mathrm{lim}$\footnote{In the literature,
$\alpha\equiv -d\log{\overline{N}(>F)}/d\log{F}=2.5s$ in terms of  the
limiting flux $F$ is often used instead of $s$ \citep[e.g.,][]{Umetsu2015A1689}.},
\begin{equation}
 s = \frac{d\log_{10}\overline{N}(z|<m)}{dm}\Bigg|_{m_\mathrm{lim}}.
\end{equation}
A count depletion (enhancement) results when $s < 0.4$ ($> 0.4$).

Following \citet{Umetsu2014clash},
we interpret the observed source-averaged magnification bias as 
\begin{equation}
 \label{eq:bmu_approx}
   \begin{aligned}
    \widehat{b}_\mu(\btheta) &\equiv
    \frac{\widehat{N}_\mu(\btheta| <m_\mathrm{lim})}{\overline{N}_\mu(<m_\mathrm{lim})}
    \approx   \Delta_\mu(\btheta)^{1-2.5s_\mathrm{eff}},\\
    \Delta_\mu(\btheta) &= \frac{\int_0^\infty\!dz\,\overline{N}_\mu(z|<m_\mathrm{lim})\Delta_\mu(\btheta,z)}
    {\int_0^\infty\!dz\,\overline{N}_\mu(z|<m_\mathrm{lim})}\\
    &\approx \left[1-\langle W\rangle_\mu
    \kappa_\infty(\btheta)\right]^2- \langle W\rangle_\mu^2|\gamma_\infty(\btheta)|^2,
   \end{aligned}
\end{equation}
where
$\overline{N}_\mu(<m_\mathrm{lim})=\int_0^\infty\!dz\,\overline{N}_{\mu}(z| <m_\mathrm{lim})$,
$s_\mathrm{eff}=d\log_{10}\overline{N}_{\mu}(<m)/dm\big|_{m_\mathrm{lim}}$,
and $\langle W\rangle_\mu$  is the source-averaged relative lensing
strength (Section \ref{subsec:Nz}).
Equation (\ref{eq:bmu_approx}) gives a good approximation for depleted
populations with $s_\mathrm{eff}\ll 0.4$
\citep[for details, see Appendix A.2 of][]{Umetsu2013}.
For simplicity, we write
$N_\mu(\btheta) = N_\mu(\btheta|<m_\mathrm{lim})$ and
$\overline{N}_\mu = \overline{N}_\mu(<m_\mathrm{lim})$.
In the weak-lensing limit, Equation (\ref{eq:bmu_approx}) reads
$\widehat{b}_\mu -1\approx (5s_\mathrm{eff}-2)\langle W\rangle_\mu \kappa_\infty$.


We azimuthally average the observed counts $N_\mu(\btheta)$ in
clustercentric, circular annuli and calculate  the surface number 
density $\{n_{\mu,i}\}_{i=1}^{\Nbin}$ of background galaxies as
\citep{Umetsu2015A1689,Umetsu2016clash}
\begin{equation}
 \label{eq:nb}
  n_{\mu,i} =
  \frac{1}{(1-f_{\mathrm{mask},i})\Omega_\mathrm{cell}}
  \sum_{m} {\cal P}_{im}N_\mu(\btheta_m)
\end{equation}
with 
$\Omega_\mathrm{cell}$ the solid angle per cell
and
${\cal P}_{im}=(\sum_m A_{mi})^{-1}A_{mi}$ the projection
matrix normalized in each annulus by $\sum_m {\cal P}_{im}=1$.
Here, $A_{mi}$ is the area fraction of 
the $m$th cell lying within the $i$th radial bin  ($0\le A_{mi} \le 1$), 
and $f_{\mathrm{mask},i}$ is the mask correction factor for
the $i$th radial bin
due to saturated objects, foreground and cluster galaxies, and bad
pixels \citep[for details, see Section 3.4 of][]{Umetsu2016clash}.

The theoretical expectation for the estimator (\ref{eq:nb}) is 
\begin{equation}
 \label{eq:nb_th}
\widehat{n}_{\mu,i} =
\overline{n}_\mu
\sum_m {\cal P}_{im}\Delta_\mu(\btheta_m)^{1-2.5s_\mathrm{eff}}
\end{equation}
with $\overline{n}_\mu =\overline{N}_\mu/\Omega_\mathrm{cell}$.


The choice of annular constraints rather than pixelated ones is mainly
because magnification constraints are by far noisier than shear
measurements, especially due to the local clustering noise
\citep[see][]{UB2008}. An optimal choice of the resolution is to have
per-element signal-to-noise ratio of the order  
unity or above. This is satisfied by azimuthally averaging noisy count
measurements, while it allows us to  estimate the variance due
to angular clustering at large clustercentric distances.

\subsubsection{CLASH Weak-lensing Magnification Data}
\label{subsubsec:clash_magbias}


We use the CLASH weak-lensing magnification measurements
obtained using flux-limited samples of red background galaxies as
published in \citet{Umetsu2014clash}.
They measured the magnification effects in $\Nbin=10$ log-spaced 
circular annuli centered on the cluster. The radial bins range from
$0.9\arcmin$ to $16\arcmin$ for all clusters,
except $0.9\arcmin$ to $14\arcmin$
for RX~J2248.7$-$4431 observed with ESO/WFI.
Our magnification analysis begins at
$\theta_\mathrm{min}=0.9\arcmin$,
which is sufficiently large compared to 
the range of effective Einstein radii for our sample
\citep{Zitrin2015clash,Umetsu2016clash}. 
The magnification profiles
$\{n_{\mu,i}\}_{i=1}^{\Nbin}$ used in the present
work are presented in Figure 2 of \citet{Umetsu2014clash}. 

Here we briefly describe the magnification analysis
performed in \citet{Umetsu2014clash}. 
Their analysis was limited to the $24\arcmin\times 24\arcmin$ region
centered on the cluster. 
They accounted for the Poisson, intrinsic
clustering, and additional systematic contributions to the total
uncertainty $\sigma_\mu$. 
The clustering noise term $\sigma_{\mu,i}^\mathrm{int}$ was estimated in
each circular annulus from the variance due to variations of the counts
along the azimuthal direction.  
Besides, a positive tail of $>\nu\sigma$ cells with $\nu=2.5$ was
removed in each circular annulus by iterative $\sigma$ clipping to
reduce the bias due to intrinsic angular clustering of red galaxies.
We checked that the clipping threshold chosen is sufficiently high
compared to the maximum variations of the magnification signal due to
halo ellipticity (typically,
$|\delta\kappa(\theta)|/\langle\kappa(\theta)\rangle\simlt 0.5$ for a
projected halo axis ratio of $\ge 0.6$).
The Poisson noise term $\sigma_{\mu,i}^\mathrm{stat}$ was estimated from 
the clipped mean counts in each annulus. 
The difference between the mean counts estimated with and without 
$\sigma$ clipping was taken as a systematic error, 
$\sigma_{\mu,i}^\mathrm{sys}=|n_{\mu,i}^{(\nu)}-n_{\mu,i}^{(\infty)}|/\nu$,
where $n_{\nu,i}^{(\nu)}$ and $n_{\mu,i}^{(\infty)}$ denote the
clipped and unclipped mean counts in the $i$th annulus, respectively. 
Finally, these errors were combined in quadrature as 
\begin{equation}
 \label{eq:sigma_mu}
 \sigma_{\mu,i}^2 = (\sigma_{\mu,i}^\mathrm{int})^2 +
  (\sigma_{\mu,i}^\mathrm{stat})^2 +(\sigma_{\mu,i}^\mathrm{sys})^2.
\end{equation}
Our magnification bias measurements are stable and insensitive
to the particular choice of $\nu$ because of the inclusion of the
$\sigma_\mu^\mathrm{sys}$ term in the error analysis.

Masking of observed sky was accounted and corrected for using the method
of \citet[][Method B of Appendix A]{Umetsu+2011}, which can be fully
automated  
once the configuration parameters of SExtractor \citep{SExtractor} are 
optimally tuned \citep{Umetsu+2011,Umetsu2014clash}.  
\citet{Chiu2016magbias} adopted this method to estimate the
masked area fraction in their magnification analysis and
found that the SExtractor configuration of \citet{Umetsu+2011} is optimal for
their data taken with Megacam on the Magellan Clay telescope.

The count normalization and slope parameters 
$(\overline{n}_\mu, s_\mathrm{eff})$ were estimated
in the cluster outskirts \citep{Umetsu2014clash}\footnote{As discussed
in \citet[][their Section 7.4.2]{Umetsu2014clash}, the 2-halo term does
not cause bias in the reconstruction, because the range of the uniform
prior on $\overline{n}_\mu$ is sufficiently wide.}.     
The mask-corrected magnification bias profile 
$b_{\mu,i}=n_{\mu,i}/\overline{n}_\mu$ is
proportional to
$(1-f_\mathrm{mask,back})/(1-f_{\mathrm{mask},i})\equiv 1+\Delta f_{\mathrm{mask},i}$
with  $f_{\mathrm{mask,back}}$
estimated in the background region \citep[see][]{Umetsu2014clash,Umetsu2016clash}. 
The effect of the mask correction is thus sensitive to the
difference of the $f_\mathrm{mask}$ values, 
which is insensitive to the particular choice of the SExtractor
configuration parameters. 
The typical variation of $\Delta f_{\mathrm{mask},i}$ across the full
radial range is $\sim 5\percent$
\citep[see also][]{Chiu2016magbias},
much smaller than the typical magnification signal
$\delta n_\mu/\overline{n}_\mu\sim -0.3$
in the innermost bin.
Accordingly, the systematic uncertainty on the 
mask correction is likely negligible.

The estimated values and errors for 
$\langle \beta\rangle_\mu$, $\overline{n}_\mu$, and $s_\mathrm{eff}$ are
summarized in Table 4 of \citet{Umetsu2014clash}.
The values of $s_\mathrm{eff}$ span the range $[0.11, 0.20]$ with a mean
of $\langle s_\mathrm{eff}\rangle = 0.153$ and  
a typical fractional uncertainty of $33\percent$ per cluster field.
We marginalize over the calibration parameters ($\langle
\beta\rangle_\mu, \overline{n}_\mu, s_\mathrm{eff}$) for each cluster in
our joint likelihood analysis of shear and magnification (Section
\ref{subsec:massrec}).

\subsection{Mass Reconstruction Algorithm}
\label{subsec:massrec}

To perform a mass reconstruction, we use the inversion algorithm
developed by \citet{Umetsu2015A1689}, who generalized the cluster
lensing mass inversion ({\sc clumi}) code of \citet{Umetsu2013} 
\citep[see also][]{Umetsu+2011}
into a 2D description of the pixelized mass distribution.
This free-form method combines a spatial shear pattern
($g_1(\btheta), g_2(\btheta)$) with azimuthally averaged
magnification measurements $\{n_{\mu,i}\}_{i=1}^{\Nbin}$,
which impose a set of azimuthally integrated constraints on the
$\Sigma$ field, thus effectively breaking the 
mass-sheet degeneracy.
The {\sc clumi}-2D algorithm takes full account of the nonlinear 
subcritical regime of gravitational lensing properties.


According to the Bayes' theorem, 
given a model $\blambda$ and observed data $\bd$, the
joint posterior probability  $P(\blambda|\bd)$ is proportional to the
product of the likelihood ${\cal L}(\blambda)\equiv P(\bd|\blambda)$ and
the prior probability $P(\blambda)$.
In our inversion problem, $\blambda$ represents a signal vector
containing the pixelized mass coefficients
$\bs=\{\Sigma_n\}_{n=1}^{\Npix}$ (Section \ref{subsec:massmodel})
and calibration nuisance parameters $\bc$ (Section \ref{subsubsec:calib}), 
so that $\blambda\equiv (\bs,\bc)$.

We write the likelihood function ${\cal L}$ for combined weak-lensing data
$\bd$ as a product of the two separate likelihoods, 
${\cal L}={\cal L}_g {\cal L}_\mu$ with ${\cal L}_g$ and 
${\cal L}_\mu$ the likelihood functions for shear and magnification,
respectively.
This implicitly assumes that the cross-covariance between shear and
magnification due to projected uncorrelated large-scale structure (LSS)
is ignored (see Sections \ref{subsubsec:lg} and \ref{subsubsec:lmu}). In our
measurements, the uncertainty is dominated by observational measurement
errors at all scales, as shown in Figure 1 of \citet{Umetsu2016clash}, so
that the contribution from this cosmic cross-term is not expected to  
significantly impact our results.
We assume that the observational errors follow a Gaussian distribution,  
so that ${\cal L}\propto \exp(-\chi^2/2)$, with $\chi^2$ the standard
misfit statistic.

\subsubsection{Shear Log-likelihood Function}
\label{subsubsec:lg}

The log-likelihood function 
$l_g\equiv -\ln{\cal L}_g$ for 2D shear data can be written 
(ignoring constant terms) as
\citep{Oguri2010LoCuSS} 
\begin{equation}
 \begin{aligned}
l_g(\blambda) =& \frac{1}{2}
 \sum_{m,n=1}^{\Npix}
 \sum_{\alpha,\beta=1}^{2}
[g_{\alpha,m}-\widehat{g}_{\alpha,m}(\blambda)]
\left({\cal W}_g\right)_{\alpha\beta,mn}\\
  &\times [g_{\beta,n}-\widehat{g}_{\beta,n}(\blambda)],
  \end{aligned}
\end{equation}
where  $\widehat{g}_{\alpha,m}(\blambda)$
is the theoretical expectation for $g_{\alpha,m}=g_\alpha(\btheta_m)$,
and $({\cal W}_g)_{\alpha\beta,mn}$ is the shear weight matrix,
\begin{equation}
  \left({\cal W}_g\right)_{\alpha\beta,mn} = M_m M_n \left(C_g^{-1}\right)_{\alpha\beta,mn}.
\end{equation}
Here, $M_m$ is a mask weight, defined such that $M_m=0$ if the $m$th cell is
masked out and $M_m=1$ otherwise, and $C_g$ is the shear covariance matrix.
We account for contributions from the shape covariance
$C_g^\mathrm{shape}$ and the cosmic covariance
$C_g^\mathrm{lss}$ due to uncorrelated LSS projected along the line of
sight as   
\begin{equation}
 (C_g)_{\alpha\beta,mn} = (C_g^\mathrm{shape})_{\alpha\beta,mn} + (C_g^\mathrm{lss})_{\alpha\beta,mn},
\end{equation}
where
$(C_g^\mathrm{lss})_{\alpha\beta,mn}=\xi^\mathrm{lss}_{\alpha\beta}(|\btheta_m-\btheta_n|)$
with $\xi^\mathrm{lss}_{\alpha\beta}=\xi^\mathrm{lss}_{\beta\alpha}$ ($\alpha,\beta=1,2$)
the cosmic shear correlation function
\citep{Hu+White2001,Oguri2010LoCuSS}.
We compute the elements of the $C_g^\mathrm{lss}$ matrix for a given
source population, using the nonlinear matter power spectrum
of \citet{Smith+2003halofit} for
the {\em Wilkinson Microwave Anisotropy Probe} ({\em WMAP}) seven-year
cosmology \citep{Komatsu+2011WMAP7}.
For each cluster, we use the effective mean source redshift
\citep[$\overline{z}_\mathrm{eff}$; Table 3 of][]{Umetsu2014clash}
estimated with our multiband photometric redshifts.

\subsubsection{Magnification Log-likelihood Function}
\label{subsubsec:lmu}

Similarly, the log-likelihood function for magnification bias data 
$l_{\mu}\equiv -\ln{\cal L}_\mu$ is written as \citep{Umetsu2015A1689}
\begin{equation}
l_\mu(\blambda)= \frac{1}{2}\sum_{i=1}^{\Nbin}
[n_{\mu,i}-\widehat{n}_{\mu,i}(\blambda)]
\left({\cal W}_\mu\right)_{ij}
[n_{\mu,j}-\widehat{n}_{\mu,j}(\blambda)],
\end{equation}
where $\widehat{n}_{\mu,i}(\blambda)$ is the theoretical expectation for the
observed counts $n_{\mu,i}$,
and
$({\cal W}_\mu)_{ij}$ is the magnification weight matrix,
${\cal W}_\mu = C^{-1}_\mu$, with $C_\mu$ the corresponding covariance matrix,
\begin{equation}
 \label{eq:C_mu}
  (C_\mu)_{ij} = \sigma_\mu^2\delta_{ij} + (C_\mu^\mathrm{lss})_{ij}.
\end{equation}
The diagonal term in Equation (\ref{eq:C_mu}) is responsible for
the observational errors, and the bin-to-bin covariance matrix
$C_\mu^\mathrm{lss}$ accounts for the 
cosmic noise contribution due to projected uncorrelated LSS,
$(C_\mu^\mathrm{lss})_{ij} = \left[(5s_\mathrm{eff}-2) \overline{n}_\mu\right]^2 (C_\kappa^\mathrm{lss})_{ij}$,
where $C_\kappa^\mathrm{lss}$ is the cosmic convergence matrix
\citep{Umetsu+2011stack}.
We compute the elements of the $C_\kappa^\mathrm{lss}$ matrix   
for a given source redshift
\citep[$\overline{z}_\mathrm{eff}$; Table 4 of][]{Umetsu2014clash} in a
similar manner to those of the $C_g^\mathrm{lss}$ matrix (Section
\ref{subsubsec:lg}).  We evaluate the $C_\mu^\mathrm{lss}$ matrix by
fixing the values of $\overline{n}_\mu$ and $s_\mathrm{eff}$ to the
observed ones (Section \ref{subsubsec:clash_magbias}).

The $l_\mu$ function sets azimuthally integrated constraints on the
projected mass distribution and  provides the otherwise unconstrained
normalization of $\Sigma(\bR)$ over a set of concentric annuli where
magnification measurements are available.    
In this algorithm, no assumption is made about the azimuthal symmetry or 
isotropy of $\Sigma(\bR)$. 
We use Monte Carlo integration to compute the projection matrix
${\cal P}_{im}$ (Equation \ref{eq:nb})
of size $\Nbin \times \Npix$, which is needed to predict 
$\{\widehat{n}_{\mu,i}(\blambda)\}_{i=1}^{\Nbin}$
for a given model $\blambda=(\bs,\bc)$.

\subsubsection{Calibration Parameters}
\label{subsubsec:calib}

We account for the calibration uncertainty in the
observational nuisance parameters,
\begin{equation}
\label{eq:calib}
\bc=(\langle W\rangle_g, f_{W,g}, \langle W\rangle_\mu,
\overline{n}_\mu, s_\mathrm{eff}).
\end{equation}
To this end, we include in our joint-likelihood analysis 
Gaussian priors on $\bc$
with mean values and errors estimated from data.

\subsubsection{Best-fit Solution and Covariance Matrix}
\label{subsubsec:cmat}

The log-posterior function
$F(\blambda) =-\ln{P(\blambda|\bd)}$ is expressed as a linear sum of the
log-likelihood and prior terms. 
For each cluster,
we find the global maximum of the joint posterior probability
distribution over $\blambda$, by minimizing $F(\blambda)$   
with respect to $\blambda$. 
In the {\sc clumi}-2D implementation of \citet{Umetsu2015A1689}, we use
the conjugate-gradient method  \citep{1992nrfa.book.....P} to find the
global solution $\widehat{\blambda}$.
We employ an  analytic expression for the gradient function $\bnabla F(\blambda)$ obtained
in the nonlinear, subcritical regime (Appendix \ref{appendix:DD}).

To quantify the reconstruction errors, we evaluate the Fisher matrix at
$\blambda=\widehat{\blambda}$ as 
\begin{equation}
 \label{eq:Fisher}
{\cal F}_{pp'} = 
\left\langle \frac{\partial^2 F(\blambda)}{\partial \lambda_p \partial \lambda_{p'}}
\right\rangle\Bigg|_{\widehat{\blambda}}
\end{equation}
where the angular brackets denote an ensemble average, and the
indices $(p,p')$ run over all model parameters $\blambda=(\bs,\bc)$.
The error covariance matrix of the reconstructed parameters
is obtained by
$C = {\cal F}^{-1}$.
We note that the reconstructed mass pixels are correlated primarily
because the relation between the shear and convergence is 
nonlocal (Equation (\ref{eq:kappa2gamma})). Additionally, the effects of
spatial averaging (Equation (\ref{eq:xi_H}))
and cosmic noise (Sections \ref{subsubsec:lg} and \ref{subsubsec:lmu})
produce a covariance between different pixels. In our analysis, the
effects of correlated errors are modeled analytically
(i.e., $\xi_H, \xi^\mathrm{lss}, C_\kappa^\mathrm{lss}$).


\subsection{Major Differences from Previous Work}
\label{subsec:diff}

The present sample of 20 CLASH clusters has been analyzed by
\citet{Umetsu2014clash} and \citet{Umetsu2016clash} using high-quality
CLASH lensing data sets.
In what follows, we summarize the major differences of our analysis from
these previous studies, which focused on reconstructing $\Sigma(R)$
profiles from azimuthally averaged lensing measurements. 

First of all, this work represents a 2D generalization of the
\citet{Umetsu2014clash} weak-lensing analysis based on their
background-selected shear catalogs.
Both studies use
identical sets of azimuthally averaged magnification constraints (Section
\ref{subsec:magbias}) as input for respective mass
reconstructions.
In addition to the measurement error and cosmic noise contributions, 
\citet{Umetsu2014clash} accounted for systematic uncertainties
$C^\mathrm{sys}$ due to the residual mass-sheet degeneracy.
This uncertainty was estimated in each $\Sigma$ bin as a difference 
between the global (joint) and local (marginal) posterior solutions.
On the other hand, owing to the large number of parameters involved
($48^2+5=2309$), we do not directly sample posterior probability
distributions (Section \ref{subsubsec:cmat}), and thus we are not able
to include the $C^\mathrm{sys}$ term in the present analysis. However, as
we will see in Section \ref{subsec:sys}, our cluster mass measurements 
are highly consistent with those of \citet{Umetsu2014clash}, with no
evidence for systematic offsets in the mass determinations.

\citet{Umetsu2016clash} combined the wide-field shear and 
magnification constraints of \citet{Umetsu2014clash}
with central {\em HST} constraints in the form of the enclosed projected
mass $M_\mathrm{2D}(<\theta)$, 
which was derived from detailed mass models of
\citet{Zitrin2015clash} based on
their joint analysis of {\em HST} strong- and weak-lensing data sets. 
The strong-lensing, weak-lensing shear and magnification constraints
were combined a posteriori to reconstruct azimuthally averaged
$\Sigma(R)$ profiles for the 20 individual clusters.
In addition to the inclusion of the {\em HST}
data, an important difference between the two studies is that
\citet{Umetsu2016clash} included the intrinsic signal covariance matrix
$C^\mathrm{int}$ \citep{Gruen2015}
in their error analysis, as well as the $C^\mathrm{sys}$ term.
Here, the $C^\mathrm{int}$ matrix accounts for the variations of the
projected cluster lensing signal due to the intrinsic scatter in 
the $c$--$M$ relation\footnote{As noted by \citet{Umetsu2016clash},
when simultaneously determining the mass and concentration for an
individual cluster, the contribution from the intrinsic $c$--$M$
scatter should be excluded from $C^\mathrm{int}$. We note that the
effect of the $c$--$M$ scatter becomes important only at
$\theta\simlt 2\arcmin$ \citep{Gruen2015}.},
the halo asphericity, and the presence of
correlated halos \citep{Gruen2015}. This contribution is particularly
important at small cluster radii, and hence in the inner {\em HST}
region \citep[see Figure 1 of][]{Umetsu2016clash}.
In the CLUMP-3D program, we explicitly account for the effects of
triaxiality (in particular, halo elongation along the line of sight) in
the mass modeling by simultaneously constraining the cluster 
mass, concentration, triaxial shape, and orientation from Bayesian
inference \citep{Sereno2017clump3d}. We defer such full triaxial
analyses to our companion papers
\citep{Chiu2018clump3d,Sereno2018clump3d}.

\section{Results}
\label{sec:results}

\subsection{Weak-lensing Mapmaking}
\label{subsec:maps}

Following the methodology outlined in Section \ref{sec:wl}, we analyze
our weak-lensing shear and magnification data sets and perform
reconstructions of the 2D lensing fields for our sample of 20 CLASH
clusters.
For magnification measurements, we have 10 azimuthally averaged
constraints $\{n_{\mu,i}\}_{i=1}^{\Nbin}$ in log-spaced clustercentric
annuli (Section \ref{subsec:magbias}), as obtained by
\citet{Umetsu2014clash}.  
To derive reduced shear maps $(g_1(\btheta),g_2(\btheta))$,
we use a top-hat window of $\theta_\mathrm{f}=0.4\arcmin$ (Section
\ref{subsec:shear}) to average
galaxy ellipticities into a regular grid of $\Npix=48\times 48$ pixels,
each with $\Delta\theta=0.5\arcmin$ spacing.
The shear grid covers a $24\arcmin\times 24\arcmin$ region centered on
the cluster (Table \ref{tab:sample}), where \citet{Umetsu2014clash}
obtained the magnification measurements.
The filter size corresponds to an effective resolution of 
$2D_\mathrm{l}\theta_\mathrm{f}\simeq 180\,\kpch$
at the median redshift of the sample, $z=0.377$. 
To avoid potential systematic errors,
we exclude from our analysis those pixels having no usable
background galaxies and the innermost central pixels 
where $\Sigma$ can be greater than or close to the critical
value $\Sigma_\mathrm{c}$, ensuring that all of the measurements are in
the subcritical regime.

\begin{figure}[!htb] 
  \begin{center}
   \includegraphics[scale=0.45, angle=0, clip]{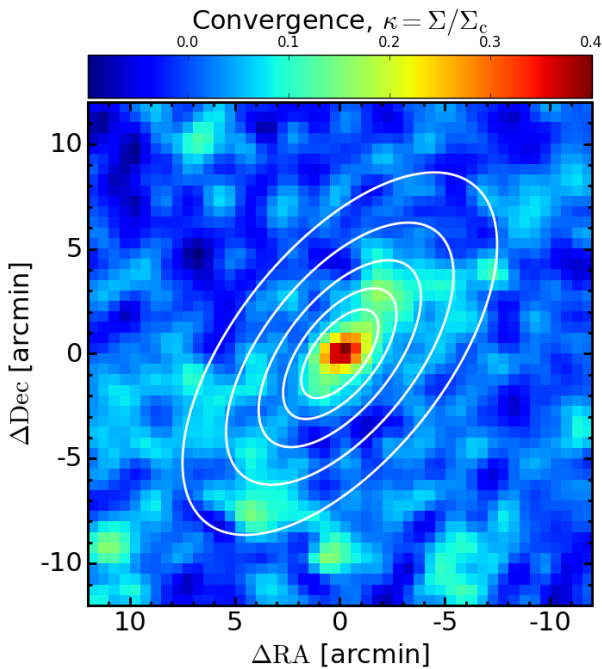} 
  \end{center}
\caption{
 \label{fig:M0329}
Example of our weak-lensing mass reconstruction shown for the cluster
 MACS~J0329.7$-$0211 at $z=0.45$. The mass map is
 $24\arcmin\times 24\arcmin$ in size (5.9\,$h^{-1}$\,proper Mpc on a side)
 and centered on the BCG. The color bar indicates the lensing
 convergence
 $\kappa(\btheta)=\langle\Sigma_\mathrm{c}^{-1}\rangle\Sigma(\btheta)$,
 scaled to the mean depth of Subaru weak-lensing observations,
 $1/\langle\Sigma_\mathrm{c}^{-1}\rangle=3.65\times 10^{15}hM_\odot$\,Mpc$^{-2}$.
For visualization purposes, the mass map is smoothed with a
 $1.2\arcmin$\,FWHM Gaussian. North is to the top, east to the
 left. Elliptical isodensity contours of the best-fit elliptical
 Navarro--Frenk--White model  
 are shown in white. The contour levels are logarithmically spaced from
 $\kappa=0.01$ to $\kappa=0.1$.
 }
\end{figure}

For each cluster, we pixelize the $\kappa_\infty$ and $\gamma_\infty$
fields on a $\Npix=48\times 48$ grid covering the central
$24\arcmin\times 24\arcmin$ region.
The model $\blambda=(\bs,\bc)$ is specified by 
$\Npix=48^2$ mass coefficients,
$\bs=\{\Sigma_n\}_{n=1}^{\Npix}$,
and a set of five calibration parameters $\bc$ (Equation (\ref{eq:calib}))
to marginalize over. 
We utilize the FFTW implementation
of fast Fourier transforms (FFTs) to compute
$\gamma_\infty(\btheta)$ from $\kappa_\infty(\btheta)$ using Equation
(\ref{eq:shear2m}).
To minimize spurious aliasing effects from the periodic boundary
condition, the  maps are zero-padded to twice the original length
in each spatial dimension
\cite[e.g.,][]{1998ApJ...506...64S,UB2008,Umetsu2015A1689}.
In Figure \ref{fig:M0329}, we show the reconstructed $\Sigma$ field
centered on MACS~J0329.7$-$0211 as an example of our weak-lensing mass
reconstruction.


\subsection{Characterizing the Cluster Mass Distribution}
\label{subsec:eNFWfit}


\begin{figure*}[!htb] 
 \begin{center}
  \includegraphics[scale=0.55, angle=0, clip]{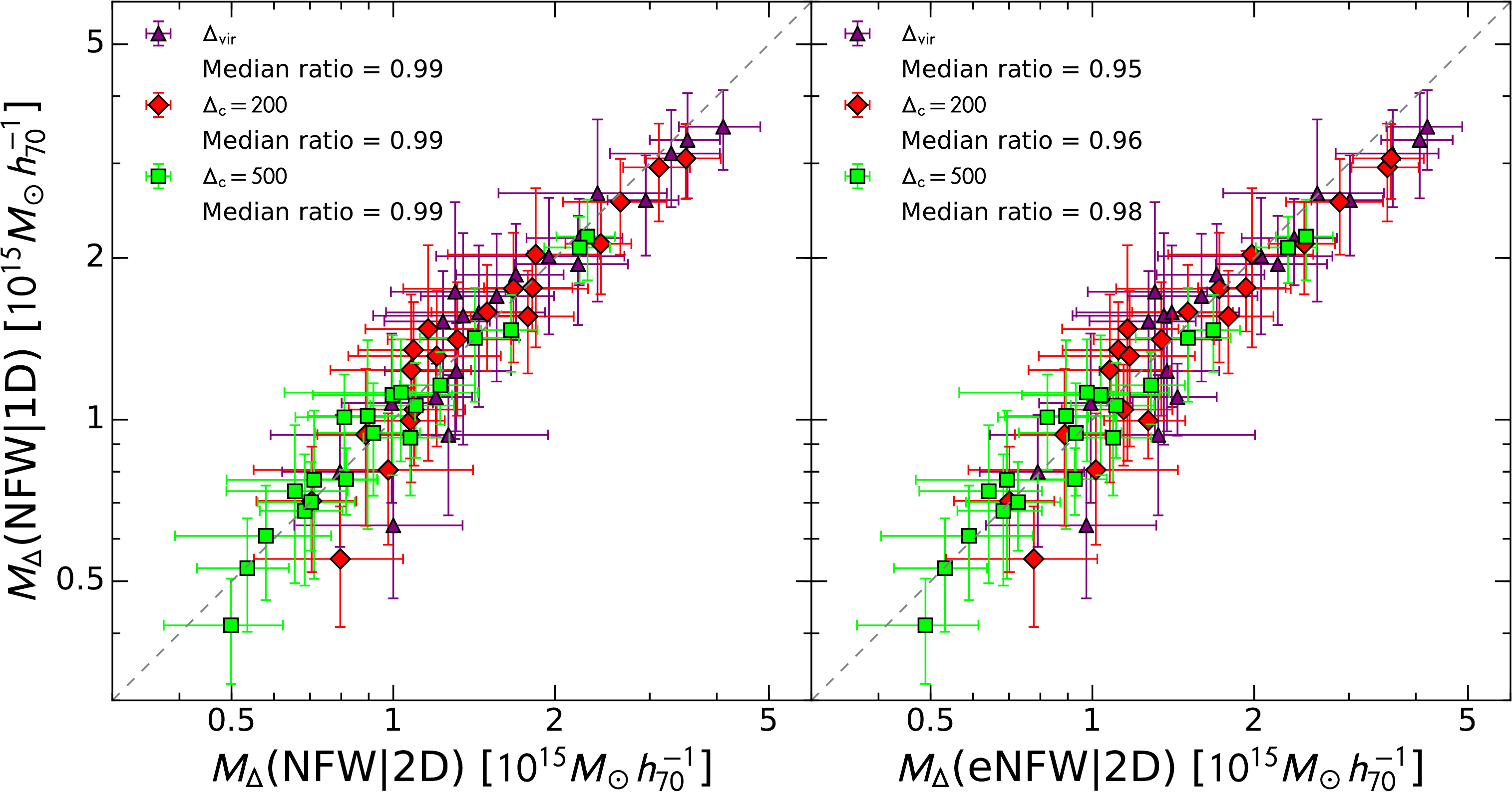} 
 \end{center}
 \caption{
 \label{fig:mcomp}
 Comparison of weak-lensing mass estimates for our sample of 20 CLASH
 clusters derived from our full 2D analysis (horizontal axes) to 
 those from the azimuthally averaged 1D analysis of
 \citet[][vertical axis]{Umetsu2014clash}.
 The left and right panels present our 2D results
 obtained using the NFW and elliptical NFW (eNFW) density
 profiles, respectively. 
 For each comparison, we measure the cluster mass for three
 characteristic overdensities,
 $\Delta_\mathrm{vir}$ (purple triangles),
 $\Delta_\mathrm{c}=200$ (red diamonds), and
 $\Delta_\mathrm{c}=500$ (green squares).
 The dashed line shows the one-to-one relation.
 The median mass ratios
 $\langle M_{\Delta}(\mathrm{1D})/M_{\Delta}(\mathrm{2D})\rangle$ are
 reported in each panel.
}
\end{figure*}

\subsubsection{Spherical and Elliptical Mass Models}

We model the radial mass distribution in galaxy clusters with a
 Navarro--Frenk--White (NFW) density profile, motivated by
 cosmological $N$-body simulations 
\citep{1996ApJ...462..563N,1997ApJ...490..493N}
as well as by direct lensing observations
\citep[e.g.,][]{Newman+2013a,Okabe+2013,Umetsu2014clash,Umetsu2016clash,Niikura2015,Okabe+Smith2016,Umetsu+Diemer2017}.
The radial dependence of the spherical NFW density profile is given by
\citep{1996ApJ...462..563N}
\begin{equation}
 \rho(r)=\frac{\rho_\mathrm{s}}{(r/r_\mathrm{s})(1+r/r_\mathrm{s})^2}
\end{equation}
with $\rho_\mathrm{s}$ the characteristic density parameter
and $r_\mathrm{s}$ the characteristic scale radius at which the
logarithmic slope of the density profile equals $-2$.
We specify the spherical NFW model with the halo mass, $M_\mathrm{200c}$,
and the concentration parameter, $c_\mathrm{200c}\equiv r_\mathrm{200c}/r_\mathrm{s}$.

The surface mass density $\Sigma(R)$ as a function of projected
clustercentric radius $R$ is given by
projecting $\rho(r)$ along the line of sight.
We employ an analytic expression given by \citet{2000ApJ...534...34W}
for the radial dependence of the projected NFW profile,
$\Sigma(R|M_\mathrm{200c},c_\mathrm{200c})$,
which provides a good approximation for the projected halo model within 
a couple of virial radii \citep[][]{Oguri+Hamana2011}
and an excellent description of the projected mass distribution in
clusters at $R\simlt r_\mathrm{200m}$
\citep{Umetsu2016clash,Umetsu+Diemer2017}. 

We follow the prescription given by
\citet{Oguri2010LoCuSS,Oguri+2012SGAS} to construct an elliptical NFW 
(eNFW hereafter) model, which can be used to characterize the
morphology of projected triaxial ellipsoids. 
To this end, we introduce the mass ellipticity, $\epsilon$, in
isodensity contours of the projected NFW profile
$\Sigma(R|M_\mathrm{200c},c_\mathrm{200c})$ as 
\citep{Evans+Bridle2009,Oguri2010LoCuSS,Umetsu+2012,Medezinski2016}
\begin{equation}
  R^2 = X'^2 (1-\epsilon) + Y'^2 /(1-\epsilon),
\end{equation}
where our definition of the ellipticity is $\epsilon = 1-q_\perp$
with $q_\perp\le 1$ the projected minor-to-major axis ratio of isodensity
contours, and
we have chosen the coordinate system $(X',Y')$ centered on the cluster halo,
such that the $X'$ axis is aligned with the major axis of the
projected ellipse.
Note that the isodensity area is $\pi R^2$, so that
$R$ represents the geometric mean radius of the isodensity
ellipse.
Accordingly, the $M_\mathrm{200c}$ and $c_\mathrm{200c}$
parameters in the eNFW model can be interpreted as respective spherical
equivalent quantities\footnote{This corresponds to a triaxial model 
with a special geometric configuration, where
$f_\mathrm{geo}\equiv e_{||}/\sqrt{q_\perp}=1$
with $e_{||}$ the 3D halo elongation parameter of \citet{Umetsu2015A1689}.}.
In this work, we adopt the observer's coordinate system in which the
$X$- and $Y$-axes are aligned with the west and north, respectively. With
this coordinate system, the position angle (PA) of the projected major
axis is measured east of north. 
An alternative definition for the projected ellipticity is
$e = (1-q_\perp^2)/(1+q_\perp^2)$
\citep[e.g.,][]{Evans+Bridle2009}.

\subsubsection{Bayesian Inference}
\label{subsubsec:mcmc}

We use a Bayesian Markov Chain Monte Carlo (MCMC) method to obtain 
an accurate inference of the eNFW parameters from our 2D weak-lensing
data (Section \ref{subsec:maps}).
In this study and subsequent companion papers
\citep{Chiu2018clump3d,Sereno2018clump3d},
we perform model fitting to the 2D surface
mass density data, rather than fitting directly to the combined shear
and magnification data sets\footnote{In principle, we can forward-model
and directly fit a model to 2D shear and magnification constraints. This
will require additional numerical integrals corresponding to the 2D
Poisson equation \citep{Keeton2001}.}. This allows consistency
checks with existing codes used in previous work 
\citep[e.g.,][]{Oguri2005,UB2008,Morandi2011A1689,Sereno+Umetsu2011,Sereno2013glszx,Sereno2017clump3d,Umetsu2015A1689}, 
that also used weak-lensing $\Sigma$ map data for 2D and 3D mass
modeling. The mass maps obtained in this work are also useful for
further studies of substructures in the context of the multiwavelength
CLUMP-3D program.

The projected eNFW model is specified by four parameters,
$\bp=(M_\mathrm{200c}, c_\mathrm{200c}, q_\perp, \mathrm{PA})$.
We use uniform prior distributions for the projected axis ratio and
position angle in the range
$0.1\le q_\perp\le 1$ and $-90^\circ\le \mathrm{PA} < 90^\circ$ \citep{Oguri2010LoCuSS}.
 Following
 \citet{Umetsu2014clash,Umetsu2016clash}, we assume log-uniform priors
 for $M_\mathrm{200c}$ and $c_\mathrm{200c}$ 
 in the range
 $0.1\le M_\mathrm{200c}/(10^{15}\Msunh) \le 10$ and
 $0.1 \le c_\mathrm{200c} \le 10$.
As found by \citet{Sereno2015cM} and \citet{Umetsu2016clash},  the
mass and concentration estimates for the CLASH sample are not sensitive
 to the choice of the priors, thanks to the deep high-quality
 weak-lensing observations. 
 The $\chi^2$ function for our
 observations is
 \begin{equation}
  \chi^2(\bp) = \sum_{m,n=1}^{\Npix}
   \left[\Sigma_m-\widehat{\Sigma}_m(\bp)\right]
   \left(C^{-1}\right)_{mn}
   \left[\Sigma_n-\widehat{\Sigma}_n(\bp)\right],
 \end{equation}
 where $\widehat{\Sigma}_m(\bp)$ denotes the surface mass density
 at the grid position $(X_m,Y_m)$ predicted by the model $\bp$.
For all clusters, we restrict the fitting to a square region of side
 $4\,\Mpch$ centered on the cluster,
where half the side length corresponds to the typical
 $r_\mathrm{200m}$ radius of CLASH clusters. 
This is to minimize the impact of the 2-halo term and local
 substructures that are abundant in cluster outskirts, which otherwise
 can lead to bias in cluster mass estimates
 \citep{Meneghetti+2010a,Becker+Kravtsov2011,Rasia+2012}.
Similarly, our previous CLASH studies performed fitting to azimuthally
 averaged lensing profiles by restricting the fitting range to
 $R\le 2\,\Mpch$ 
 \citep{Umetsu2014clash,Umetsu2016clash,Merten2015clash}. 

We also fit the data with a spherical NFW
 halo ($q_\perp=1$) using the same log-uniform priors on
 $M_\mathrm{200c}$ and 
 $c_\mathrm{200c}$, in order to examine the consistency of our results
 with \citet{Umetsu2014clash}.

In Table \ref{tab:sample}, we list the marginalized posterior
constraints on each of the model parameters
$\bp=(M_\mathrm{200c}, c_\mathrm{200c}, q_\perp, \mathrm{PA})$ for 20
individual clusters of our sample.
In this work, we employ the robust biweight estimators of
\citet{1990AJ....100...32B} for the center location ($C_\mathrm{BI}$)
and scale ($S_\mathrm{BI}$) of the marginalized 1D posterior distributions
\citep[e.g.,][]{Stanford1998,Sereno+Umetsu2011,Biviano2013}.
The biweight estimator is insensitive to and stable against outliers, as
it assigns higher weight to points that are closer to the center of the
distribution \citep{1990AJ....100...32B}.
In the table, we also report, for each cluster, the minimum
$\chi^2$ value per degree of freedom (dof) as an indicator of goodness
of fit. Here, the number of dof is defined as the difference between the
number of mass pixels within the fitting region and the number of free
parameters.  
We find that the minimum $\chi^2$/dof values for our sample
range from 0.41 (MACS~J0744.9$+$3927) to 1.32 (RX~J1347.5$-$1145), with
a median of 0.80. This indicates that complex morphologies in the
projected cluster mass distribution (e.g., substructures and deviations
from elliptical isodensity contours) are not statistically significant
in individual clusters, and the eNFW model provides an adequate
description of our 2D weak-lensing data.  

In Appendix \ref{appendix:eNFW}, we show, for each of the clusters, the
2D marginalized posterior distributions of the eNFW parameters, with
the contours enclosing $68\percent$ and $95\percent$ of the   
posterior probability (Figure \ref{fig:eNFWfit}). For each
parameter, we also present the 1D marginalized distribution, in which
the $C_\mathrm{BI}$ location is marked with a vertical line.
We see from Figure \ref{fig:eNFWfit} that the mass and concentration
parameters
are well constrained by the data, except for MACS~J1931.8$-$2635 and
MACS~J0429.6$-$0253, for which the posterior distribution on
$c_\mathrm{200c}$ is largely informed by the prior,
in the sense that the likelihood extends outside of the prior range.
Overall, the 2D weak-lensing constraints on the halo shape parameters
($q_\perp,\mathrm{PA}$) are not strongly degenerate with halo mass 
and concentration, as found by \citet{Oguri2010LoCuSS}.

\section{Discussion}
\label{sec:discussion}

\subsection{Systematic Errors}
\label{subsec:sys}

We have accounted for various sources of errors associated with the
weak-lensing shear and magnification measurements (Sections
\ref{subsec:shear} and \ref{subsec:magbias}). All of these errors are
encoded in the measurement uncertainties $(\sigma_g, \sigma_\mu)$ and
the cosmic noise errors $(C_g^\mathrm{lss},C_\mu^\mathrm{lss})$,
both of which contribute to the covariance matrix $C$ of the mass
reconstruction (Appendix \ref{appendix:DD}).

\citet{Umetsu2016clash} quantified unaccounted sources
of systematic errors in the CLASH weak-lensing measurements by
considering the following effects: 
(1) dilution of the weak-lensing signal by residual contamination from 
cluster members ($2.4\percent \pm 0.7\percent$),
(2) photo-$z$ bias in the mean depth estimates
($0.27\percent$; Section \ref{subsec:back}),
and (3) shear calibration uncertainty ($5\percent$; Section
\ref{subsubsec:shearmeas}). 
These errors 
add up to $5.6\percent$ in quadrature, which is translated into the
cluster mass uncertainty as $5.6\percent/\Gamma\simeq 7\percent$ with 
$\Gamma\simeq 0.75$ the typical value of the
logarithmic derivative of the weak-lensing signal with respect to
cluster mass \citep{Melchior2017des}.

Alternatively, measuring the shear and magnification effects
independently provides an empirical consistency check of weak-lensing 
measurements. 
Performing a shear--magnification consistency test,
\citet{Umetsu2014clash} found the systematic uncertainty  
in the overall mass calibration to be $8\percent$.
Following the CLASH program
\citep{Umetsu2014clash,Umetsu2016clash,Merten2015clash,PennaLima2017}, 
we conservatively adopt this value as the systematic uncertainty in the 
ensemble mass calibration.   



In Figure \ref{fig:mcomp}, we compare our mass estimates of 20
individual clusters from the present 2D weak-lensing analysis,
$M_{\Delta_\mathrm{c}}(\mathrm{2D})$, 
with those from the 1D weak-lensing analysis of 
\citet{Umetsu2014clash}, $M_{\Delta_\mathrm{c}}(\mathrm{1D})$,
shown for three characteristic overdensities.
Since the two studies use the same data sets, this comparison allows us
to assess the robustness and consistency of weak-lensing mass
determinations from different inversion methods.
The mass estimates of \citet{Umetsu2014clash} were obtained assuming
the spherical NFW profile, with the same priors on $M_\mathrm{200c}$ and
$c_\mathrm{200c}$ as in Section \ref{subsec:eNFWfit}.
Results are shown separately for our NFW and eNFW mass estimates in the
left and right panels, respectively. No aperture correction is applied
in all cases.
%

As shown in the left panel of Figure \ref{fig:mcomp},
the $\Mvir$, $M_\mathrm{200c}$, and $M_\mathrm{500c}$ masses
estimated from direct NFW fits to the 2D mass distribution $\Sigma(\bR)$
agree within $1\percent$ with those from the $\Sigma(R)$ profile of
\citet{Umetsu2014clash}.   
The right panel gives a comparison of our eNFW mass 
estimates with the NFW results of \citet{Umetsu2014clash}, showing that 
the eNFW masses are on average $5\percent$, $4\percent$, and $2\percent$
smaller for 
$\Delta_\mathrm{vir}$,
$\Delta_\mathrm{c}=200$,
and $\Delta_\mathrm{c}=500$, respectively,
than the NFW ones inferred from the azimuthally averaged 1D analysis.
In summary, the mass estimates for our cluster sample derived with 1D
and 2D inversion methods \citep{Umetsu2014clash,Umetsu2016clash} agree well
within the overall calibration uncertainty of $8\percent$, indicating that
systematic effects due to azimuthal averaging applied to the shear data
\citep{Umetsu2014clash}, as well as to details of the inversion
procedures (Section \ref{subsec:diff}), are not significant.

\subsection{Cluster Ellipticity}
\label{subsec:q}

\subsubsection{CLASH Ensemble Distribution}
\label{subsubsec:q_clash}


\begin{figure}[!htb] 
 \begin{center}
  \includegraphics[scale=0.5, angle=0, clip]{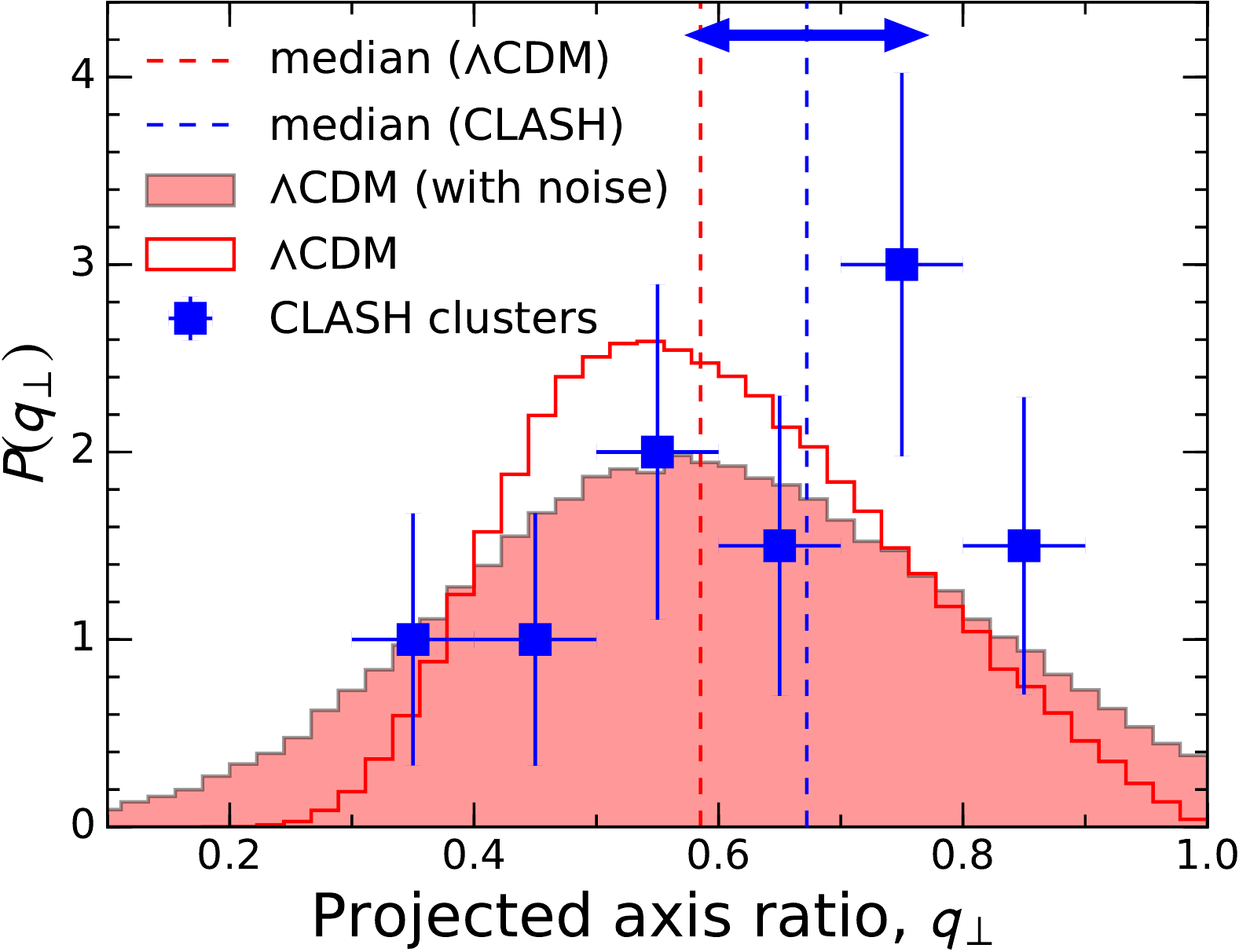} 
 \end{center}
 \caption{
 Distribution of projected axis ratios $q_\perp$ for our
 sample of 20 CLASH clusters. The blue squares with error bars show
 the distribution
 constructed from point estimates of individual cluster posteriors
 (Table \ref{tab:sample}). 
 The blue dashed vertical line corresponds to the median 
 $\langle q_\perp\rangle = 0.67\pm 0.07$ of the distribution (Table
 \ref{tab:q}), with the
 $1\sigma$ error indicated by the double-headed arrow.
 The red shaded histogram represents the theoretical expectation
 based on the triaxial halo model of \citet{Bonamigo2015} assuming
 random orientations of the clusters,
 where the $q_\perp$ distribution predicted for each cluster has been
 convolved with a Gaussian of width equal to the observed uncertainty
 $\sigma(q_\perp)$. 
 The red open histogram shows the theoretical distribution without the
 Gaussian convolution.
The expected median value is $\langle q_\perp\rangle = 0.59$ (red dashed
 vertical line) in both
 cases with and without the Gaussian 
convolution, consistent with our measurement at the $1\sigma$ level.  
\label{fig:q}
}
\end{figure}

\input{table2.tex}

We examine here the ensemble distribution of projected axis ratios
using the results from the Bayesian inference of the eNFW parameters.
Figure \ref{fig:q} shows the resulting distribution function for our 
sample of 20 clusters (blue squares), 
constructed from posterior point estimates ($C_\mathrm{BI}$) of
the $q_\perp$ parameter (Table \ref{tab:sample}). The median axis
ratio for our sample, measured within a radial scale of 2\,$\Mpch$, is  
$\langle q_\perp\rangle=0.67\pm 0.07$, or
$\langle \epsilon\rangle=0.33\pm 0.07$ and
$\langle e\rangle=0.38\pm 0.08$
in terms of the projected halo ellipticity (Table \ref{tab:q}), where
the errors were estimated by bootstrap resampling the cluster sample. 
To check at which radius the constraint on the halo ellipticity
effectively comes from, we calculate the mass-weighted, projected
clustercentric radius $R_\mathrm{eff}$ for our sample, averaged within
the central $2\,\Mpch$ region. Using the best-fit NFW model based on the
stacked lensing analysis by \citet{Umetsu2016clash}, we find
$R_\mathrm{eff}\simeq 0.89\Mpch$.

For a consistency check, we also create a composite probability
distribution function (PDF) of $q_\perp$ by stacking the marginal
posterior distributions (Figure \ref{fig:eNFWfit}) for all clusters in
our sample. This yields a median axis ratio of
$\langle q_\perp\rangle=0.65\pm 0.05$, or
$\langle \epsilon\rangle=0.35\pm0.05$ and
$\langle e\rangle=0.41\pm 0.07$,
where the errors are based on 50,000 random samples that are drawn from the
posterior distributions for individual clusters.
These results are in good agreement with those from the posterior point
estimates.

{\red It is important to note that the shape of the clusters in the
CLASH sample is expected to be rounder on average due to a high fraction
of relaxed clusters \citep[Section
\ref{subsec:data};][]{Meneghetti2014clash}. 
Accordingly, there could be a bias toward higher values of the projected
axis ratio in our sample.} 
For our X-ray-selected subsample (excluding the four high-magnification
clusters with complex morphologies), we find the median axis ratio to be 
$\langle q_\perp\rangle=0.72\pm 0.07$ from the posterior point
estimates and $\langle q_\perp\rangle=0.68\pm 0.06$ from the stacked PDF.
Although the amplitude of the shift is not statistically significant,
the direction of the shift is consistent with the effect of the CLASH
X-ray-selection function (Section \ref{subsec:data}).

\citet{Donahue2016clash} found the typical axis ratio for all 25 CLASH 
 clusters in the {\em Chandra} X-ray brightness distribution to be
 $0.88\pm 0.06$ {\red within an aperture radius of} $350\,\kpch$
 ($\sim 1.1r_\mathrm{2500c}$),
 consistent with the value of $0.90\pm 0.06$ inferred from the SZE maps
 observed with Bolocam operating at 140\,GHz
 \citep{Sayers2013pressure,Czakon2015}.
This comparison for the CLASH sample, albeit at different radial scales,
 indicates that  
 the shape of the projected mass distribution as measured from weak
 lensing is more elongated than the gas distribution.
  This is qualitatively consistent with the theoretical expectation that
 the intracluster gas in hydrostatic equilibrium is rounder than the
 underlying matter distribution \citep{Lee+Suto2003}.
However, we note that the values of cluster morphological parameters for
 the X-ray and SZE maps could be different if measured over
 larger radial scales (see the discussion in Section
 \ref{subsubsec:q_theory}).  
 On the other hand, our ground-based weak-lensing data do not
 sufficiently resolve morphological structures
 {\red within a radial scale of} $350\,\kpch$
 ($\sim 1.6\arcmin$ at the median sample redshift of $z=0.377$) as
 they are limited by the small number density of 
 background galaxies \citep{Umetsu2014clash}.

\subsubsection{Comparison with $\Lambda$CDM Predictions} 
\label{subsubsec:q_theory}


\begin{figure}[!htb] 
 \begin{center}
  \includegraphics[scale=0.5, angle=0, clip]{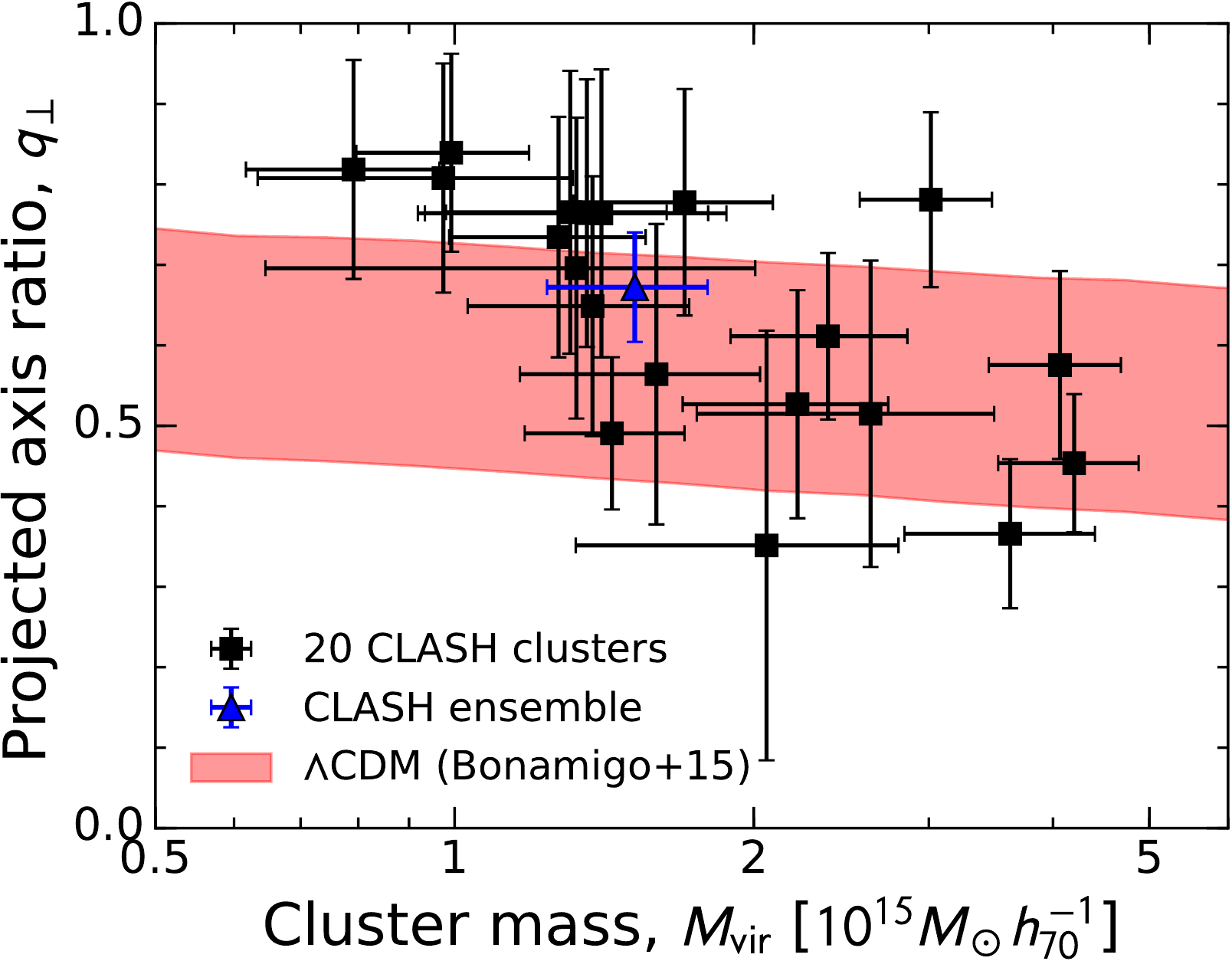} 
 \end{center}
 \caption{
Projected halo axis ratio $q_\perp$ plotted against
 the virial mass $\Mvir$  for our sample of 20 CLASH clusters (black 
 squares). The blue triangle represents the median values for the
 sample, 
 $\langle q_\perp\rangle=0.67\pm 0.07$ at
 $\langle\Mvir\rangle=(15.2\pm 2.8)\times 10^{14}\Msun$. The red shaded
 region represents the $1\sigma$ range of the theoretical distribution
 at $z=0.377$
 predicted with the triaxial model of \citet{Bonamigo2015} assuming
 random orientations of the clusters. 
\label{fig:qM}
}
\end{figure}

These results can be compared with predictions from
$\Lambda$CDM $N$-body simulations, in which DM halos are often modeled 
as triaxial ellipsoids  
\citep[e.g.,][]{Jing+Suto2000,Bett2007,Bonamigo2015,Despali2014,Despali2016,Vega2017}\footnote{A
triaxial halo in projection is seen as elliptical isodensity contours \citep{Stark1977}.}.
Here we restrict the comparison to predictions
for the shape of the DM distribution measured within the virial radius,
which is close to the maximum fitting radius of our analysis
($2\,\Mpch \sim r_\mathrm{200m}\sim 1.1\rvir$).
In particular, the triaxial model of \citet{Bonamigo2015} is of
particular interest because they extended the original work of
\citet{Jing+Suto2000} to a wider mass range with higher precision.
\citet{Bonamigo2015} selected and analyzed those halos for which the
offset between the center of mass and geometrical center of the ellipsoid is
less than $5\percent$ of their virial radius. The fraction of selected
clusters is $\sim 50\percent$ at $z=0$ \citep[see Figure 2
of][]{Bonamigo2015}.
On the other hand, applying stringent relaxation selection criteria
typically results in a much smaller fraction of selected halos (e.g.,
$\sim 15\percent$ at $z=0.25$ as found by \citet{Meneghetti2014clash}).

In Figure \ref{fig:q}, we show the theoretical expectation
$P(q_\perp)$ for our composite clusters (red shaded histogram) obtained
using the fitting formula of \citet{Bonamigo2015}, which describes the
intrinsic distribution of triaxial axis ratios for
{\red their DM halos.}  
Here, the predicted distribution has been constructed as
follows:
first, we compute for each cluster the intrinsic axis ratio distribution
as a function of halo virial mass $\Mvir$ and redshift
\citep{Bonamigo2015}  by accounting for the uncertainty in the mass  
determination. Next, we construct the distribution of the projected axis
ratios by projecting triaxial halos onto the sky plane
assuming random orientations.
The $q_\perp$ distribution predicted for each cluster is then convolved
with a Gaussian of width equal to the observed uncertainty
$\sigma(q_\perp)$.
The composite PDF is finally obtained by adding the renormalized
distributions of all 20 clusters. 
In Figure \ref{fig:q}, we also show
the theoretical distribution $P(q_\perp)$ without the Gaussian
convolution (red open histogram).
Because of the projection effect, the overall shape of the distribution
of projected axis ratio is broader and shifted to values
(rounder) higher than those of the intrinsic minor-to-major axis ratio
\citep{Suto2016}. 
The expected median value is $\langle q_\perp \rangle = 0.59$ in both
cases with and without the Gaussian convolution, in agreement with our
measurement within the uncertainty.
We reiterate that our sample is expected to contain a
high fraction of relaxed clusters ($\sim 60\percent$ in our full
composite sample, compared to $\sim 70\percent$ in the X-ray-selected
subsample; see Section \ref{subsec:data}), and hence the average
projected axis ratio is likely biased high to some degree. In
particular, if we restrict our comparison to the X-ray-selected
subsample, the observed median projected axis ratio (Table \ref{tab:q})
is higher at the $1.8\sigma$ level than predicted from the triaxial
model of \citet{Bonamigo2015}.
 
We plot in Figure \ref{fig:qM} the projected axis ratio $q_\perp$
as a function of the virial mass $\Mvir$ for our sample of 20
CLASH clusters (black squares). The blue triangle in Figure \ref{fig:qM}
represents the (unweighted) median values for the sample,
$\langle q_\perp\rangle=0.67\pm 0.07$ at
$\langle\Mvir\rangle=(15.2\pm 2.8)\times 10^{14}\Msun$.
Again, our CLASH weak-lensing measurements are in good agreement with
$\Lambda$CDM expectations at the median sample redshift of $z=0.377$.
All the 20 CLASH clusters lie within the $2\sigma$ distribution
predicted with the triaxial model of \citet{Bonamigo2015} assuming
random orientations of the clusters.

\citet{Suto2016} studied the mass and radial dependence and the redshift 
evolution of the non-sphericity of cluster-size halos using DM-only
simulations. 
They found that the average 3D minor-to-major axis ratio of simulated
halos has a strong radial dependence as a function of enclosed mass,
indicating that the internal structure of halos deviates from a
self-similar geometry of concentric ellipsoidal surfaces.
\citet{Suto2016} thus constructed the PDF of the projected axis
ratio, $P(q_\perp)$, directly from the projected density distributions
of simulated halos, without involving triaxial modeling\footnote{We note
that although their typical halo mass  
($\Mvir=2\times 10^{14}\Msunh$ at $z=0.2$) is considerably smaller than
that of our sample, $\Mvir\sim 11\times 10^{14}\Msunh$, this is not
critical because the mean projected axis ratio exhibits little mass
dependence over the relevant mass interval 
\citep[Figure 11 of][]{Suto2016}.}.
Using their fitting formula for the PDF at $z=0.4$,
the mean, standard deviation, and median are found to be 0.57, 
0.17, and 0.58, respectively.
The predicted median agrees with our full-sample result at the
$1.3\sigma$ level. 
On the other hand, when halos in multicomponent systems are excluded
from their analysis, 
the mean, standard deviation, and median of the PDF are 0.59, 0.16, and
0.60, respectively.  The observed median axis ratio for the CLASH
X-ray-selected subsample is $1.6\sigma$ above this predicted median.

More recently, it was shown by \citet{Suto2017} that
baryonic physics operating in the cluster central region, such as
cooling and feedback effects, can have a substantial impact on the
non-sphericity of cluster halos up to half the virial radius, even
though these baryonic effects have little impact on the spherically
averaged DM density profile. They found that the DM distribution becomes
more spherical, depending on the distance from the cluster center, when
the effects of baryons are included.
Since our sample comprises highly relaxed and highly disturbed clusters
(Section \ref{subsec:data}), a more quantitative comparison with 
theoretical expectations would require a detailed modeling of baryonic 
physics by accounting for the selection function.
In fact, Figure \ref{fig:qM} shows a slight tendency for lower-mass
clusters to have projected axis ratios that are higher than the
predicted distribution. This tendency is qualitatively consistent with
the combination of the CLASH selection function and the baryonic
effects. 
Nevertheless, our analysis is currently limited by the small number of
clusters.



\subsubsection{Comparison with Other Observational Studies}
 
Our CLASH lensing results can be compared to the
weak-lensing measurements obtained by the LoCuSS collaboration 
\citep{Oguri2010LoCuSS}, who performed a 2D shear-fitting analysis
on a sample of 25 X-ray-luminous clusters at
$\langle z\rangle \sim 0.2$.
\citet{Oguri2010LoCuSS}
modeled the projected mass distribution in each individual cluster
assuming a single eNFW halo, as done in our work. They fitted an
eNFW profile to the grid-averaged 2D reduced shear field.
For a subset of 18 clusters that were specifically chosen to give good 
ellipticity constraints,
they found a $\sim 7\sigma$ detection of the mean cluster ellipticity,
$\langle\epsilon\rangle = 0.46\pm 0.04$, for their clusters with 
$\langle\Mvir\rangle\sim 10\times 10^{14}\Msun$.
Their mean ellipticity is higher than, but consistent within the errors
with, our full-sample measurement (Table \ref{tab:q}).
{\red This difference is not statistically significant, but
could be due in part to the CLASH X-ray selection function (Section
\ref{subsec:data}).}

Similarly, \citet{Oguri+2012SGAS} found from their 2D shear analysis of
25 strong-lensing-selected clusters that the stacked cluster ellipticity
is nearly constant, $\langle\epsilon\rangle\sim 0.45$,
with cluster radius within their errors.

{\red
On group/cluster scales, several authors have constrained the halo
ellipticity from stacked weak-lensing measurements
\citep{Evans+Bridle2009,Clampitt+Jain2016,vanUitert+2017,Shin+2018}}
using quadrupole shear estimators and their variants
\citep[e.g.,][]{Natarajan+Refregier2000,Adhikari2015,Clampitt+Jain2016}.
Their stacked weak-lensing measurements span a range of mean $q_\perp$ 
values from  $0.48^{+0.14}_{-0.09}$ \citep{Evans+Bridle2009} to $\sim 0.78$
\citep{Clampitt+Jain2016}, broadly consistent with $\Lambda$CDM
predictions.
We note that, in their approach, one probes halo quadrupoles with
respect to the major axis of the light distribution (e.g., BCGs) chosen
as a reference orientation. On the
other hand, we have directly measured the shape and orientation of
individual cluster halos from deep high-quality weak-lensing data
\citep{Umetsu2014clash}.
{\red
We present in Section \ref{subsubsec:wlquad} our stacked quadrupole
shear measurement for our full sample of 20 CLASH clusters using the
shape of the {\em Chandra} X-ray brightness distribution (see Section
\ref{subsubsec:alignments_clash}) as a reference orientation.}

\subsection{Cluster Misalignment Statistics}
\label{subsec:PA}

\subsubsection{Alignments of CLASH Clusters}
\label{subsubsec:alignments_clash}


\begin{figure*}[!htb] 
 \begin{center}
  $
  \begin{array}
   {c@{\hspace{0.1in}}c@{\hspace{0.1in}}c}
    \includegraphics[scale=0.48, angle=0, clip]{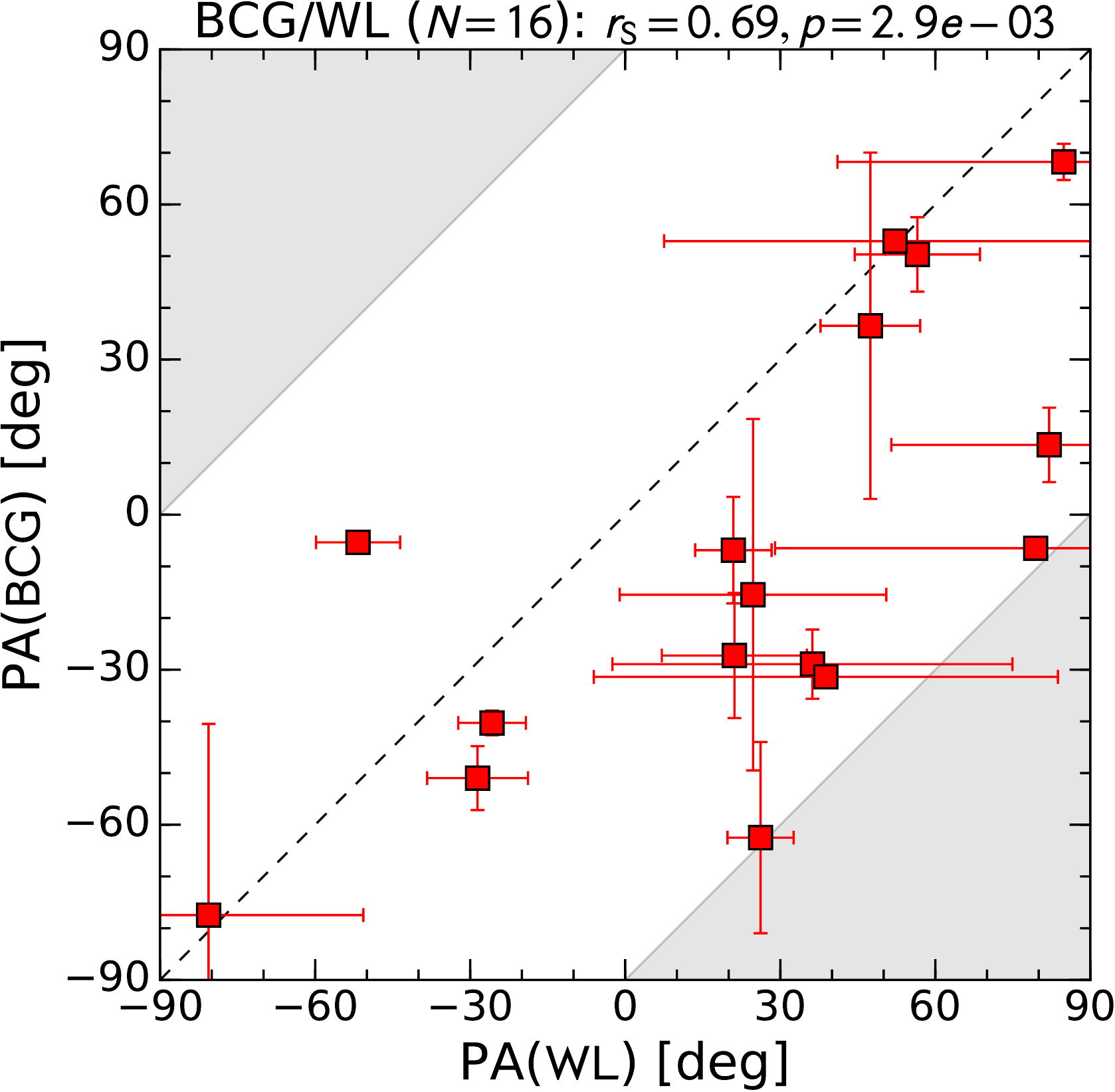} & 
    \includegraphics[scale=0.48, angle=0, clip]{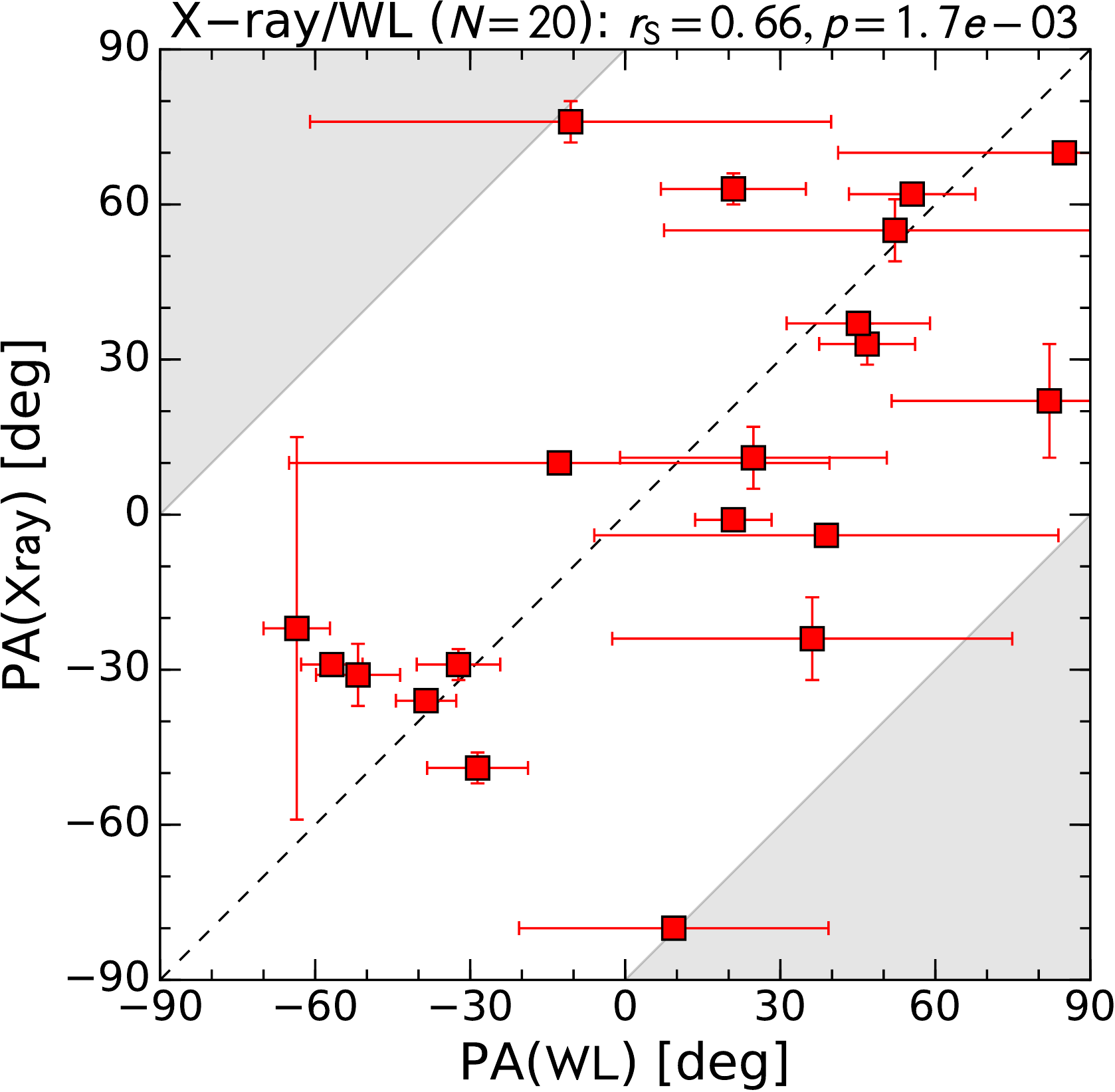}
    \vspace{0.1in}
  \end{array}
  $
  $
  \begin{array}
  {c@{\hspace{0.1in}}c@{\hspace{0.1in}}c}
   \includegraphics[scale=0.48, angle=0, clip]{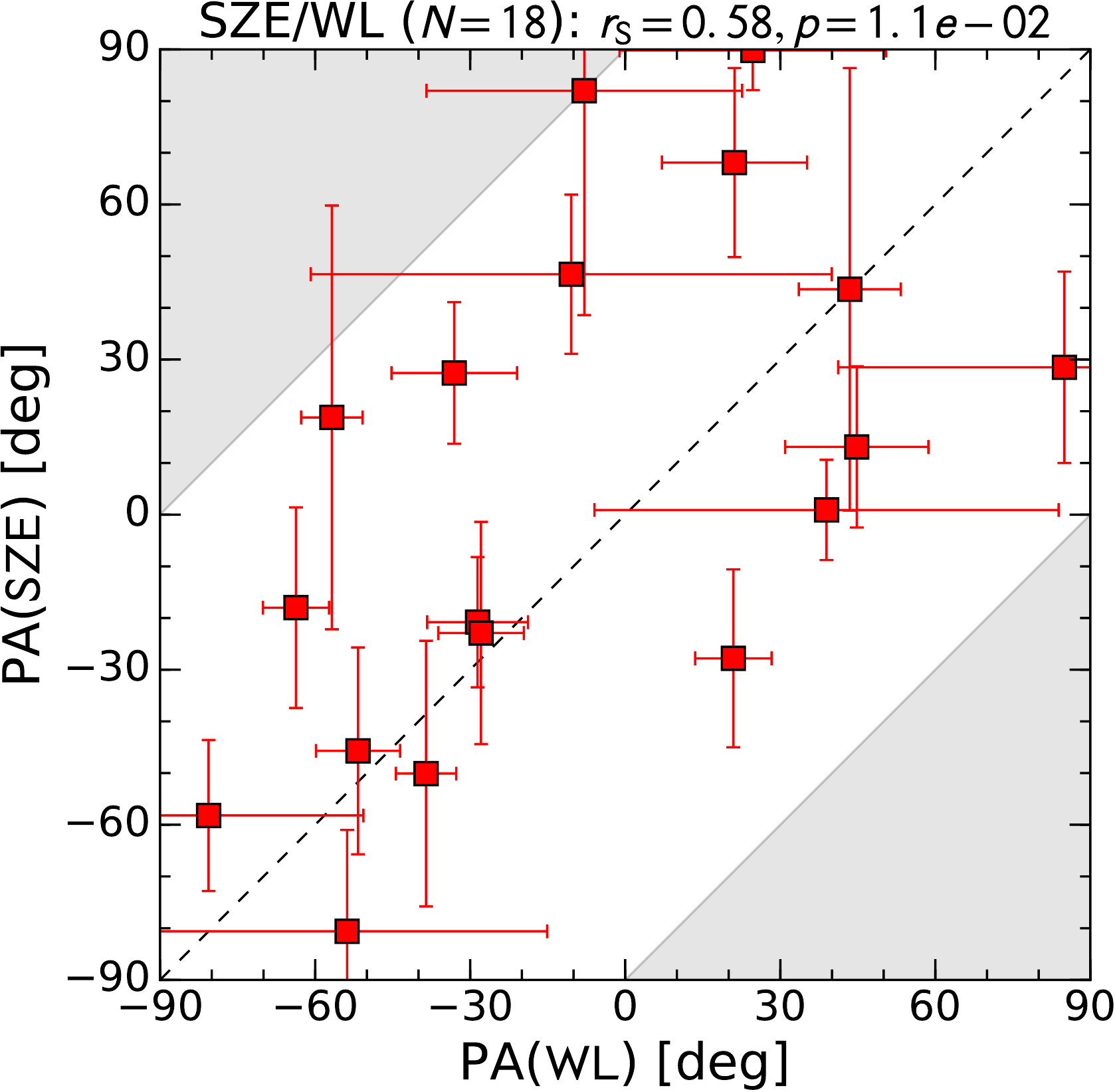}&  
   \includegraphics[scale=0.48, angle=0, clip]{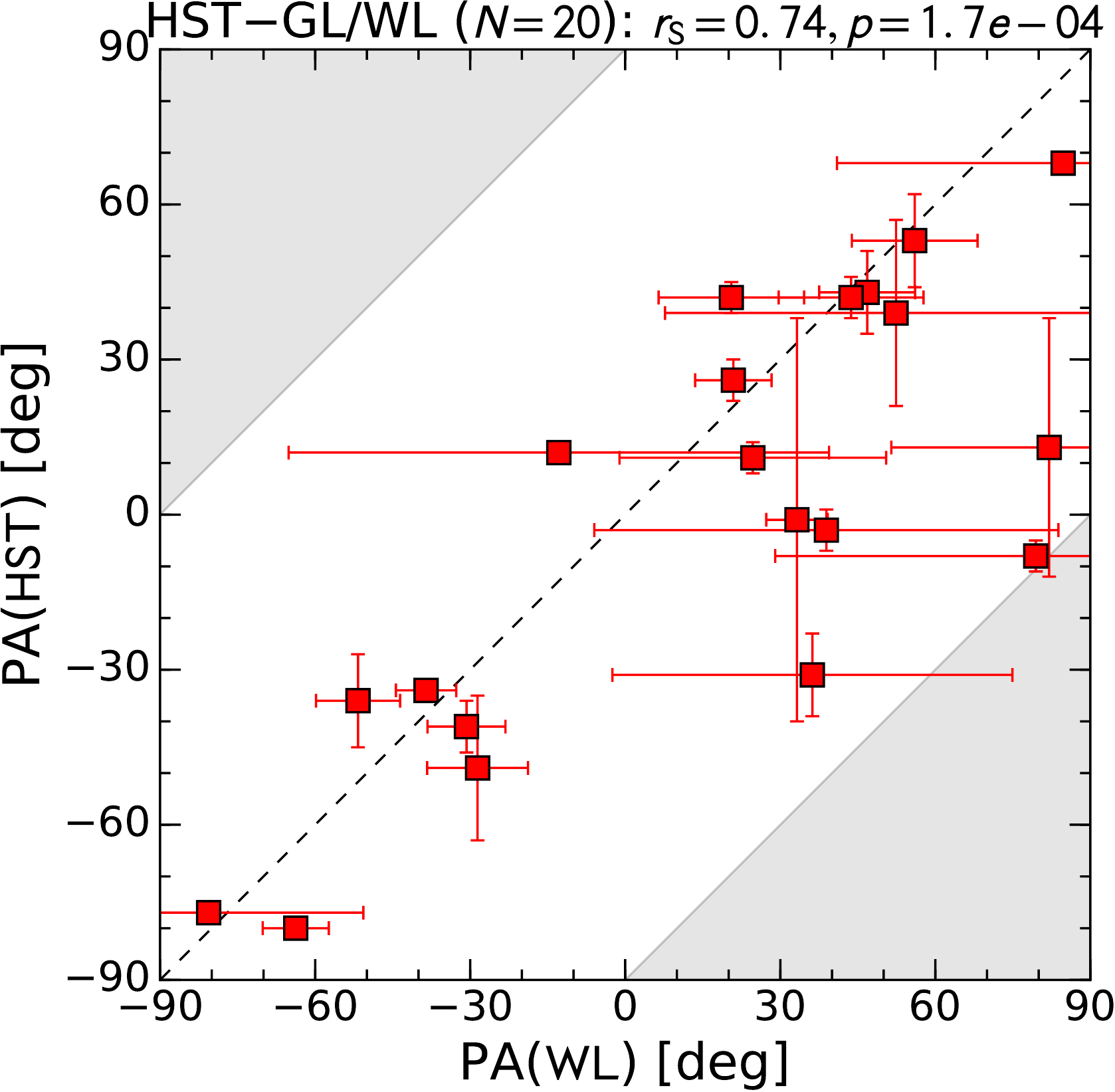}
  \end{array}
  $
 \end{center}
 \caption{\label{fig:PA}
{
Position angle (PA) in degrees of the weak-lensing (WL) major axis
 plotted against that of the brightest cluster galaxy (BCG; upper left)
 and those for the X-ray (upper right), thermal Sunyaev--Zel'dovich
 effect (SZE; lower left), and
 {\em HST} strong+weak lensing ({\em HST}-GL; lower right) maps.}
 PAs are measured east of north and defined in 
 the range $[-90,+90)$\,degrees. Absolute misalignment angles relative
 to weak lensing, $|\Delta\mathrm{PA}|$,
 are constrained in the range $[0, 90]$\,degrees. For clusters
 that fall within the gray shaded areas, their PAs are shifted by
 $90^\circ$
 \citep[e.g., Figure 6 of][]{Oguri2010LoCuSS},
 so as to fit in the proper region with
 $|\Delta\mathrm{PA}|\le 90^\circ$. 
 The dashed line represents the
 one-to-one relation.
 The Spearman rank correlation coefficient $r_\mathrm{P}$ and the
 corresponding $p$-value for testing the null hypothesis are reported
 in each panel (lower probabilities are more significant).
 Note that although there is a high degree of
 correlation between the BCG and weak-lensing PAs, the distribution
 of $\Delta\mathrm{PA}=\mathrm{PA(BCG)}-\mathrm{PA(WL)}$ is not
 symmetric about 
 $\Delta\mathrm{PA}=0$, with a median offset of
 $\langle\Delta\mathrm{PA}\rangle=-25^\circ\pm 14^\circ$ (Table \ref{tab:PA}).
}
\end{figure*}

\citet{Donahue2015clash,Donahue2016clash} presented a detailed study of 
morphologies and alignments of BCGs, intracluster gas, and mass at small
cluster radii for the CLASH sample.
\citet{Donahue2015clash} measured PAs for all 20
X-ray-selected CLASH clusters from the rest-frame ultraviolet (UV) and
near-infrared (NIR) light distributions of the BCGs using CLASH {\em
HST} data (see their Table 3). 
Similarly, \citet{Donahue2016clash} examined the morphological
properties of all 25 CLASH clusters with {\em Chandra} X-ray data
\citep{Donahue2014clash}, and compared them with those inferred from
Bolocam SZE maps of the intracluster gas
\citep{Sayers2013pressure,Czakon2015} and those from mass maps based on
{\em HST} strong and weak lensing
\citep[hereafter {\em HST}-GL;][]{Zitrin2015clash}.
The lensing maps used are based on parametric modeling assuming that
light traces mass for the galaxy-scale mass components
\citep[see][]{Meneghetti2017}. 
They found a strong correlation between PAs as
measured from the X-ray, SZE, and {\em HST}-GL maps
inside a consistent metric aperture of radius
$350\,\kpch$ ($\sim 0.2r_\mathrm{vir}$ for these clusters).
They also found a strong alignment of the cluster shapes
at this scale,
as measured from the X-ray, SZE, and {\em HST}-GL maps,
with that of the NIR BCG light at 10\,kpc scales.
In particular, they obtained a median misalignment angle of
$|\Delta\mathrm{PA}|\simeq 11^\circ$
between the BCG and X-ray orientations for the 20 X-ray-selected CLASH 
clusters. 

Now we compare the PAs determined from our 2D weak-lensing analysis (WL)
to those from three baryonic tracers, namely the BCGs, X-ray maps, and
SZE maps, as well as from the {\em HST}-GL maps.
To this end, we use the BCG PAs derived from the {\em HST} NIR 
images \citep[Table 3 of][]{Donahue2015clash}, and
the X-ray, SZE, and {\em HST}-GL PAs measured within an aperture radius
of $350\,\kpch$ \citep[Tables 3, 6, and 5 of][]{Donahue2016clash}.
To be conservative, we estimate errors on the BCG PAs from differences
between the {\em HST} UV and NIR measurements of
\citet{Donahue2015clash}.
There are 16, 20, 18, and 20 clusters available for BCG/WL, X-ray/WL,
SZE/WL, and {\em HST}-GL/WL comparisons, respectively.
The typical uncertainty in the weak-lensing PA measurements is 
 $\sim 20^\circ$ per cluster,
 comparable to that in the SZE measurements,
 while those in the BCG, X-ray, and {\em HST}-GL measurements are much   
 smaller, $\sim 10^\circ$, $\sim 4^\circ$, and $\sim 4^\circ$ per
 cluster, respectively.

In Figure \ref{fig:PA}, the PAs of the clusters determined from our
weak-lensing analysis are plotted against those from the BCGs, X-ray
maps, SZE maps, and {\em HST}-GL maps.
 To quantify the degree of correlation between PAs of different tracers,
 we calculate the Spearman rank correlation 
 coefficient $r_\mathrm{P}$ and the corresponding probability ($p$) for
 the null hypothesis of random orientations. The test indicates a
 similarly good correlation in all four of these comparisons,
 where a low probability indicates high significance of correlation:
 $r_\mathrm{S}= 0.69$ ($p=2.9\times 10^{-3}$) for BCG/WL,  
 $r_\mathrm{S}= 0.66$ ($p=1.7\times 10^{-3}$) for X-ray/WL, 
 $r_\mathrm{S}= 0.58$ ($p=1.1\times 10^{-2}$) for SZE/WL PAs,
 and
 $r_\mathrm{S}= 0.74$ ($p=1.7\times 10^{-4}$) for {\em HST}-GL/WL PAs.


\begin{figure*}[!htb] 
 \begin{center}
  $
  \begin{array}
   {c@{\hspace{0.1in}}c@{\hspace{0.1in}}c}
    \includegraphics[scale=0.48, angle=0, clip]{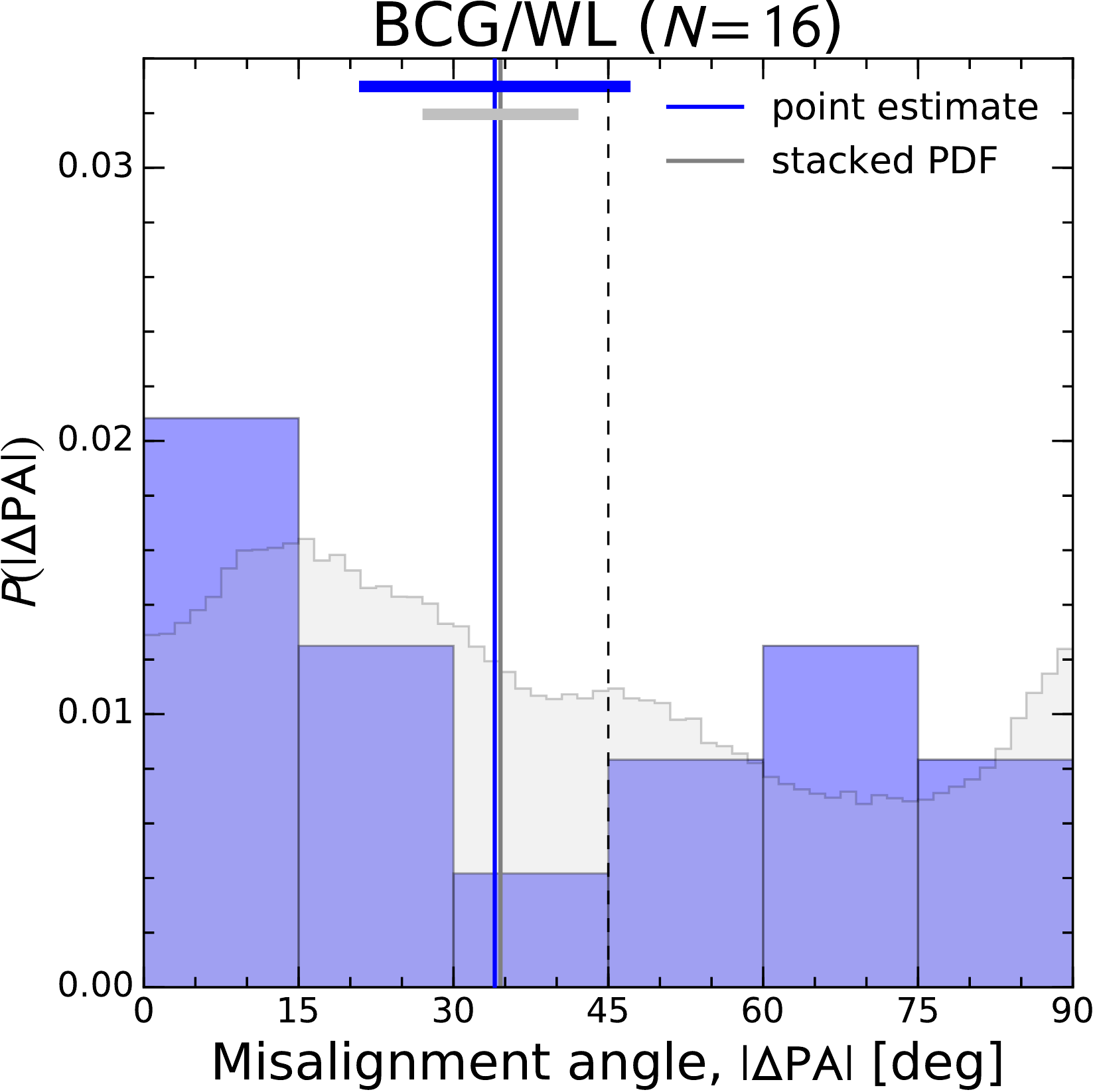} & 
    \includegraphics[scale=0.48, angle=0, clip]{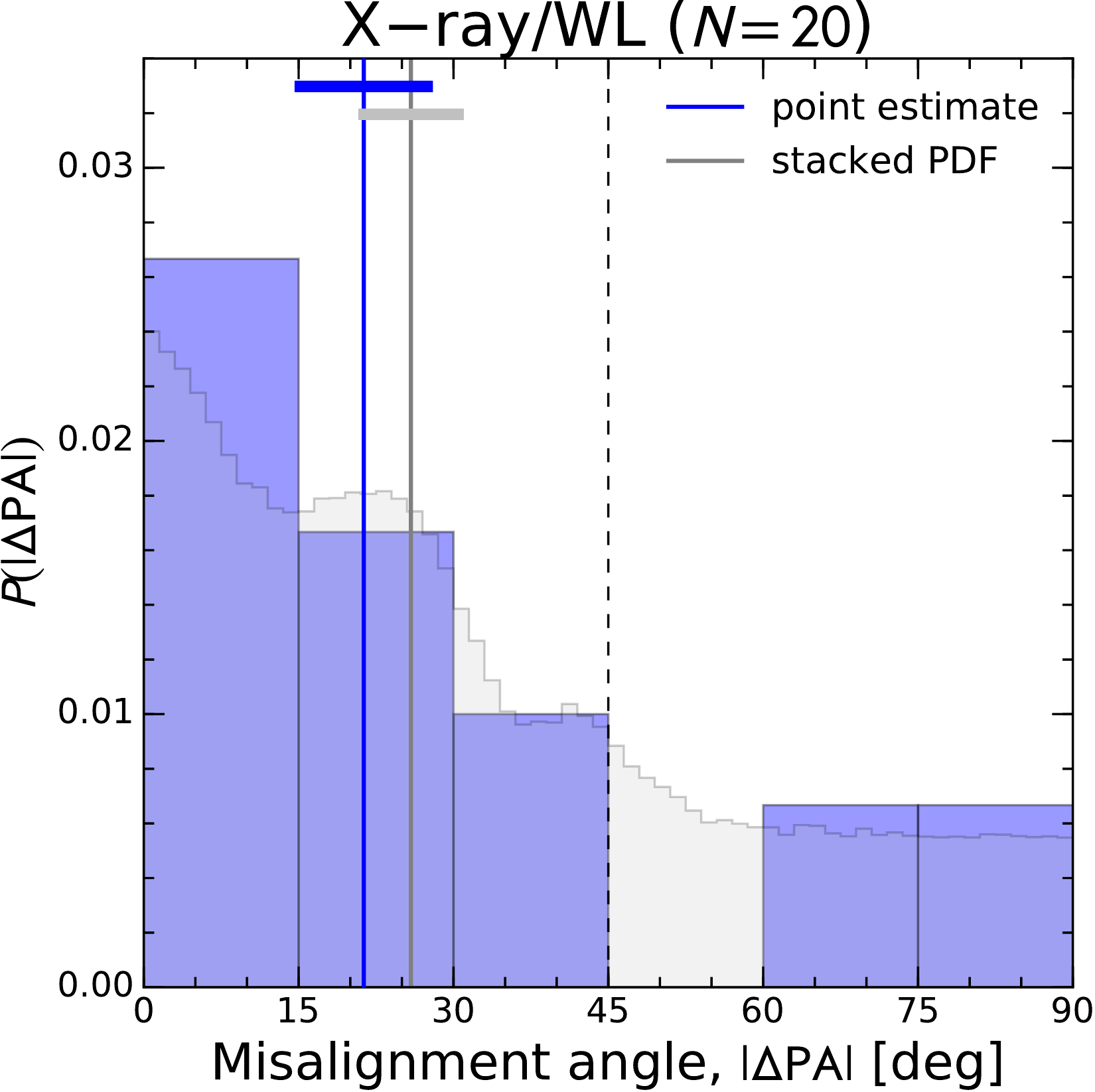}
   \vspace{0.1in}
  \end{array}
  $
  $
  \begin{array}
  {c@{\hspace{0.1in}}c@{\hspace{0.1in}}c}
   \includegraphics[scale=0.48, angle=0, clip]{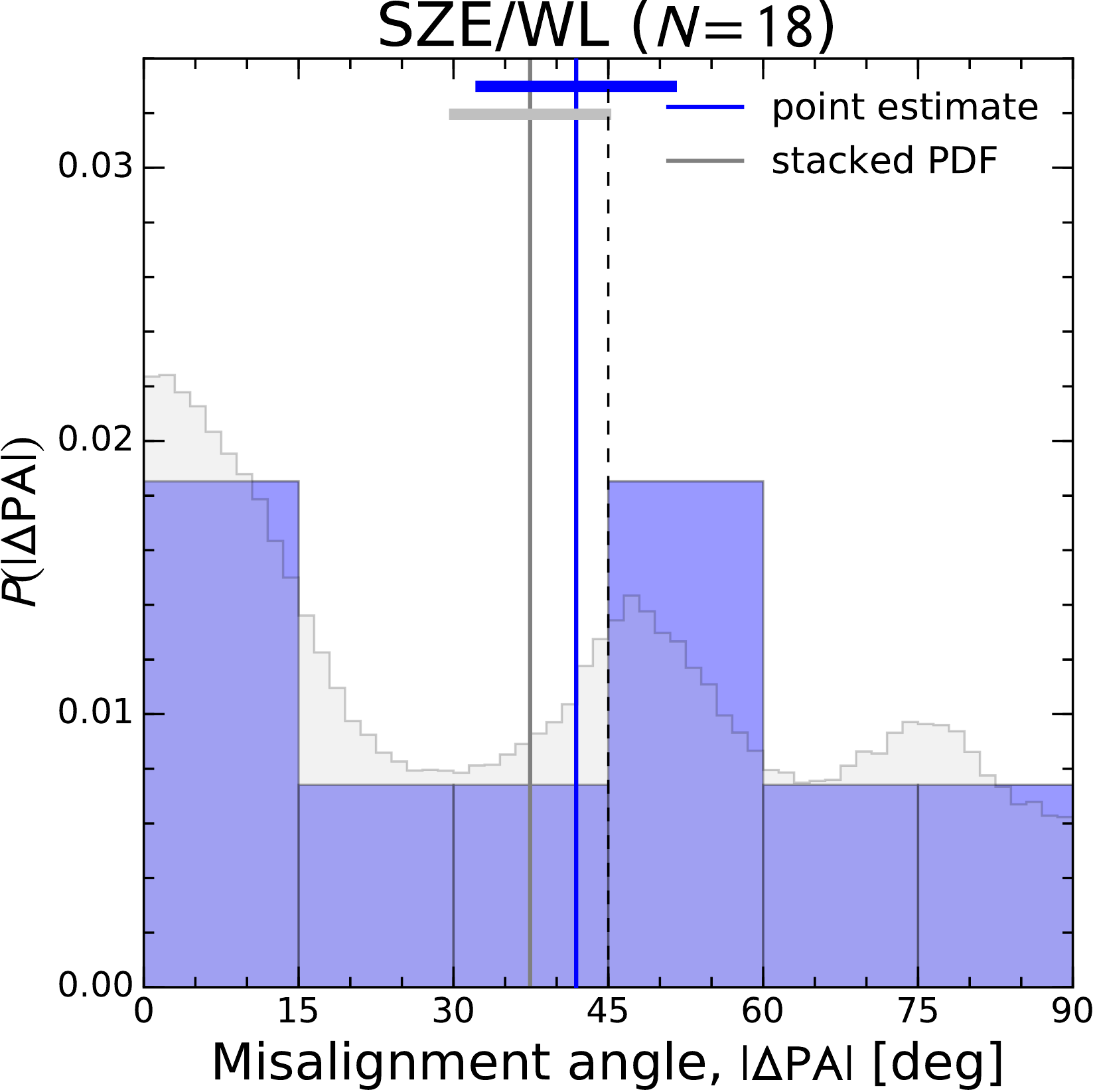}&  
   \includegraphics[scale=0.48, angle=0, clip]{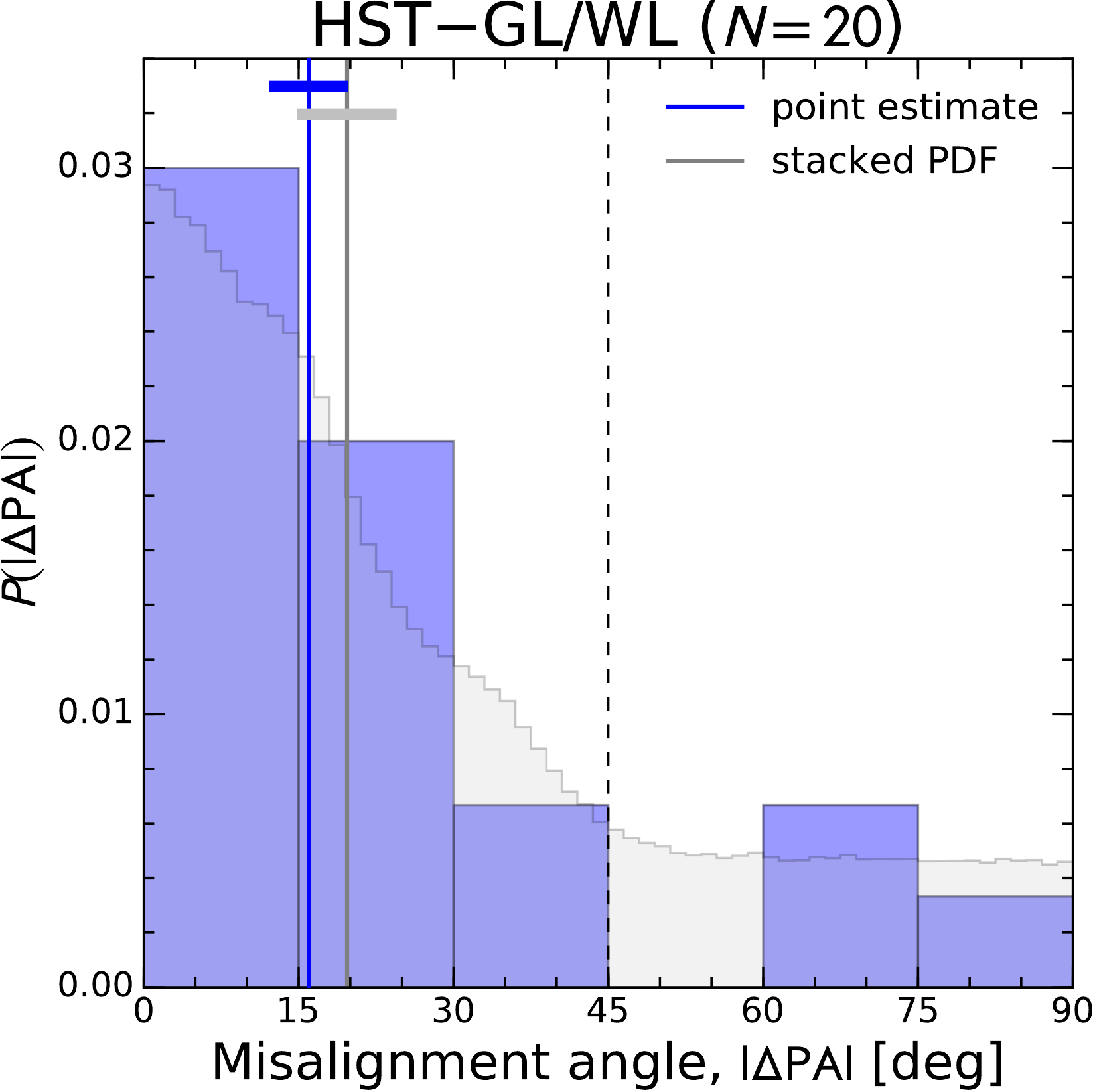}
  \end{array}
  $
 \end{center}
 \caption{\label{fig:misalignment}
 {Histogram of the absolute misalignment angles $|\Delta\mathrm{PA}|$
 between the BCG/WL (upper left), X-ray/WL (upper right), SZE/WL (lower
 left), and {\em HST}-GL/WL (lower right) major axes.} 
 In each panel, the blue and gray shaded histograms show the
 distributions constructed from posterior point estimates
 ($C_\mathrm{BI}$) and from the stacked composite PDF,
 respectively. 
 For each histogram, the median value of the distribution is indicated
 by a vertical line, and its $1\sigma$ error by a thick horizontal line. 
In the absence of any alignment, the expected median value is
 45\,degrees (vertical dashed line).} 
 \end{figure*}

\input{table3.tex}

Next, we quantify the alignment of the different components in the clusters 
by constructing the probability distributions of the absolute misalignment
angles $|\Delta\mathrm{PA}|$
\citep[e.g.,][]{Faltenbacher2009,Zhang2013alignment,West2017}
with respect to the weak-lensing halo shape. 
In the absence of any alignment, $|\Delta\mathrm{PA}|$
 follows a uniform distribution between $0^\circ$ and $90^\circ$,
 with a mean (median) of $\langle|\Delta\mathrm{PA}|\rangle=45^\circ$.
Values of $\langle|\Delta\mathrm{PA}|\rangle < 45^\circ$ indicate that
the PAs of the two distributions are, on average, aligned
 parallel with each other.

We show in Figure \ref{fig:misalignment} the histogram distributions of
$|\Delta\mathrm{PA}|$  
between the BCG/WL, X-ray/WL, SZE/WL, and {\em HST}-GL/WL orientations
derived for our sample.
Table \ref{tab:PA} lists, for the respective comparisons, the median
values of $|\Delta\mathrm{PA}|$ and the corresponding significance
probabilities based on the posterior point estimates of
weak-lensing PAs.
As evident from Table \ref{tab:PA} and Figure \ref{fig:misalignment},
among the three baryonic tracers studied here, the X-ray morphology is 
best aligned with the weak-lensing mass distribution. For this
comparison, we find a median misalignment angle of
$\langle|\Delta\mathrm{PA}|\rangle=21^\circ\pm 7^\circ$ (Table \ref{tab:PA}),
corresponding to $3.6\sigma$ significance with respect to the
null hypothesis of random orientations.
However, these median values are biased high relative to
their intrinsic values because they are estimated from the noisy
distributions. If the intrinsic PA difference
$\Delta\mathrm{PA}=\mathrm{PA(Xray)}-\mathrm{PA(WL)}$ follows a
(truncated) Gaussian distribution with a zero mean and a dispersion 
$\sigma_\mathrm{int}$,  
the median of the intrinsic distribution of $|\Delta\mathrm{PA}|$ is 
$\simeq \sigma_\mathrm{int}/1.483$. In
the presence of noise, the apparent dispersion from observations
is increased. We correct for the effect of this noise bias assuming that
the PA errors follow a Gaussian distribution. 
Adopting the typical uncertainties in the PA measurements, we find the
bias-corrected median misalignment angle to be
$\langle|\Delta\mathrm{PA}|\rangle = 16^\circ\pm 7^\circ$.

For the BCG/WL and SZE/WL comparisons, we find weaker alignment signals 
($<45^\circ$), which are consistent with a null detection within the
errors. A weak constraint on the SZE/WL alignment is in line with
expectations accounting for the large errors in both measurements.
On the other hand, the weak signal in the BCG/WL alignment appears to be
contradictory to the high degree of correlation found between the
BCG/WL PAs (Figure \ref{fig:PA})\footnote{Similar results are found when the
Spearman's rank correlation coefficient is used instead of
$r_\mathrm{P}$, as it is invariant under constant shifts.}.
 This is largely because the distribution of
 $\Delta\mathrm{PA}=\mathrm{PA(BCG)}-\mathrm{PA(WL)}$ is not symmetric
 about 
 $\Delta\mathrm{PA}=0$ (Figure \ref{fig:PA}), with a median offset of 
 $\langle\Delta\mathrm{PA}\rangle=-25^\circ\pm 14^\circ$.
For the X-ray/WL, SZE/WL, and {\em HST}-GL/WL comparisons, the median 
offsets are found to be
$\langle\Delta\mathrm{PA}\rangle=-3^\circ\pm 9^\circ$, 
$7^\circ\pm 17^\circ$,
and
$-12^\circ\pm 6^\circ$,  respectively (Table \ref{tab:PA}).

On the other hand, a strong alignment is found between the WL and {\em 
HST}-GL shapes at different radial scales, with a median misalignment
angle of $\langle|\Delta\mathrm{PA}|\rangle=16^\circ\pm 4^\circ$. This 
corresponds to a significance level of $7.7\sigma$ with respect to the
null hypothesis.
Applying the noise-bias correction yields
$\langle|\Delta\mathrm{PA}|\rangle=8^\circ\pm 4^\circ$.
This is consistent with the results of \citet{Despali2016}, who found
from $N$-body simulations that, for cluster-scale halos, the innermost
and outermost mass ellipsoids are aligned with each other within
$10^\circ$. 
We emphasize that this strong alignment signal has been found despite
using two independent data sets ({\em HST} versus ground-based
observations) and substantially different modeling methods.


\subsubsection{Weak-lensing Quadrupole Shear Measurement}
\label{subsubsec:wlquad}


\begin{figure}[!htb] 
 \begin{center}
   \includegraphics[scale=0.28, angle=0, clip]{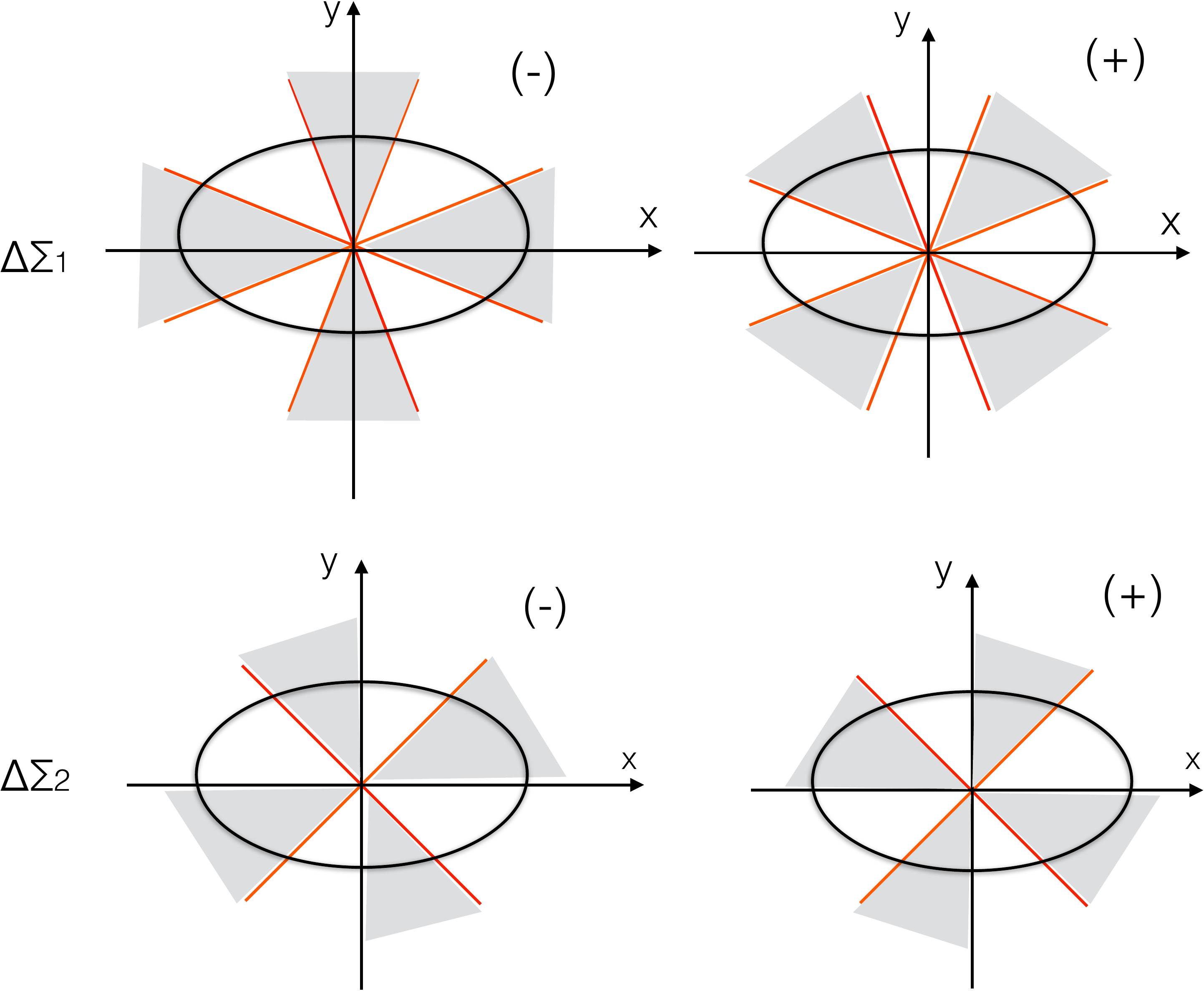}
 \end{center}
 \caption{\label{fig:group}
 Illustration of the Cartesian quadrupole shear estimators of
 \citet{Clampitt+Jain2016}. 
 The $x$-axis of the Cartesian coordinate system is aligned with the
 major axis of the X-ray brightness distribution, assumed to be aligned
 with the major axis of the underlying mass distribution.
 We group together the Cartesian first and second shear components in
 same-sign regions of $\cos{4\theta}$ and $\sin{4\theta}$ (gray shaded
 regions), respectively, and define four quadrupole shear components,
 namely, 
 $\Delta\Sigma_{1}^{(-)}$ (upper left),
 $\Delta\Sigma_{1}^{(+)}$ (upper right),
 $\Delta\Sigma_{2}^{(-)}$ (lower left), and
 $\Delta\Sigma_{2}^{(+)}$ (lower right). 
} 
\end{figure}


\begin{figure*}[!htb] 
  \begin{center}   
  $
  \begin{array}
   {c@{\hspace{0.5in}}c@{\hspace{0.5in}}c}
    \includegraphics[scale=0.45, angle=0, clip]{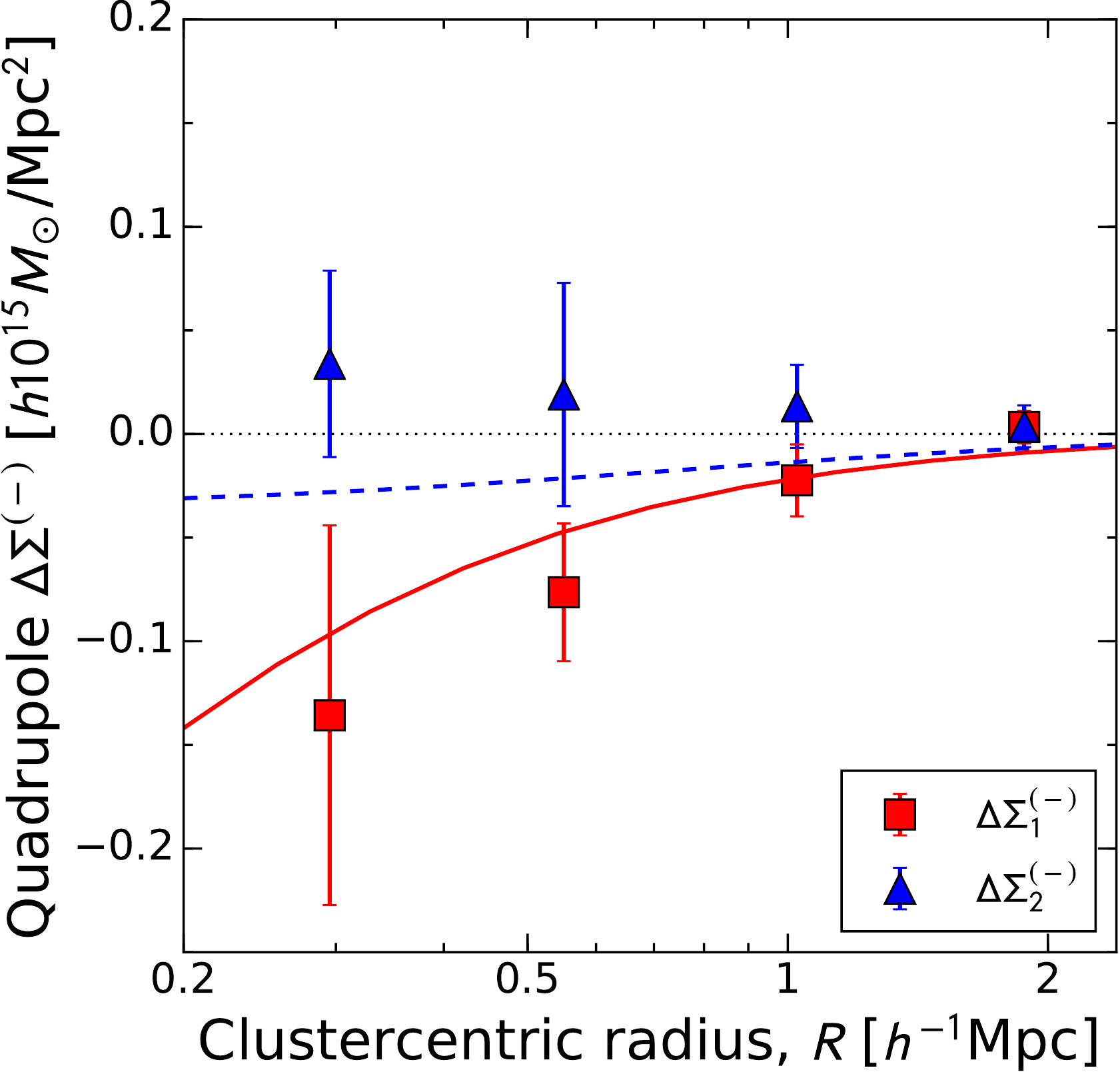} & 
    \includegraphics[scale=0.45, angle=0, clip]{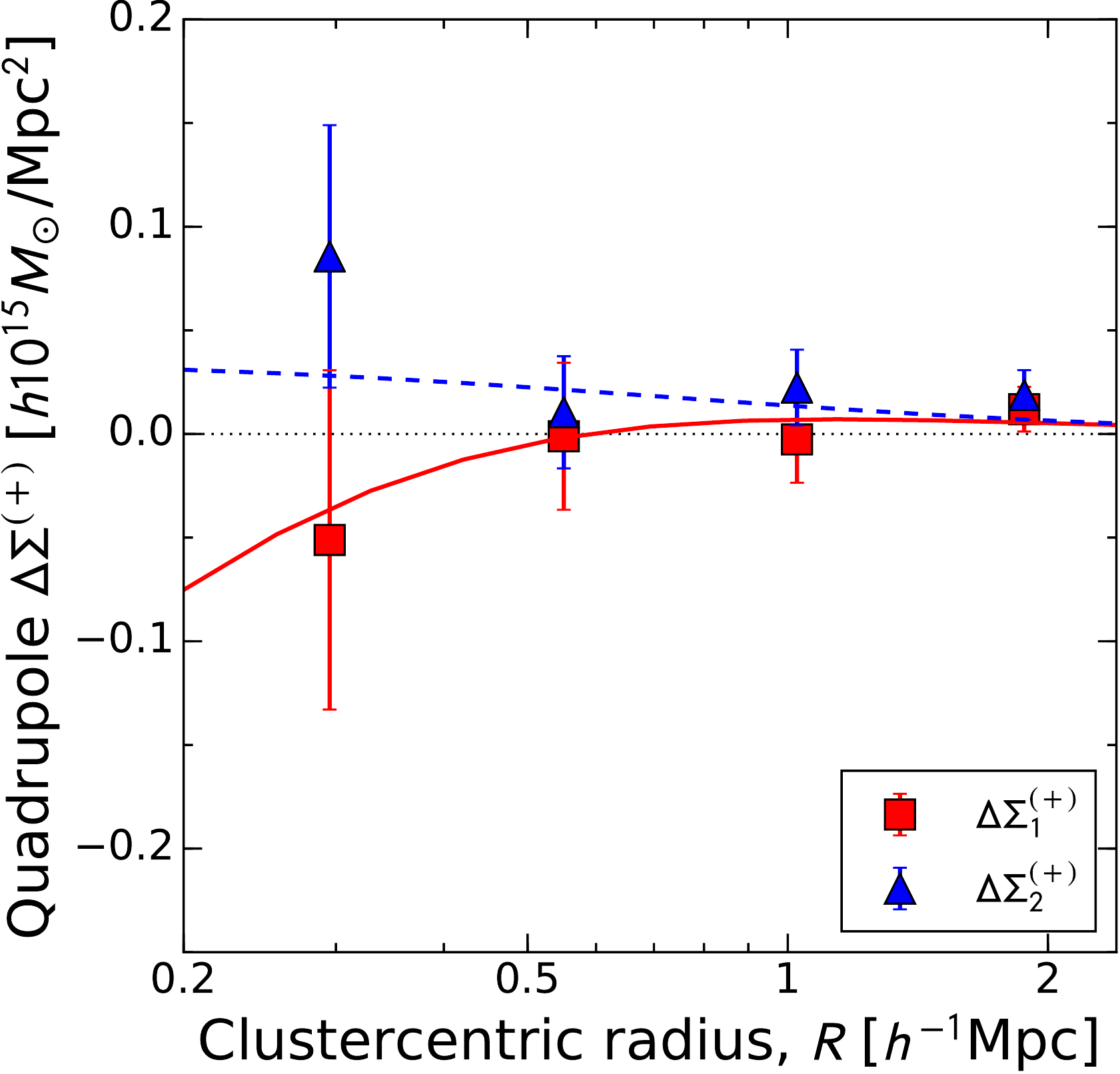}
   \vspace{0.1in}
  \end{array}
  $
  \end{center}
\caption{\label{fig:wlquad_clash}
Stacked quadrupole shear profiles for our sample of 20 CLASH clusters
 measured with respect to the X-ray major axis of each 
 cluster. Left panel: the observed   
 $\llangle\Delta\Sigma_1^{(-)}\rrangle$ (red squares) and
 $\llangle\Delta\Sigma_2^{(-)}\rrangle$ (blue triangles)
 profiles shown along with the best-fit elliptical NFW (eNFW)
 model. Right panel: the same as the left panel, but showing the results
 for the 
 $\llangle\Delta\Sigma_{1,2}^{(+)})\rrangle$ profiles. The best-fit
 model was obtained from a simultaneous eNFW fit to the four quadrupole
 shear profiles.
 }
\end{figure*}

{\red Here} we present a complementary quadrupole shear analysis to test 
the consistency of our cluster ellipticity measurements for
our sample of 20 CLASH clusters. To this end, we employ the
Cartesian estimators of \citet{Clampitt+Jain2016} that null the purely 
tangential, monopole shear contribution. Specifically, we 
measure the stacked quadrupole shear signal with respect to a coordinate 
system with the $x$-axis aligned with the X-ray major axis of each
cluster.  
We adopt the same sign convention for the Cartesian $g_1$ and $g_2$
components as defined in \citet[][see their Figure
1]{Clampitt+Jain2016} and use $\theta$ to denote the azimuthal angle
relative to $x$-axis.
Following \citet{Clampitt+Jain2016}, we group together the first and
second shear components of background galaxies in the regions 
where $\cos{4\theta}$ and $\sin{4\theta}$ have the same sign (Figure
\ref{fig:group}), respectively, and define the following estimator: 
\begin{equation}
 \label{eq:wlquad}
   \begin{aligned}
    \Delta\Sigma_\alpha^{(s)}(R)&=\Sigma_\mathrm{c}\left[\sum_{k} w_{(k)}\,g_{\alpha,k}\right]
    \left[\sum_{k} w_{(k)}\right]^{-1},
   \end{aligned}
\end{equation}
where we have introduced the notation in analogy to
the tangential shear, which probes the differential surface
mass density $\Delta\Sigma$ \citep[e.g.,][]{Umetsu2014clash}.
Here,
$\Sigma_{\mathrm{c}}\propto \langle\beta\rangle_g^{-1} D_\mathrm{l}^{-1}$
is the source-averaged critical surface density for a given cluster
(Section \ref{subsec:lensing}),
$w_{(k)}$ is the statistical weight for the $k$th background galaxy
(Section \ref{subsubsec:shear}),
and
$k$ runs over all background galaxies that fall in the specified bin,
different for each shear component 
$\alpha$ and sign $s$ \citep{Clampitt+Jain2016}:
$\alpha=1$, $s=-$, $-\pi/8\le \theta_j<\pi/8$;
$\alpha=1$, $s=+$, $\pi/8\le \theta_j<3\pi/8$;
$\alpha=2$, $s=-$, $0\le \theta_j<\pi/4$;
$\alpha=2$, $s=+$, $\pi/4\le \theta_j<\pi/2$.
For each case, the summation in Equation (\ref{eq:wlquad}) also includes
background galaxies lying in symmetrical regions shifted by $\pi/2$,
$\pi$, and $3\pi/2$, as illustrated in Figure \ref{fig:group}.

In this work, we first measure for each cluster the quadrupole shear
profiles $\Delta\Sigma_{\alpha}^{(s)}(R)$ from the background-selected shear
catalog according to Equation (\ref{eq:wlquad}), 
and then stack all clusters together by
\begin{equation}
 \begin{aligned}
 \llangle\Delta\Sigma_\alpha^{(s)}(R)\rrangle &=
  \left[\sum_n W_n\, \Delta\Sigma_{\alpha,n}^{(s)}(R)\right]
  \left[\sum_n W_n\right]^{-1},\\
  W_n &= 1/[\sigma^{(s)}_{\alpha,n}(R)]^2,
 \end{aligned}
\end{equation}
where
$\llangle...\rrangle$ denotes the sensitivity-weighted average over the
cluster sample \citep{Umetsu2014clash,Umetsu2016clash},
$n$ runs over all 20 clusters in our sample, and
$\sigma^{(s)}_{\alpha,n}(R)$ is the uncertainty of
$\Delta\Sigma_{\alpha,n}^{(s)}(R)$ estimated from bootstrap resampling
of the background galaxies.
We estimate the errors of the stacked
$\llangle\Delta\Sigma^{(s)}_{\alpha}(R)\rrangle$ profiles by bootstrap
resampling the cluster sample. The resulting stacked quadrupole profiles
for our cluster sample are shown in Figure \ref{fig:wlquad_clash}. 

\citet{Clampitt+Jain2016} modeled the quadrupole shear signal using a
multipole expansion of the surface mass density of elliptical halos
\citep{Adhikari2015}. However, this method can only be applied to 
the case with a small halo ellipticity, so that the higher-order terms
can be safely ignored.

In order to accurately model the observed signal and to make a
direct comparison with our 2D cluster ellipticity measurements (Section
\ref{subsec:q}), 
we forward-model the stacked quadrupole shear profiles by 
assuming an eNFW halo with the major axis aligned with the
X-ray major axis, that is,
$\Delta\mathrm{PA}=\mathrm{PA(Xray)}-\mathrm{PA(WL)}=0$.
Therefore, any misalignment $|\Delta\mathrm{PA}|>0$ will lead to
dilution of the quadrupole signal and hence underestimation of the halo 
ellipticity. 
We use a Bayesian MCMC approach (see Section \ref{subsubsec:mcmc}) to
simultaneously fit an eNFW model to the four stacked quadrupole
profiles, namely
$\llangle\Delta\Sigma_1^{(-)}\rrangle$,
$\llangle\Delta\Sigma_1^{(+)}\rrangle$,
$\llangle\Delta\Sigma_2^{(-)}\rrangle$,
and
$\llangle\Delta\Sigma_2^{(+)}\rrangle$, each measured in four radial bins
(Figure \ref{fig:wlquad_clash}).
We use a uniform prior distribution for the projected axis ratio in the
range $0.1\le q_\perp\le 1$ (Section \ref{subsec:q}).
We marginalize over the mass and concentration parameters using
Gaussian priors of
$M_\mathrm{200c}=(10.9\pm 0.7)\times 10^{14}\Msunh$
and
$c_\mathrm{200c}=3.3\pm 0.2$ based on the joint strong-lensing,
weak-lensing shear and magnification analysis of
\citet{Umetsu2016clash}. Marginalized posterior constraints
($C_\mathrm{BI}\pm S_\mathrm{BI}$) on the
projected axis ratio are obtained as
 $q_\perp=0.67\pm 0.10$ (Figure \ref{fig:wlquad_clash}), or
 $\epsilon=0.33\pm 0.10$ and
 $e=0.38\pm 0.12$ in terms of the halo ellipticity. These are in
 excellent agreement with the results from the 2D weak-lensing analysis
 of 20 individual clusters (Table \ref{tab:q}), supporting the
 robustness of our results and indicating that the effect of dilution
 due to X-ray/WL misalignment is not significant for our cluster sample.

\subsubsection{Comparison with Previous Cluster-scale Alignment Studies}
 
\citet{Okumura2009} measured intrinsic ellipticity correlation functions
for a large sample of luminous red galaxies (LRGs) in the redshift range
0.16--0.47 selected from the Data Release 6 (DR6) of
the Sloan Digital Sky Survey (SDSS), finding a clear signal up to a
scale of 30$\,\Mpch$. 
To model the observed ellipticity correlation,
they populated galaxies into DM
halos in cosmological $N$-body simulations using a halo occupation
distribution approach. In this context, the fraction of synthetic 
central LRGs is $93.7\percent$, and they are hosted by cluster-scale DM
halos with $M>8\times 10^{13}M_\odot$. The ellipticity correlation is
predicted to have an amplitude that is about four times higher than
their measurement when assuming a perfect alignment between the central
LRGs and their host DM halos. Assuming a Gaussian misalignment between
the major axes of central LRGs and host halos, they found a misalignment 
dispersion of 
$\sigma(\Delta\mathrm{PA})=35.4^{+4.0}_{-3.3}$\,degrees, or
an absolute median of $\langle|\Delta\mathrm{PA}|\rangle= 24^{+3}_{-2}$\,degrees.
A similar
constraint was derived by \citet{Okumura+Jing2009} from the
gravitational shear--intrinsic ellipticity correlation function measured
using the LRG sample. Their constraints on the misalignment angle are in
agreement with our BCG/WL results within the errors (Table
\ref{tab:PA}).

\citet{Huang2016alignment} studied central galaxy
alignments with respect to the spatial distribution of member galaxies
using the red-sequence Matched-filter Probabilistic Percolation
(redMaPPer) cluster catalog based on the SDSS DR8, finding an absolute
mean misalignment angle of $\sim 35^\circ$, corresponding to an absolute
median angle of $\langle|\Delta\mathrm{PA}|\rangle\sim 30^\circ$. This
is in agreement with our estimate for the median misalignment angle between
the BCG/WL major axes (Table \ref{tab:PA}) if assuming that cluster
member galaxies are an unbiased probe of the underlying mass distribution.

{\red
\citet{vanUitert+2017} presented a stacked quadrupole shear analysis of
$\sim 2600$ galaxy groups selected from the Galaxy And Mass Assembly
survey using the galaxy shear catalog from the Kilo Degree Survey. On
small scales ($<250$\,kpc), they found the major axis of the BCG to be
the best proxy of the orientation of the 
underlying mass distribution. On larger scales, a much weaker
correlation was found between the orientations of the BCG
and the mass distribution, while the distribution of member galaxies
provides a better proxy for the orientation of the overall mass
distribution in groups. 

More recently,
\citet{Shin+2018} studied the stacked halo ellipticity for a sample of
$\sim 10^4$ SDSS redMaPPer clusters.
Stacking the quadrupole shear signal along the major axis of the cluster
member distribution, they found a mean axis ratio of 
$0.558\pm 0.086\pm 0.026$ (statistical followed by systematic
uncertainty). This agrees well with the mean axis ratio of the member
distribution, $0.573\pm 0.002\pm 0.039$, indicating that cluster 
galaxies trace the shape of the cluster mass distribution within their
errors. On the other hand, they found an rms misalignment angle of
$30^\circ\pm 10^\circ$ between the  central galaxy and the cluster mass
distribution.  
}

\section{Summary}
\label{sec:summary}

In this paper, we have presented direct reconstructions of the
2D matter distribution in 20 high-mass clusters (Table
\ref{tab:sample}) selected from the CLASH survey
\citep{Postman+2012CLASH}, 
by performing a joint weak-lensing analysis of 2D shear
and azimuthally averaged magnification measurements.
This complementary combination allows for a complete analysis of the
field, effectively breaking the mass-sheet degeneracy.    

We have simultaneously constrained the structure and morphology of each
individual cluster by assuming an elliptical NFW halo.
 We have shown that spherical mass estimates
 of the clusters from azimuthally averaged weak-lensing
 measurements in previous work \citep{Umetsu2014clash} are in excellent
 agreement with our results from the full 2D analysis (Figure \ref{fig:mcomp}).  
 This indicates that systematic effects due to the azimuthal averaging
 applied to the weak-lensing data \citep{Umetsu2014clash}, as well as to
 the details of the inversion procedures (Section \ref{subsec:diff}), are
 not significant in our mass determinations.

 Combining all 20 clusters in our sample, we have measured the
 ellipticity of weak-lensing halos at the $5\sigma$
 significance level {\red within a radial scale of}
 $2\Mpch \sim 1.1\rvir$.
 The median projected axis ratio for the sample is constrained to be
 $\langle q_\perp\rangle=0.67\pm 0.07$
 (Section \ref{subsec:q}), 
which is in agreement with theoretical predictions of
 $\langle q_\perp\rangle=0.59$--$0.60$ from   
 recent $N$-body simulations \citep{Bonamigo2015,Suto2016} based on the
 standard collisionless $\Lambda$CDM model.
 However, we note that we expect the average axis ratio of the CLASH
 sample to be high due to a high fraction of relaxed clusters
 \citep{Meneghetti2014clash}. 
 Hence, there could be a bias toward higher values of $q_\perp$ in our
 sample. A more quantitative comparison with theoretical expectations
 would thus require a detailed modeling of baryonic physics, accounting
 for the selection function.
 
 We have studied the misalignment statistics of the BCG, X-ray, SZE, and
 {\em HST}-lensing morphologies
 \citep{Donahue2015clash,Donahue2016clash} with respect to our
 wide-field weak-lensing maps (Section \ref{subsec:PA}). 
 Among the three baryonic tracers studied in this paper (i.e., BCGs,
 X-ray maps, and SZE maps), we find that the X-ray morphology is best
 aligned with the weak-lensing mass distribution 
 (Table \ref{tab:PA} and Figure \ref{fig:misalignment}),
 with a median misalignment angle of 
 $\langle|\Delta\mathrm{PA}|\rangle=21^\circ\pm 7^\circ$. This
 represents a $3.6\sigma$ significance with respect to the null hypothesis  
 of random orientations.
 Adopting the typical uncertainties in the PA measurements and
 correcting for the effect of noise bias, we constrain the intrinsic 
 misalignment angle between the X-ray and weak-lensing major axes to be  
 $\langle|\Delta\mathrm{PA}|\rangle = 16^\circ\pm 7^\circ$. 
A strong alignment is found between the weak-lensing and
{\em HST}-lensing major axes determined at different radial
 scales, with a median misalignment angle of  
$\langle|\Delta\mathrm{PA}|\rangle=16^\circ\pm 4^\circ$, corresponding
 to $7.7\sigma$ significance.
After applying the noise-bias correction, we find
$\langle|\Delta\mathrm{PA}|\rangle=8^\circ\pm 4^\circ$.
This strong alignment signal has been found despite using two
independent data sets ({\em HST} versus ground-based 
observations) and substantially different modeling methods.

We also conducted a complementary stacked weak-lensing analysis of the 
20 clusters using the quadrupole shear estimators of
\citet{Clampitt+Jain2016}. Assuming that the X-ray brightness
distribution is aligned with the projected mass distribution, we have
obtained stacked constraints on the eNFW $q_\perp$ parameter of
$0.67\pm 0.10$ (Section \ref{subsubsec:wlquad}), in
 excellent agreement with the results from our 2D weak-lensing analysis.
 This consistency supports the robustness of our results and suggests
 again a tight alignment between the intracluster gas and DM.

{\red We note that while this paper was under review for publication, a
paper by \citet{OkabeT2018} appeared on the arXiv preprint service. They
studied projected alignments of stellar, gas, and DM distributions in
40 cluster halos with $M_\mathrm{200c} > 5\times 10^{13}M_\odot$ 
using cosmological hydrodynamical simulations. They showed that the
total matter distribution is tightly aligned with the X-ray
brightness distribution, with a level of misalignment that is consistent 
with our results, supporting our findings.}

Our observations support scenarios in which different cluster
components, from the innermost region of BCGs to large intracluster 
scales of DM halos, share a similar orientation and formation
history \citep{West2017}.
Our results represent a first critical step in performing a
non-spherical cluster analysis in combination with multiprobe data sets
\citep{Umetsu2015A1689,Sereno2017clump3d}, an aim of the CLUMP-3D program.
In our companion papers
{\red \citep{Chiu2018clump3d,Sereno2018clump3d},}
we explicitly account for the effects of triaxiality in forward-modeling 
our multiprobe data sets,
simultaneously constraining the cluster mass, concentration, triaxial
shape, and orientation from Bayesian inference.
Extending this analysis 
with large, well-controlled samples of clusters from ongoing and planned
surveys, such as the XXL survey \citep{Pierre2016xxl,Pacaud2016xxl},
the Subaru Hyper Suprime-Cam survey
\citep{Miyazaki2018hsc,Miyazaki2018wl,Mandelbaum2018shear,Oguri2018camira}, the Dark
Energy Survey, and the {\em WFIRST} and {\em Euclid} missions,
will be a significant step
forward in understanding the tidal and evolutionary effects of
surrounding LSS on the intracluster mass distribution.


\acknowledgments

This work was made possible by the availability of high-quality 
weak-lensing data produced by the CLASH survey. We are grateful to the
CLASH team who enabled us to carry out this work.
We thank the anonymous referee for constructive suggestions and comments.
We thank Masamune Oguri for making his simulated Subaru Suprime-Cam
observations available to us.
K.U. acknowledges support from the
Ministry of Science and Technology of Taiwan (grants
MOST 103-2112-M-001-030-MY3 and
MOST 106-2628-M-001-003-MY3)
and from the Academia Sinica Investigator Award.
M.S. and S.E. acknowledge financial support from the contracts ASI-INAF
I/009/10/0, NARO15 ASI-INAF I/037/12/0, ASI 2015-046-R.0 and ASI-INAF
n.2017-14-H.0.
Support for D.G. was provided by NASA through Einstein Postdoctoral
Fellowship grant number PF5-160138 awarded by the Chandra X-ray  Center,
which is operated by the Smithsonian Astrophysical Observatory for NASA
under contract NAS8-03060.
T.O. acknowledges support from the Ministry of Science and Technology of 
Taiwan under the grant MOST 106-2119-M-001-031-MY3.
{\red
M.M., M.S., S.E., and J.S. acknowledge support from the Italian Ministry of
Foreign Affairs and International Cooperation, Directorate General for
Country Promotion (Project "Crack the lens").}
J.S. was supported by NSF/AST-1617022.

\input{ms.bbl}
\clearpage

\begin{appendix}

\section{Shape Measurement Test}
\label{appendix:test}

We tested the reliability of our shape measurements of faint background
galaxies by closely following the approach introduced by the STEP program
\citep{2006MNRAS.368.1323H,2007MNRAS.376...13M}. To this end, we used
a set of simulated sky images \citep{Oguri+2012SGAS} that closely match the
characteristics of our Subaru/Suprime-Cam weak-lensing observations,
especially in terms of the angular size of data and the range of
signal-to-noise ratios of objects.
The simulations optimized for our weak-lensing analysis allow us to have
sufficient galaxies detected with high significance, $\nu_g>30$ (Section
\ref{subsubsec:shearmeas}), with which to test our approach to the shear
calibration. Besides, the simulations cover a wide range of input signal
strengths up to $|g|\simeq 0.3$ as found in the inner regions of
clusters, and thus are suitable for cluster weak-lensing studies.
The simulations assume constant shear across each simulated
image, ignoring higher-order lensing effects that are present in
the nonlinear regime close to the Einstein radius
\citep{2005ApJ...619..741G,HOLICs2,Schneider2008}, which is carefully
avoided in our analysis.
This simulation set has been used by several authors to test their
analysis pipelines for cluster weak-lensing work based on
Subaru/Suprime-Cam data
\citep[e.g.,][]{Umetsu+2010CL0024,Umetsu+2012,Oguri+2012SGAS,Okabe+2013,Okabe+Smith2016}. 


\begin{figure}[!htb] 
  \begin{center}
   \includegraphics[scale=0.5, angle=0, clip]{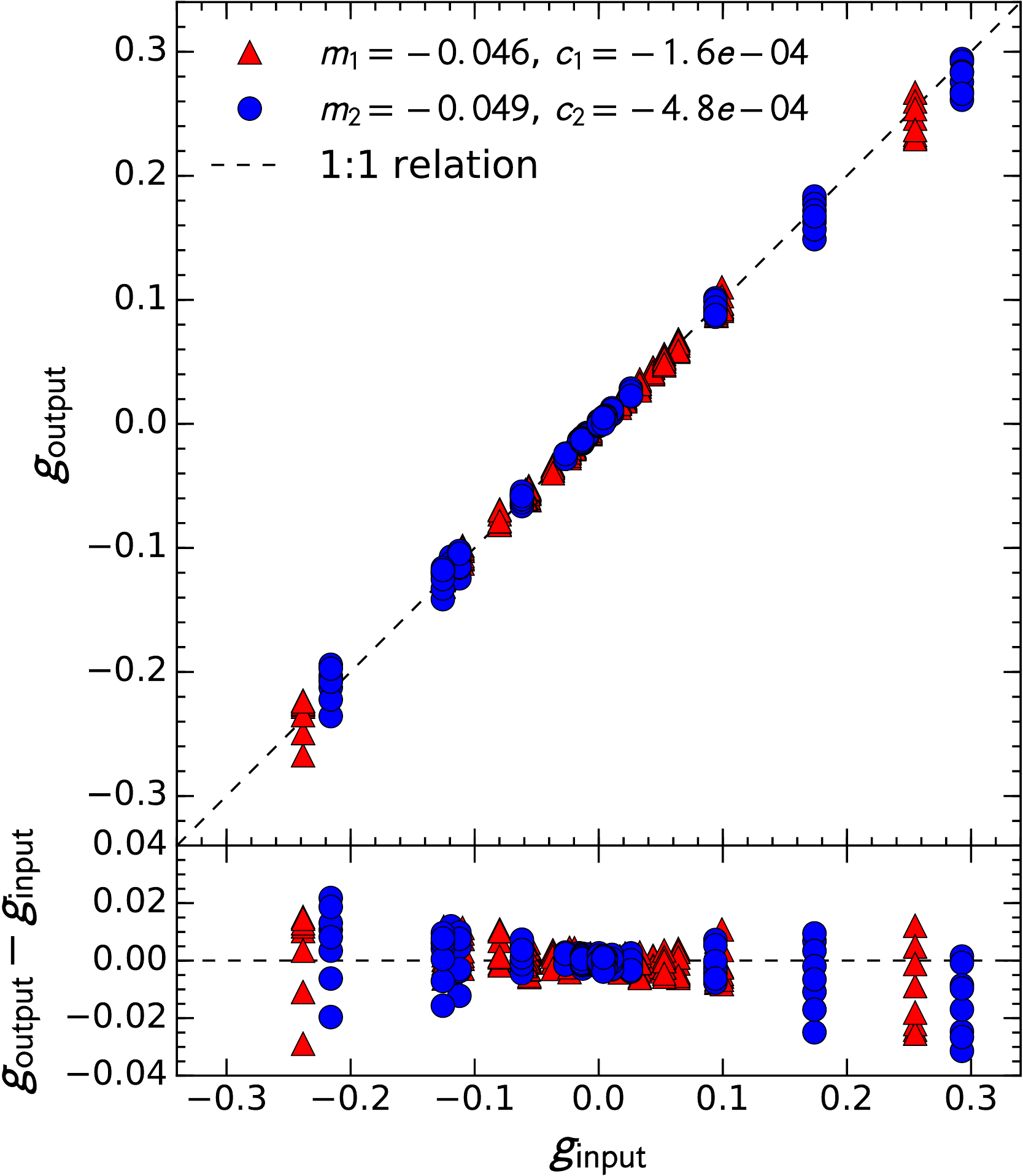}
  \end{center}
 \caption{
Results of the shear measurement test based on simulated
 Subaru/Suprime-Cam images, showing the recovered shear signal
 ($g_\mathrm{output}$) as a function of the input signal ($g_\mathrm{input}$).
 Red triangles and blue circles show the results for $g_1$ and $g_2$, respectively.
The lower panel shows the deviations from the input values,
 $g_\mathrm{output}-g_\mathrm{input}$. The dashed line shows the
 one-to-one relation.   
 \label{fig:test1}}
\end{figure}


\begin{figure*}[!htb] 
  \begin{center}
   $
   \begin{array}
    {c@{\hspace{0.05in}}c@{\hspace{0.05in}}c}
    \includegraphics[scale=0.40, angle=0, clip]{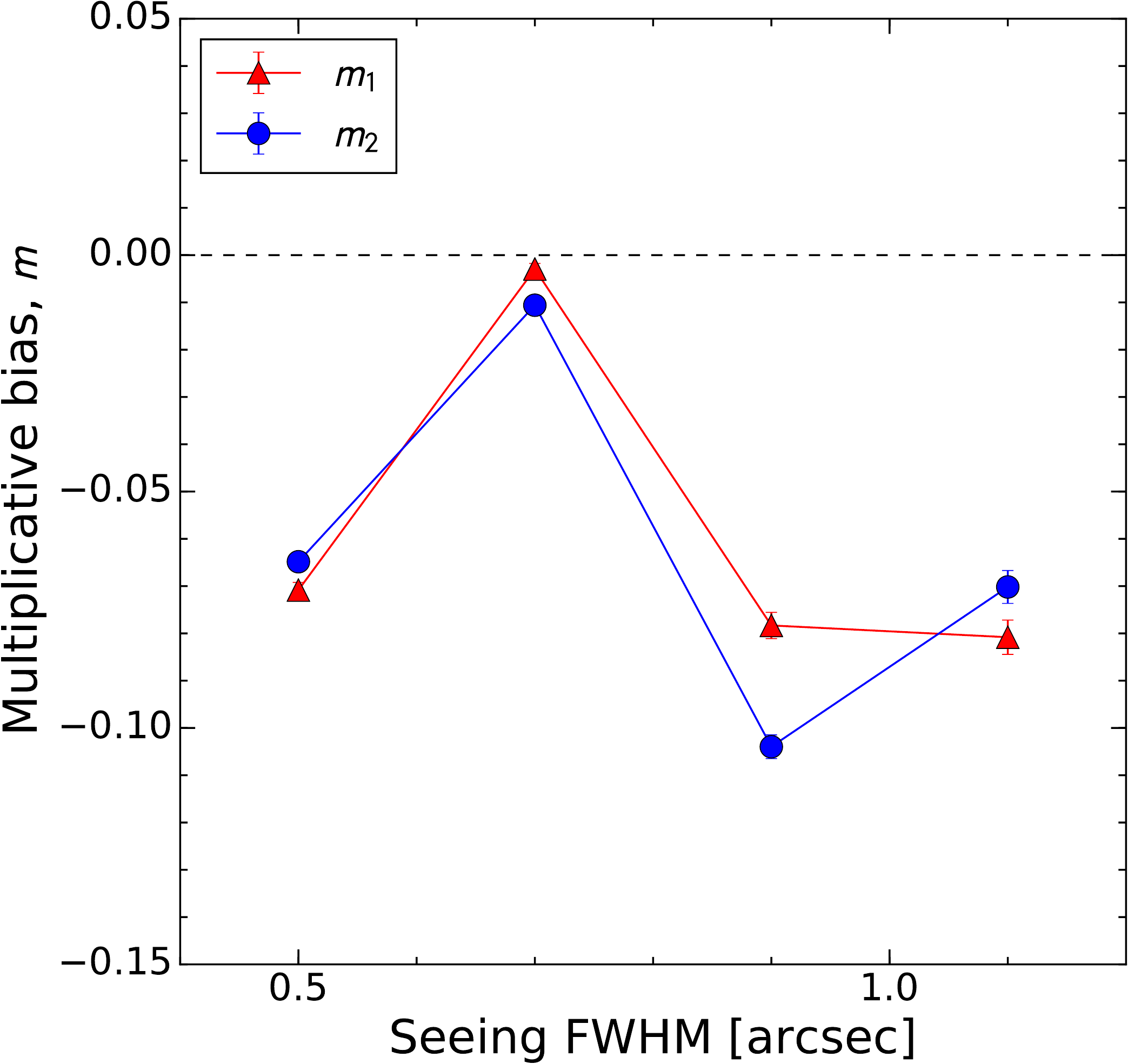}& 
    \includegraphics[scale=0.40, angle=0, clip]{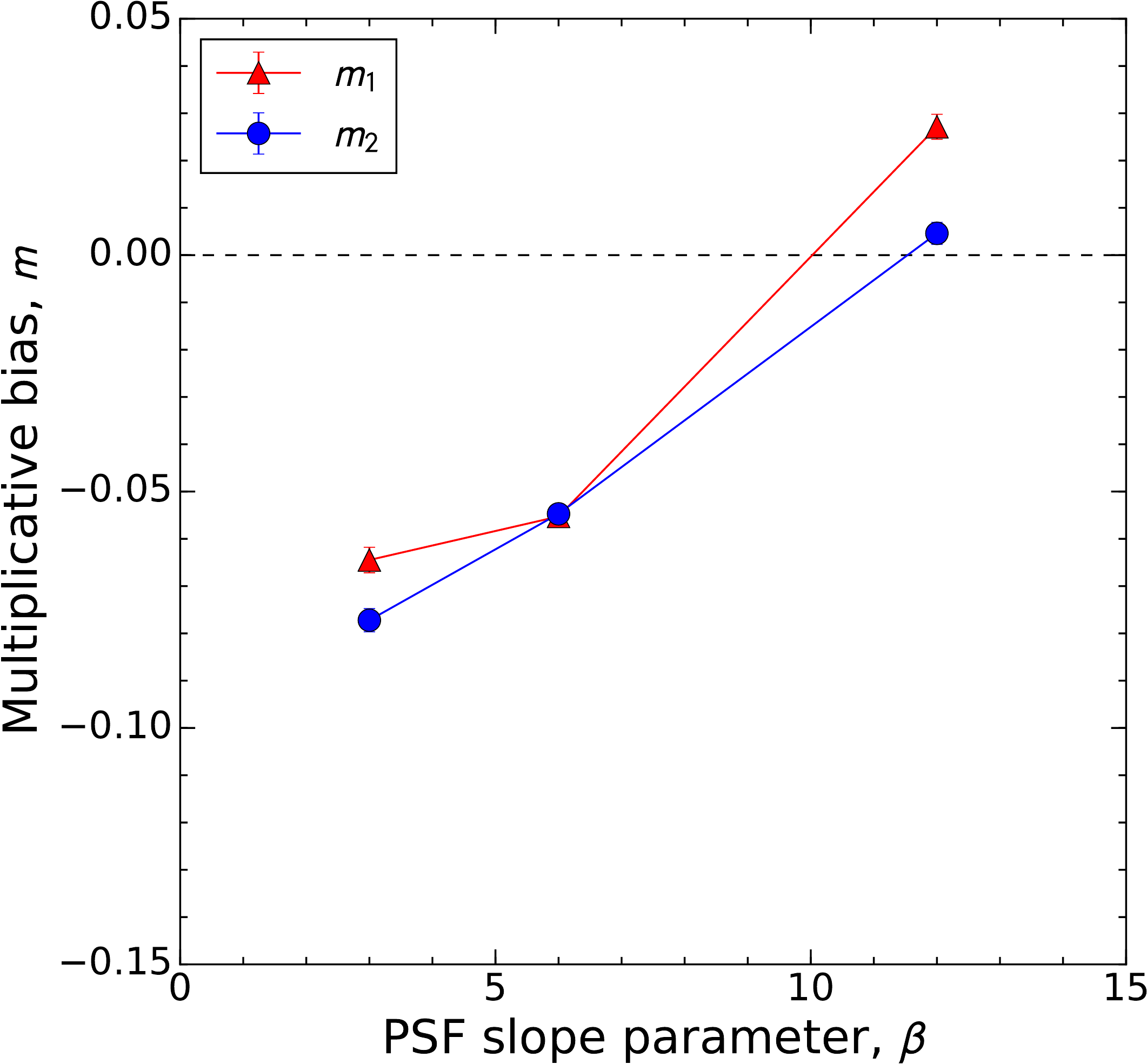}
   \end{array}
   $
  \end{center}
 \caption{Multiplicative shear calibration bias $m_\alpha$ as a function
 of the seeing FWHM and the PSF outer slope parameter $\beta$ obtained using
 imaging simulations that match the characteristics of our
 Subaru/Suprime-Cam weak-lensing data. Red triangles and blue squares
 (with error bars)
 denote $m_1$ and $m_2$, respectively.
 \label{fig:test2}}
\end{figure*}

As described in \citet{Oguri+2012SGAS} \citep[see
also][]{Okabe+Smith2016}, a series of simulated images containing
stars and sheared galaxies was created
using the software packages {\sc stuff} \citep{Bertin2009} and 
{\sc glafic}
\citep{Oguri2010glafic}.
Each model galaxy is characterized by the sum of bulge and disk
components, with S\'ersic profile indices of $n=4$ and $1$,
respectively. 
Galaxy images were convolved with an elliptical PSF model based on the
Moffat profile $\Sigma(R)\propto [1+(R/a)^2]^{-\beta}$
{\red
\citep{Oguri2010glafic},}
with seeing in the range $0.5\arcsec\le \mathrm{FWHM}\le 1.1\arcsec$
and the Moffat power-law index $\beta$ in the range $3\le \beta\le 12$.
A large number of fits frames ($10\mathrm{K}\times 8\mathrm{K}$\,pixels
with $0.2\arcsec$\,pixel$^{-1}$ sampling) matching the Suprime-Cam
field of view were produced
using the {\sc glafic} software.
A total of 160 mock Suprime-Cam images were analyzed using the CLASH
weak-lensing analysis pipeline of \citet{Umetsu2014clash} (Section
\ref{subsubsec:shearmeas}). The results of the shear measurement test
are summarized in Figures \ref{fig:test1} and \ref{fig:test2}.
Averaging over all of the analyzed images, we obtain a multiplicative shear 
calibration bias of $m_1=-0.046$ and $m_2=-0.049$ and an additive bias
of $c_1=-1.6\times 10^{-4}$ and $c_2=-4.8\times 10^{-4}$ (Figure
\ref{fig:test1}; see Equation \ref{eq:gcal}). 
As shown in Figure \ref{fig:test2}, the degree of multiplicative bias
$m_\alpha$ depends on the seeing conditions and the PSF properties to
some degree, so that 
the variation with the PSF properties limits the shear calibration
accuracy to $\delta m_\alpha\sim 0.05$.

\section{Response and Fisher Matrix}
\label{appendix:DD}

In this appendix, we derive analytic expressions for the gradient and
the Fisher matrix of the log-likelihood function for our joint shear and 
magnification weak-lensing analysis in the nonlinear  subcritical regime
of gravitational lensing.
For simplicity of notation, we drop the subscripts ($\infty, g, \mu$)
(Section \ref{sec:basics}) and simply use the symbols $(\kappa,\gamma)$
to denote the lensing fields. Note, in actual calculations, we account for the fact
that the shear and magnification data have different depths,
$\langle W\rangle_g\ne \langle W\rangle_\mu$.
We also use the dimensionless convergence
$\bkappa=\Sigma_{{\mathrm c}}^{-1}\bSigma$,
instead of $\bSigma$, to denote our model vector, so that $\blambda=(\bkappa,\bc)$.
The expectation value of an observable quantity is denoted by a hat symbol.

In the subcritical nonlinear regime of lensing,
the gradient of the log-posterior function $F(\blambda)$ 
(Section \ref{subsec:massrec}) with respect to
$\kappa_n$  ($n=1,2,..., \Npix$) is expressed as
\begin{equation}
   \begin{aligned}
    \nabla_n F(\blambda) &= \nabla_n l_g(\blambda) + \nabla_n l_\mu(\blambda),\\
    \nabla_{n}l_{g}(\blambda)
    &=
    \sum_{k,l=1}^{N_\mathrm{pix}}
    \sum_{\alpha,\beta=1}^2
    ({\cal W}_{g})_{\alpha\beta,kl}
    {\cal S}_{\alpha,kn}
    (\widehat{g}_{\beta}-g_{\beta})_l,\\    
    \nabla_n l_\mu(\blambda) &=
    \sum_{i,j=1}^{\Nbin} 
    ({\cal W}_\mu)_{ij}
    {\cal R}_{in}
    (\widehat{n}_{\mu}-n_{\mu})_j,
   \end{aligned}
\end{equation}
where ${\cal S}_{\alpha,kn} \equiv \nabla_n(\widehat{g}_{\alpha,k})$
and $R_{in}\equiv \partial\widehat{n}_{\mu,i}/\partial\kappa_n$ are
response matrices given by
\begin{equation}
 \begin{aligned}
  {\cal S}_{\alpha,kn} &=
  \frac{1}{1-\kappa(\btheta_k)}\left[
  {\cal D}_{\alpha}(\btheta_k-\btheta_n)+\delta_{kn}\widehat{g}_{\alpha}(\btheta_k)
  \right],\\
  {\cal R}_{in} &= \sum_{k=1}^{\Npix} {\cal P}_{ik}
  \frac{\partial \widehat{n}_{\mu}(\btheta_k)}{\partial \kappa_n}
  =(5s_\mathrm{eff}-2)\sum_{k=1}^{N_\mathrm{pix}}
{\cal P}_{ik} \frac{\widehat{n}_\mu(\btheta_k)}{\Delta(\btheta_k)} \left(
\left[1-\kappa(\btheta_k)\right]\delta_{kn}+\sum_{\alpha=1}^{2}\gamma_{\alpha}(\btheta_k){\cal D}_{\alpha}(\btheta_k-\btheta_n)
  \right).
  \end{aligned}
\end{equation}
In the weak-lensing limit ($\kappa,|\gamma|\ll 1$), the response
matrices reduce to
\begin{equation}
 \begin{aligned}
  {\cal S}_{\alpha,kn} &=
{\cal D}_{\alpha}(\btheta_k-\btheta_n),\\
  {\cal R}_{in} &= 
  (5s_\mathrm{eff}-2)\overline{n}_\mu
  {\cal P}_{in}.
    \end{aligned}
\end{equation}
We also calculate the derivatives of $F(\blambda)$ with respect to 
the calibration parameters $\bc$.

Similarly, an analytic expression for the Fisher matrix ${\cal F}$ can be
derived.
In particular, the Fisher matrix elements $F_{mn}$ ($m,n=1,2,...,\Npix$)
are given by
\begin{equation}
 \label{eq:F}
  \begin{aligned}
{\cal F}_{mn} &= 
\left\langle \frac{\partial^2 l_g(\blambda)}{\partial \kappa_m \partial \kappa_n}
					\right\rangle\Bigg|_{\widehat{\blambda}}
+
\left\langle \frac{\partial^2 l_\mu(\blambda)}{\partial \kappa_m \partial \kappa_n}
										  \right\rangle\Bigg|_{\widehat{\blambda}}\\
 &=
\sum_{k,l=1}^{\Npix}\sum_{\alpha,\beta=1}^2 ({\cal W}_g)_{\alpha\beta,kl}
   {\cal S}_{\alpha,km} {\cal S}_{\beta,ln}
   +
 \sum_{i,j=1}^{\Nbin}
 ({\cal W}_\mu)_{ij}
 {\cal R}_{im} {\cal R}_{jn}.
   \end{aligned}
\end{equation}

\section{Marginalized Posterior Constraints on Elliptical NFW Parameters}
\label{appendix:eNFW}

We show individual cluster constraints on the eNFW model parameters
$(M_\mathrm{200c}, c_\mathrm{200c}, q_\perp, \mathrm{PA})$
for our sample obtained from joint weak-lensing data sets of 2D
gravitational shear and azimuthaly averaged magnification measurements,
showing marginalized 1D and 2D posterior distributions for each
cluster. For each parameter, the blue  solid line denotes the central
location ($C_\mathrm{BI}$) of the marginalized 1D distribution (see
Table \ref{tab:sample}).   


\begin{figure*}[!htb] 
 \begin{center}
 $
 \begin{array}
  {c@{\hspace{1.in}}c@{\hspace{1.in}}c}
  \includegraphics[width=0.4\textwidth,angle=0,clip]{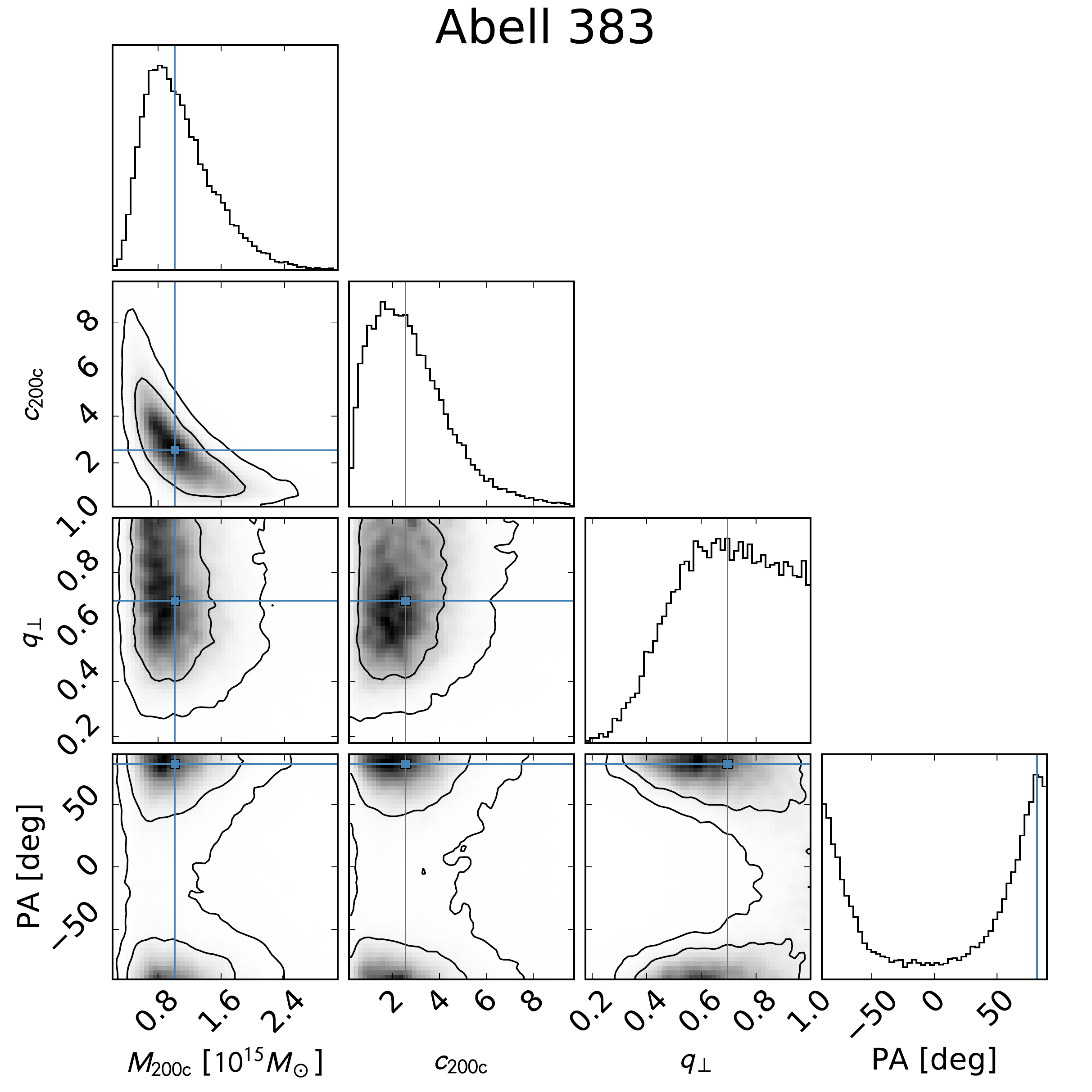} &
  \includegraphics[width=0.4\textwidth,angle=0,clip]{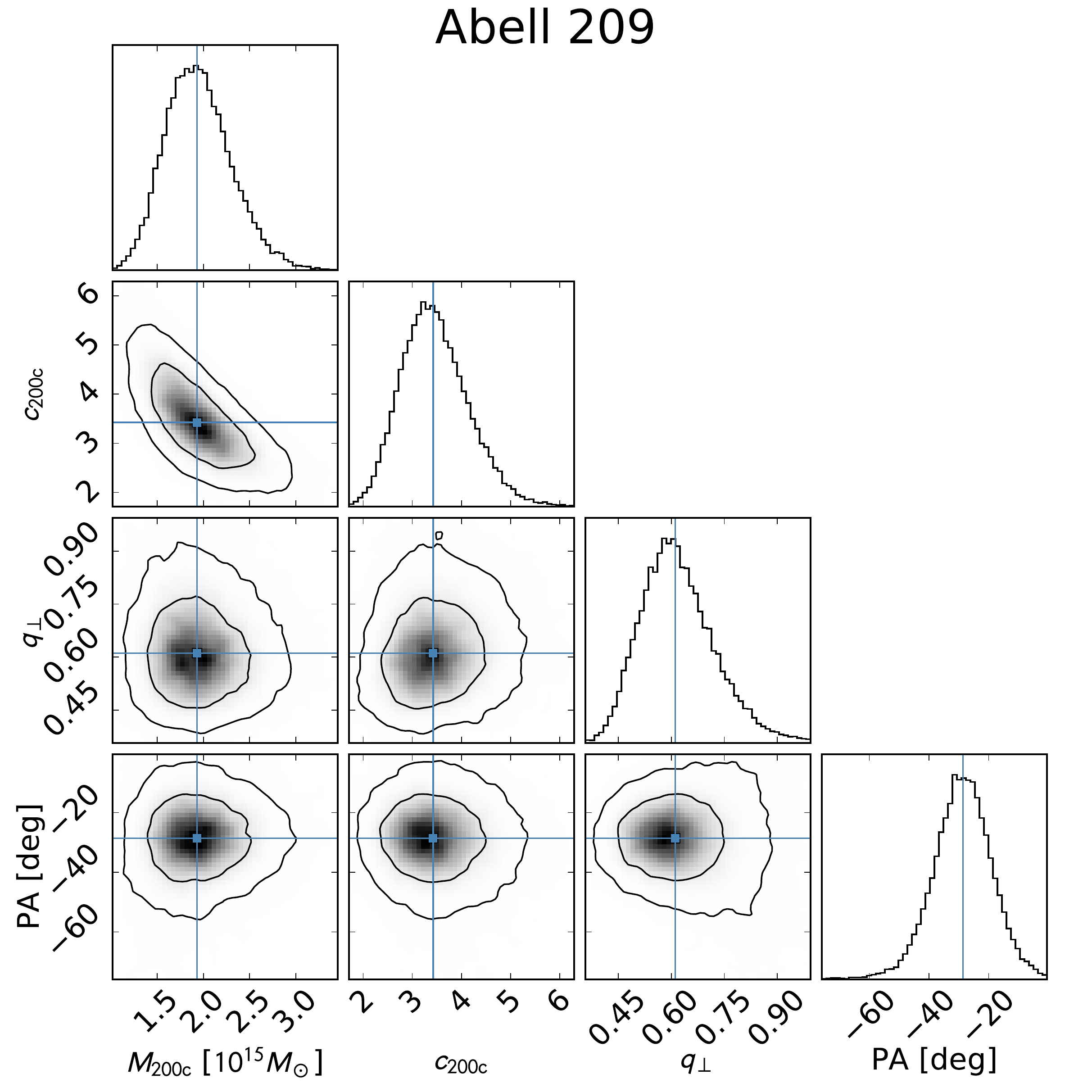} \\
 \end{array}
 $
 $
 \begin{array}
  {c@{\hspace{1.in}}c@{\hspace{1.in}}c}
  \includegraphics[width=0.4\textwidth,angle=0,clip]{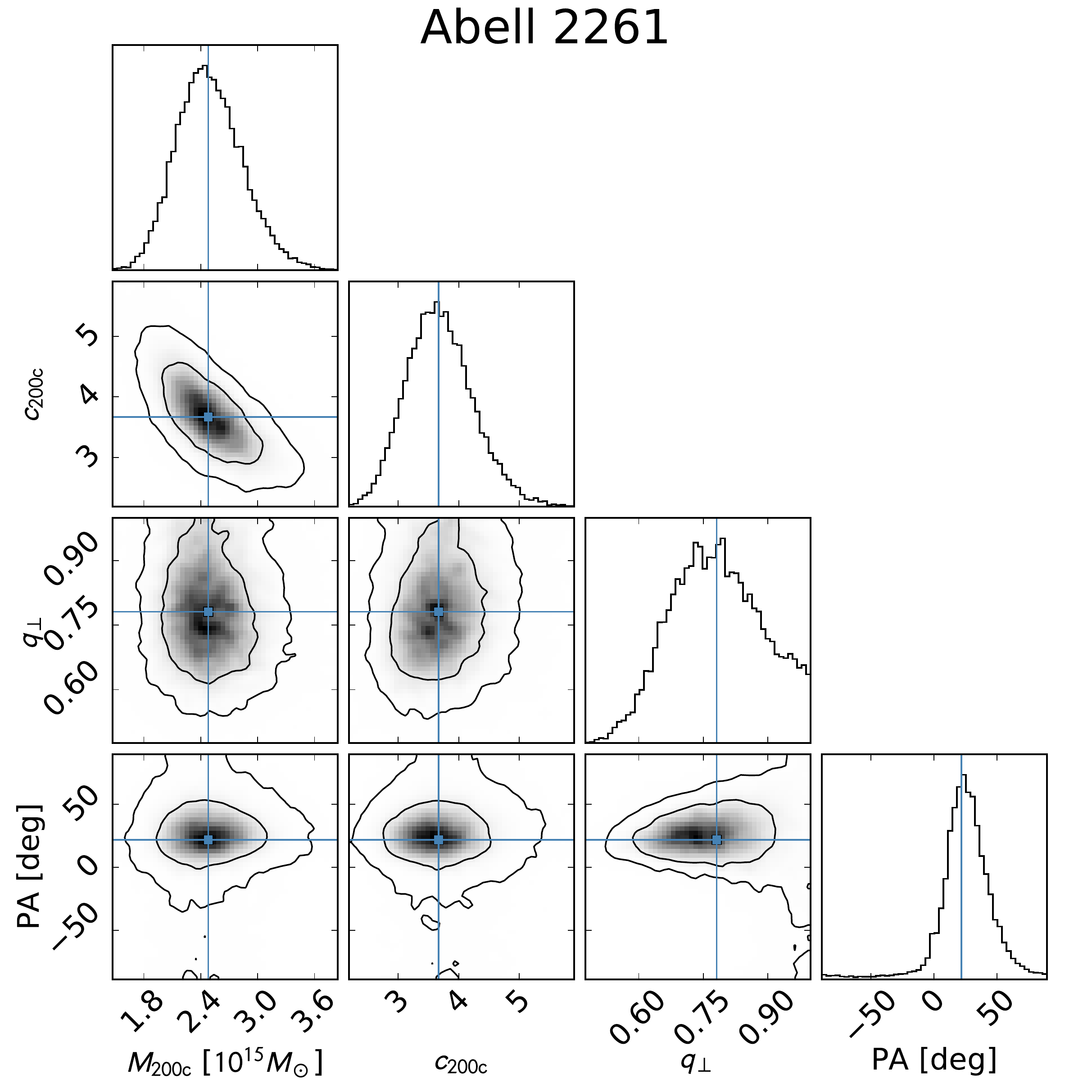} &
  \includegraphics[width=0.4\textwidth,angle=0,clip]{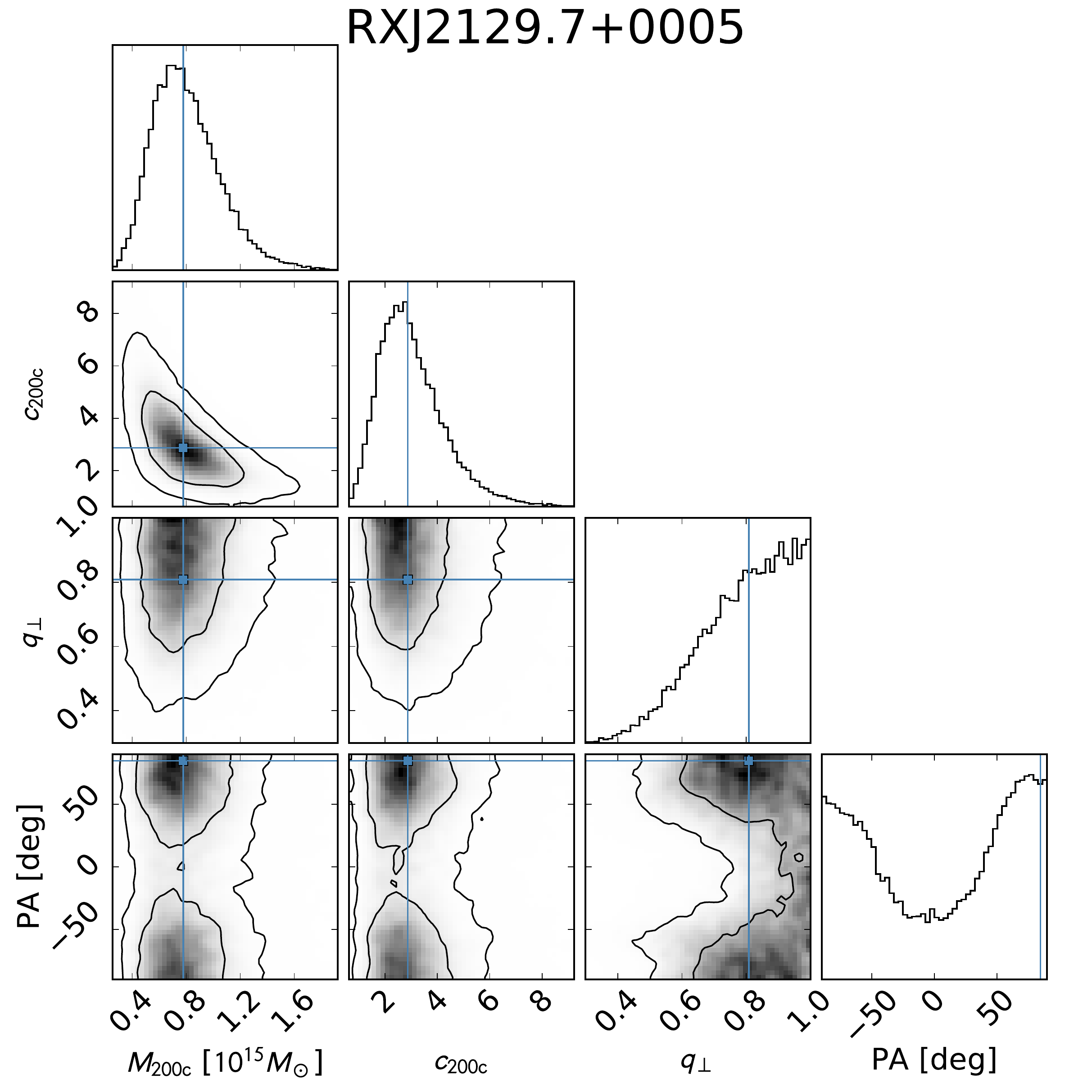} \\
 \end{array}
 $
 $
 \begin{array}
  {c@{\hspace{1.in}}c@{\hspace{1.in}}c}
  \includegraphics[width=0.4\textwidth,angle=0,clip]{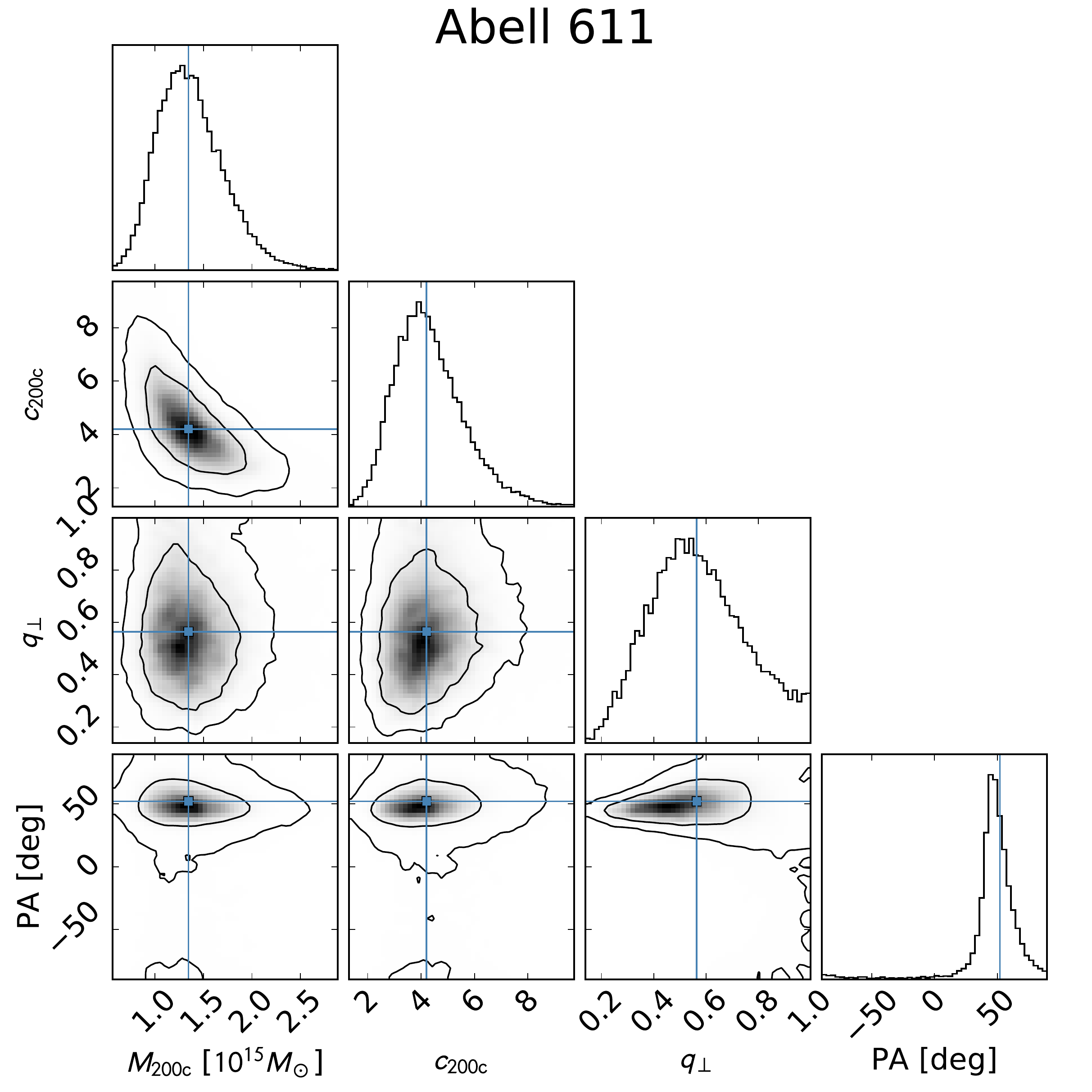} &
  \includegraphics[width=0.4\textwidth,angle=0,clip]{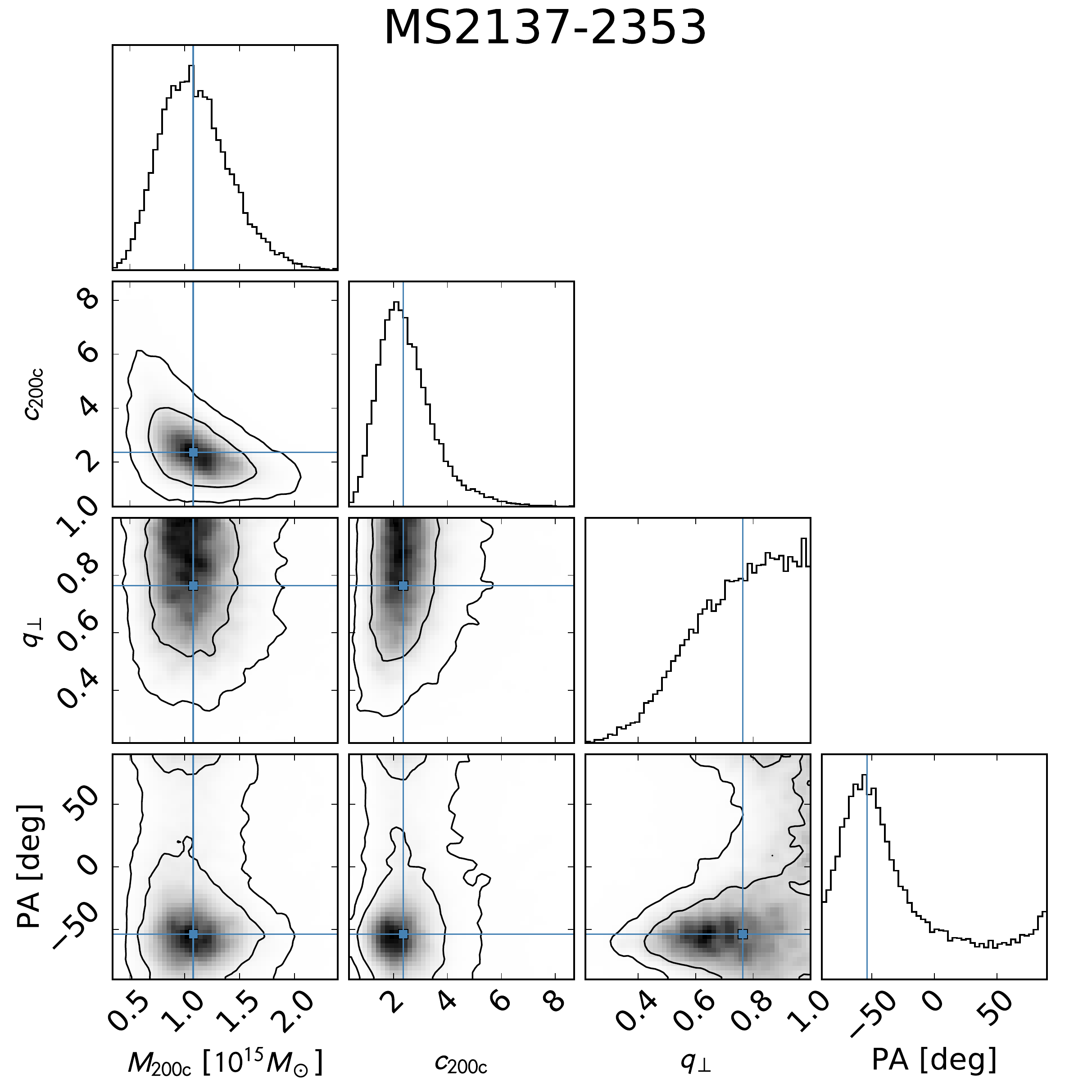} 
 \end{array}
 $
 \end{center}
 \caption{
 Constraints on the eNFW model parameters
 ($M_\mathrm{200c}, c_\mathrm{200c}, q_\perp, \mathrm{PA}$) for
 20 individual CLASH clusters obtained using joint weak-lensing data sets
 of 2D gravitational shear and azimuthally averaged magnification
 measurements, showing marginalized 1D (histograms) and 2D (68\percent\
 and 95\percent\ confidence level contour plots) posterior distributions. 
For each parameter, the blue solid line shows the biweight central
 location ($C_\mathrm{BI}$) of the marginalized 1D distribution. 
 \label{fig:eNFWfit}}
\end{figure*}

\begin{figure*}[!htb] 
\addtocounter{figure}{-1}
 \begin{center}
 $
 \begin{array}
  {c@{\hspace{1.in}}c@{\hspace{1.in}}c}
  \includegraphics[width=0.4\textwidth,angle=0,clip]{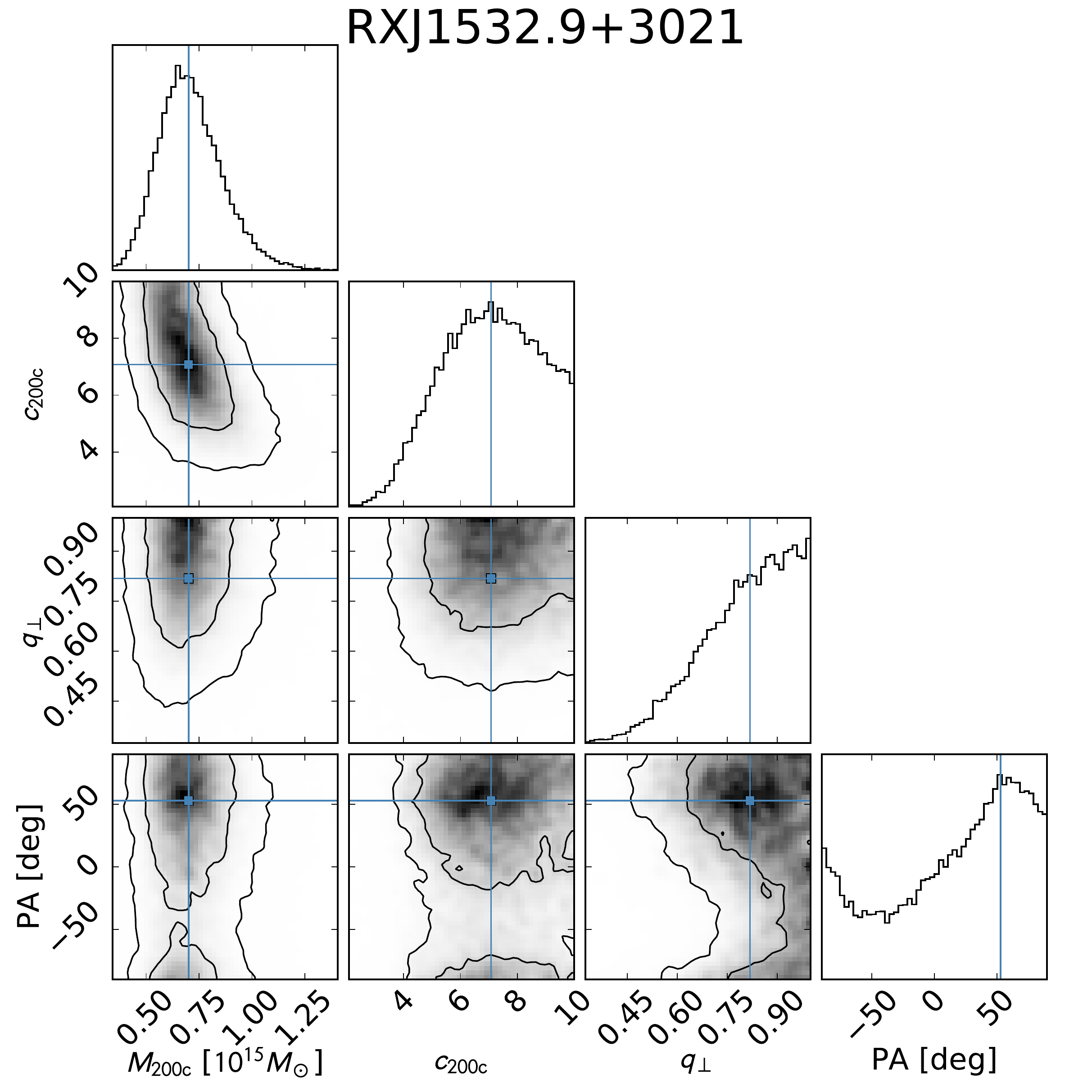}&
  \includegraphics[width=0.4\textwidth,angle=0,clip]{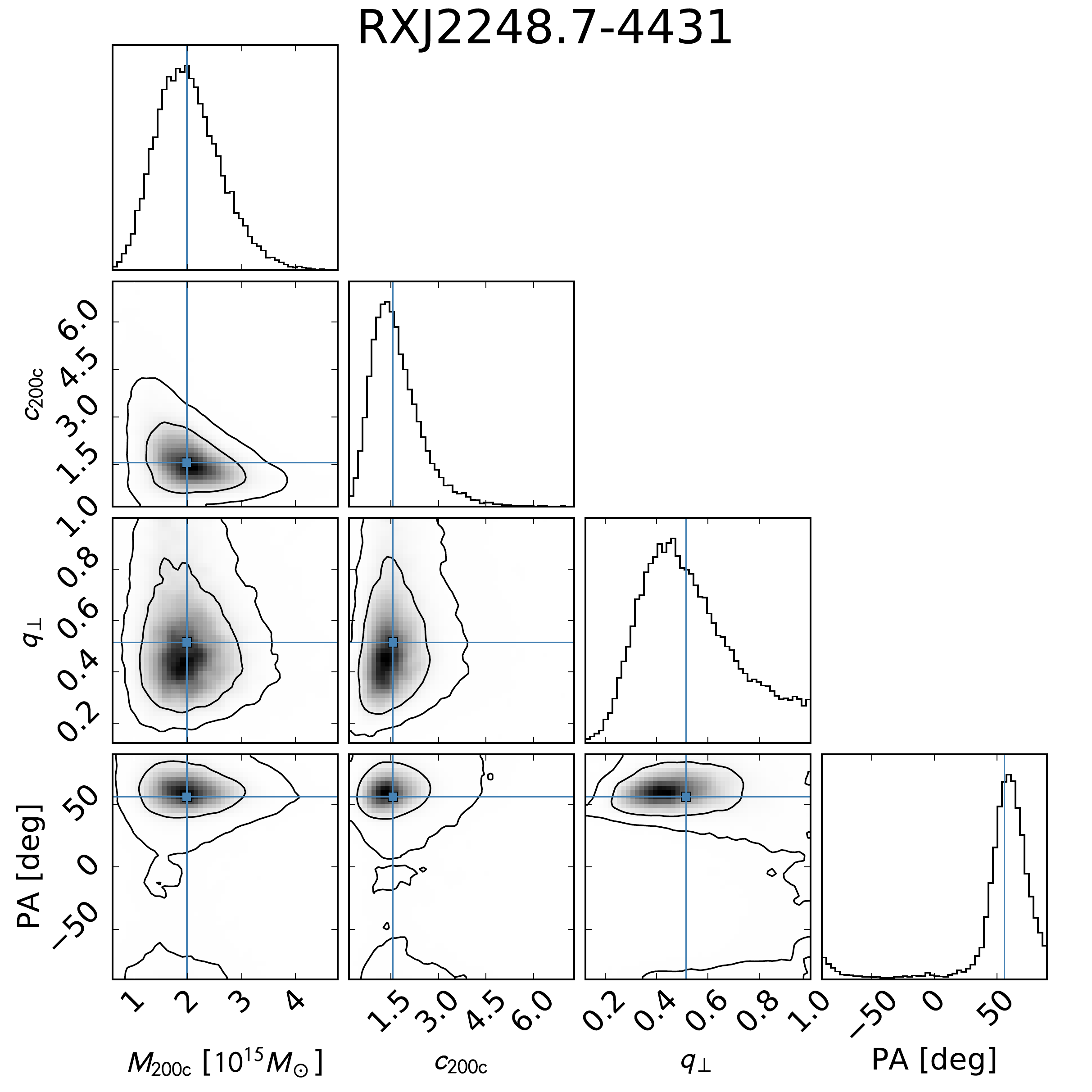}
 \end{array}
 $
 $
 \begin{array}
  {c@{\hspace{1.in}}c@{\hspace{1.in}}c}
   \includegraphics[width=0.4\textwidth,angle=0,clip]{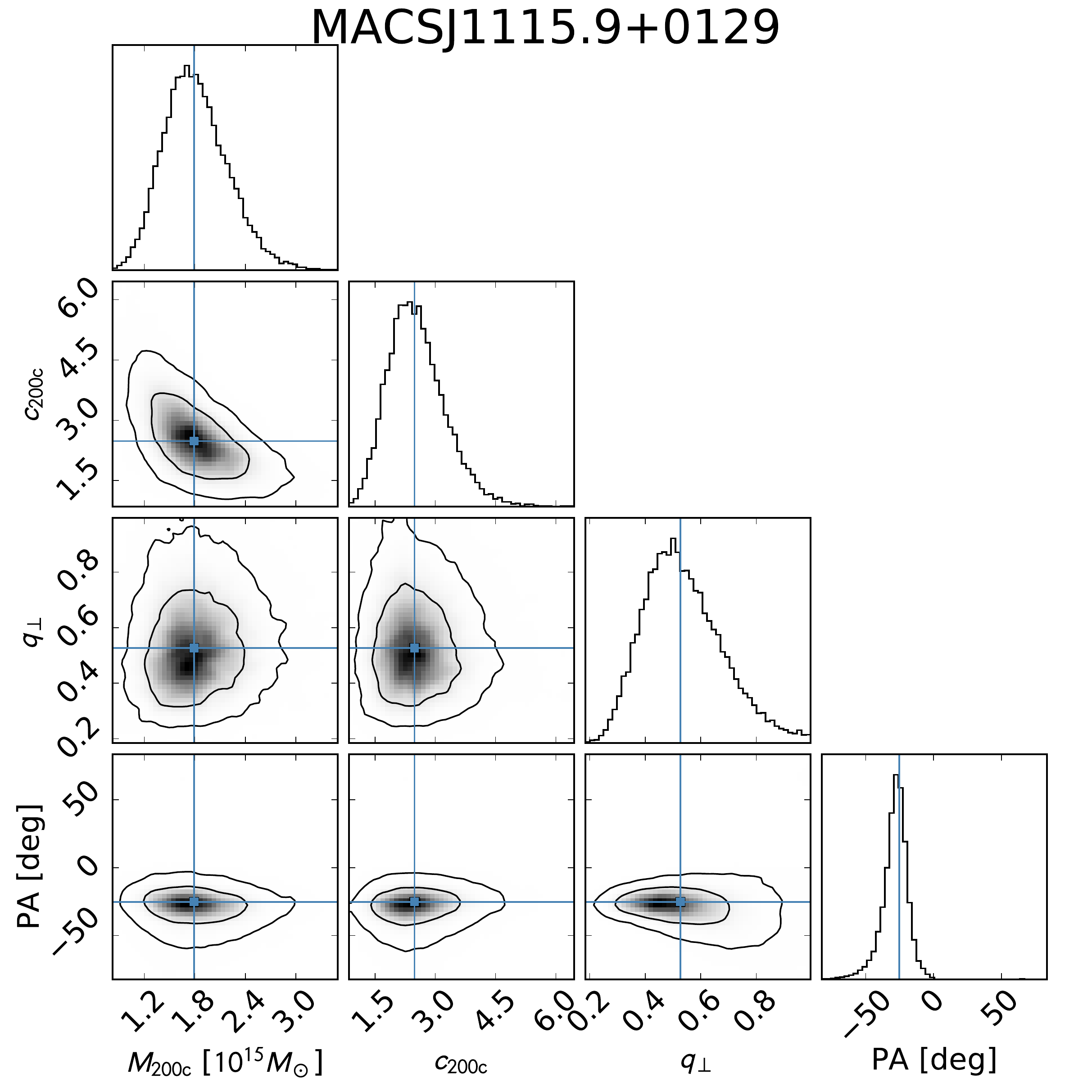}& 
   \includegraphics[width=0.4\textwidth,angle=0,clip]{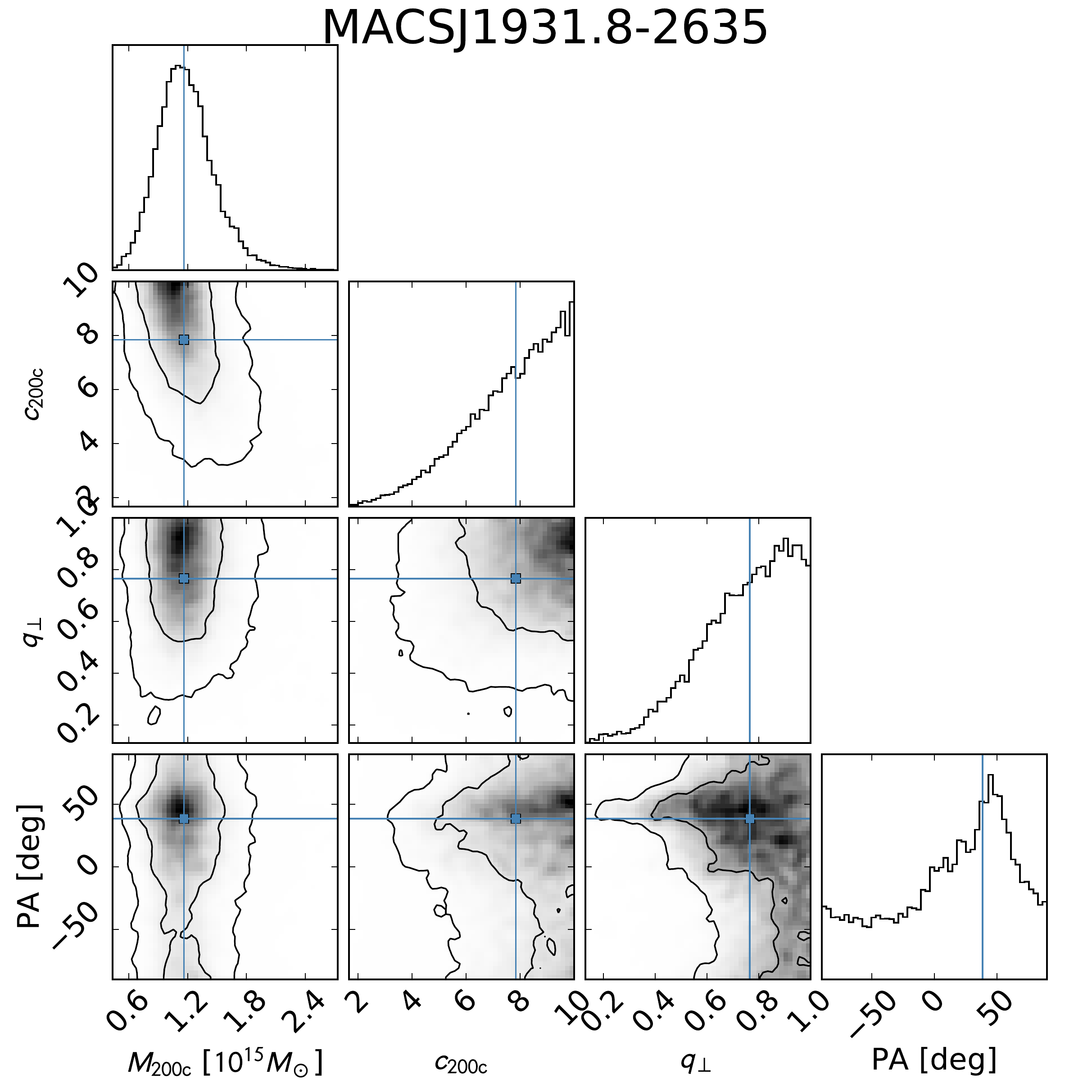}\\
 \end{array}
 $
 $
 \begin{array}
  {c@{\hspace{1.in}}c@{\hspace{1.in}}c}
   \includegraphics[width=0.4\textwidth,angle=0,clip]{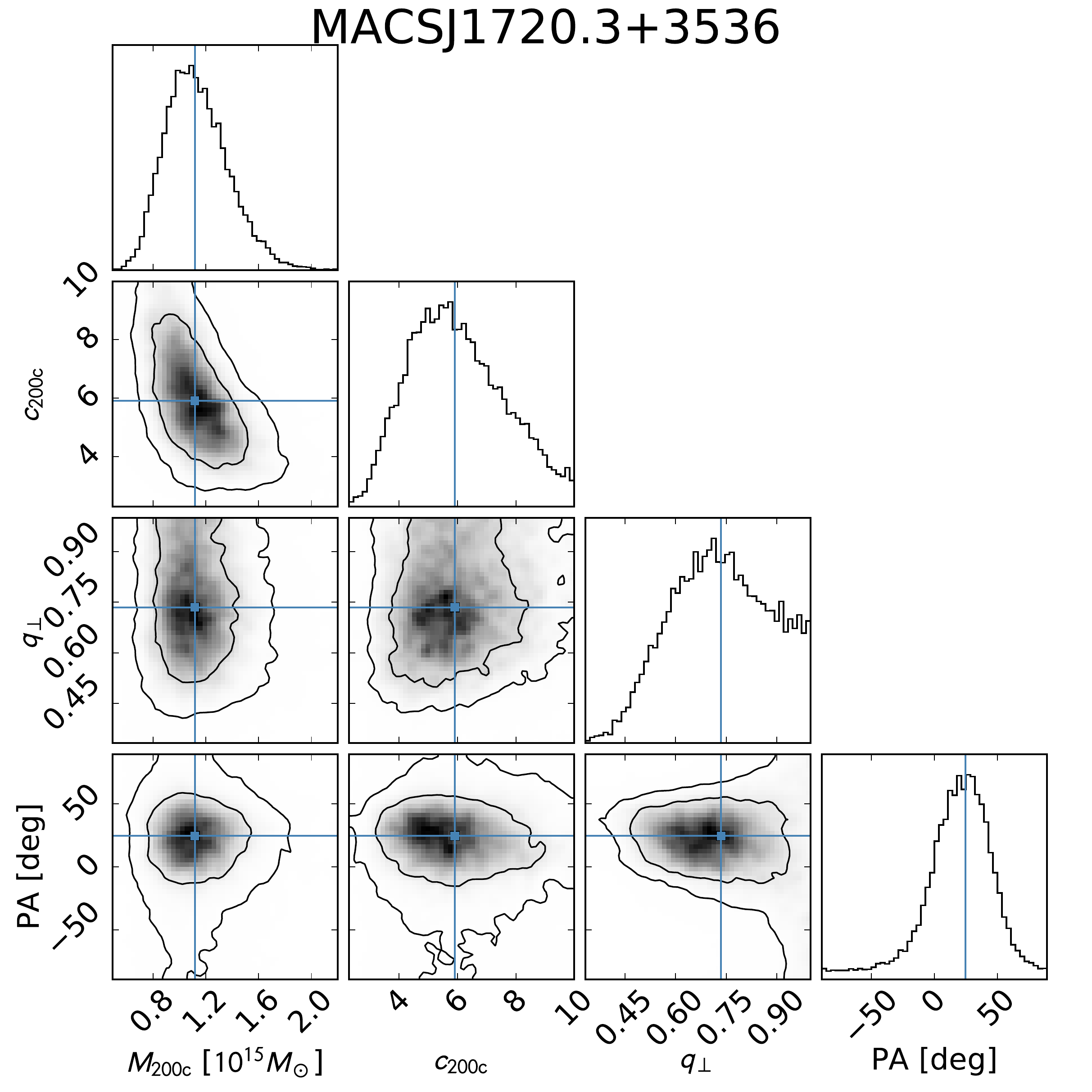}&  
   \includegraphics[width=0.4\textwidth,angle=0,clip]{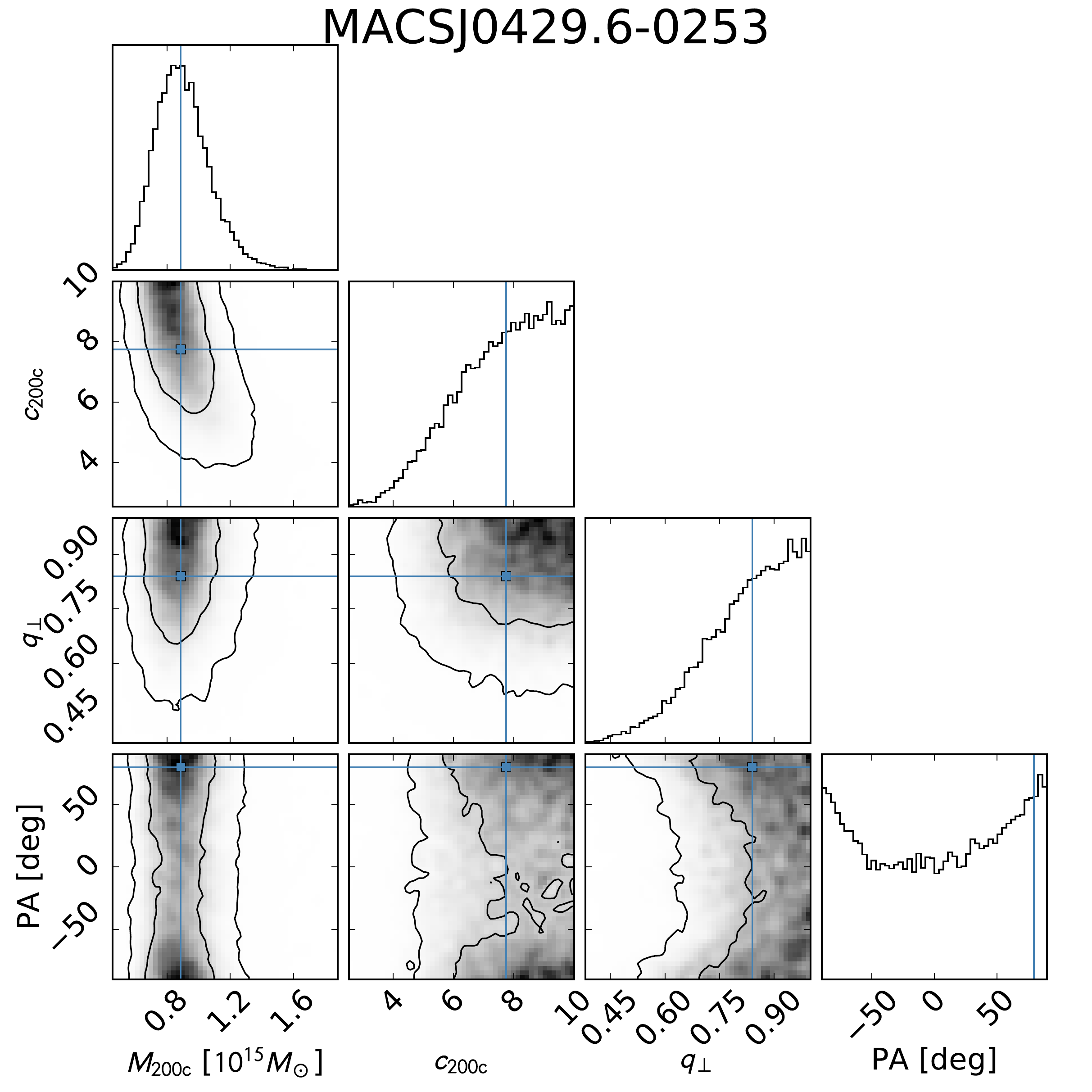}
 \end{array}
 $
 \end{center}
\caption{Continued: Posterior constraints on eNFW model parameters of
 CLASH clusters.}
 \end{figure*}

\begin{figure*}[!htb] 
\addtocounter{figure}{-1}
 \begin{center}
 $
 \begin{array}
  {c@{\hspace{1.in}}c@{\hspace{1.in}}c}
  \includegraphics[width=0.4\textwidth,angle=0,clip]{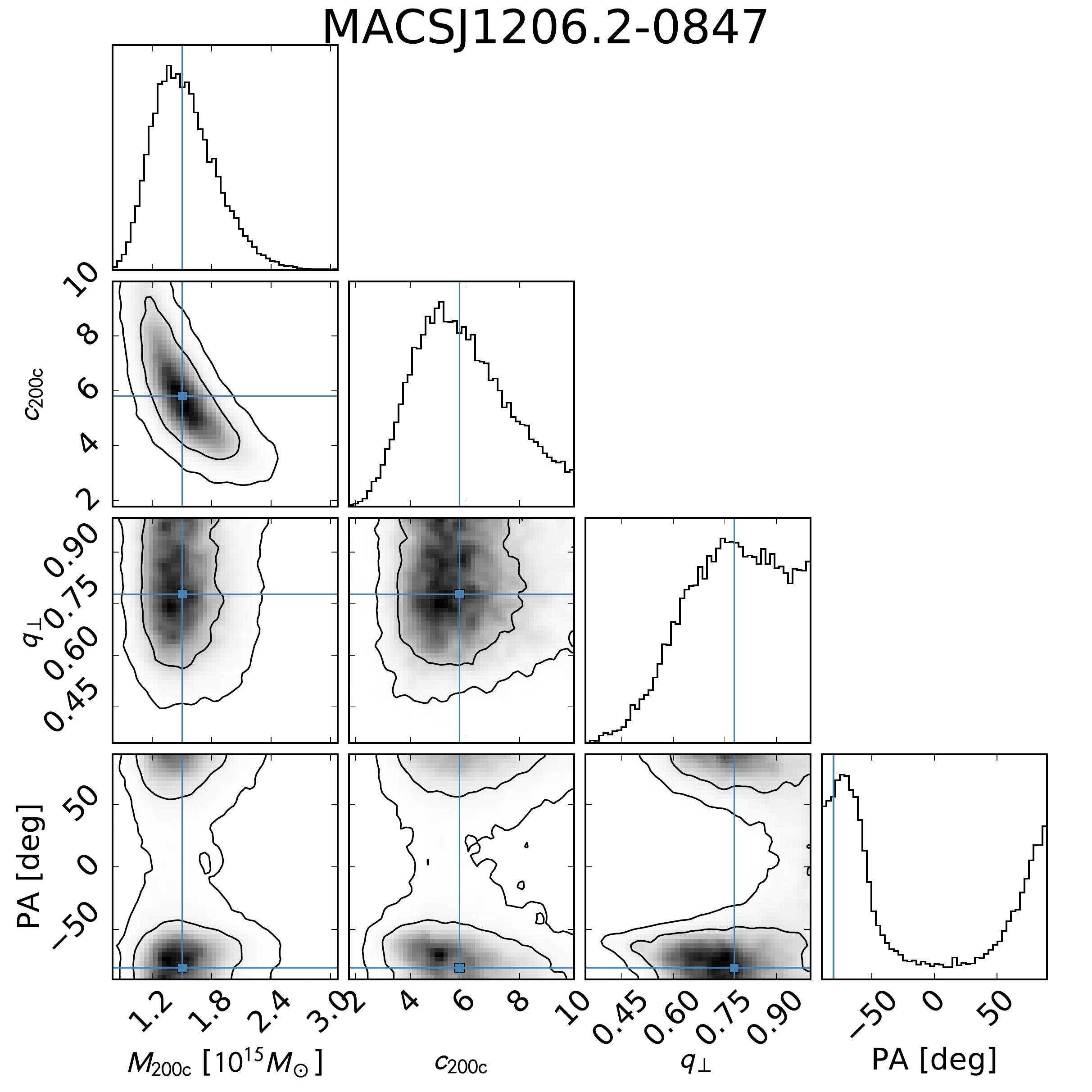}&
  \includegraphics[width=0.4\textwidth,angle=0,clip]{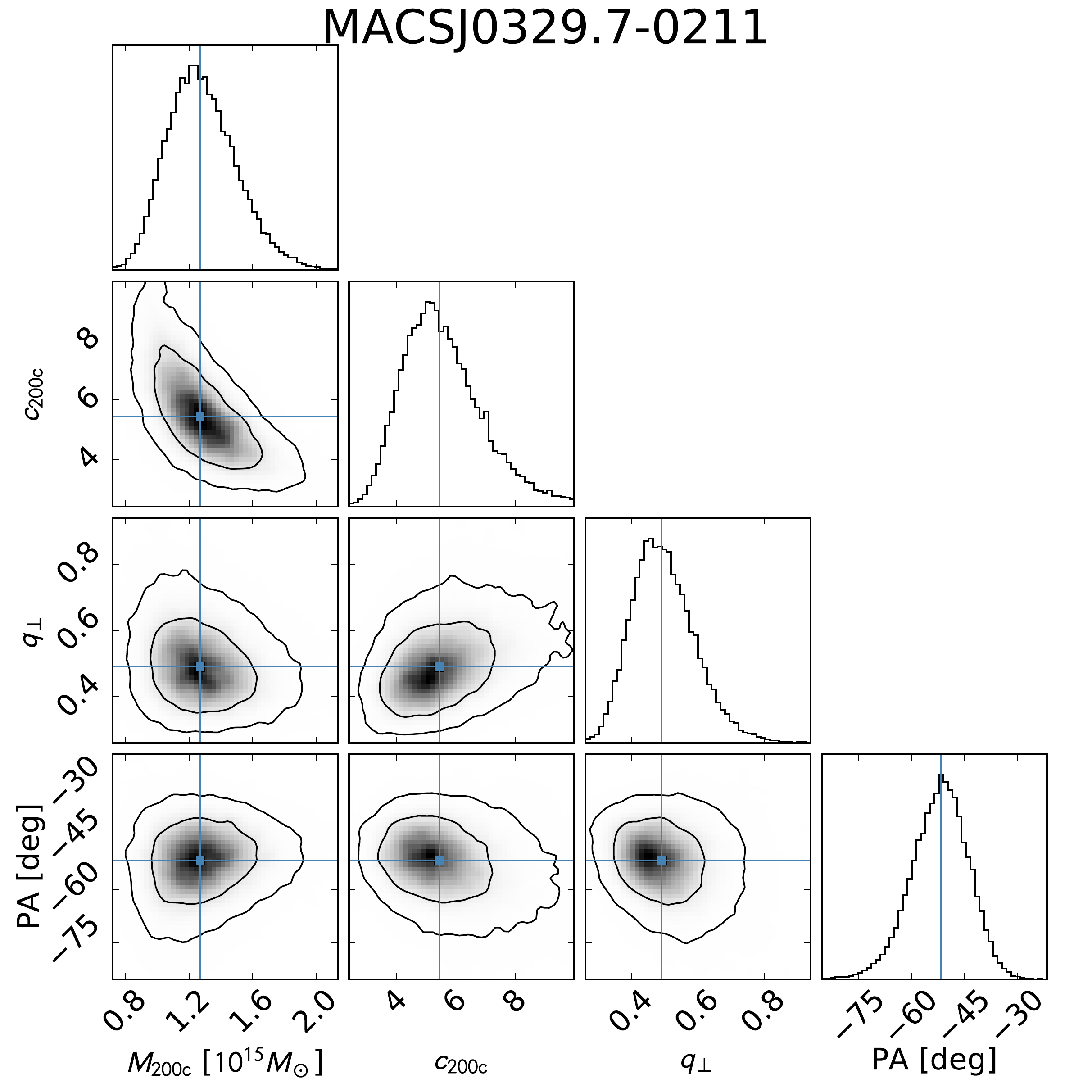}
 \end{array}
 $
 $
 \begin{array}
  {c@{\hspace{1.in}}c@{\hspace{1.in}}c}
   \includegraphics[width=0.4\textwidth,angle=0,clip]{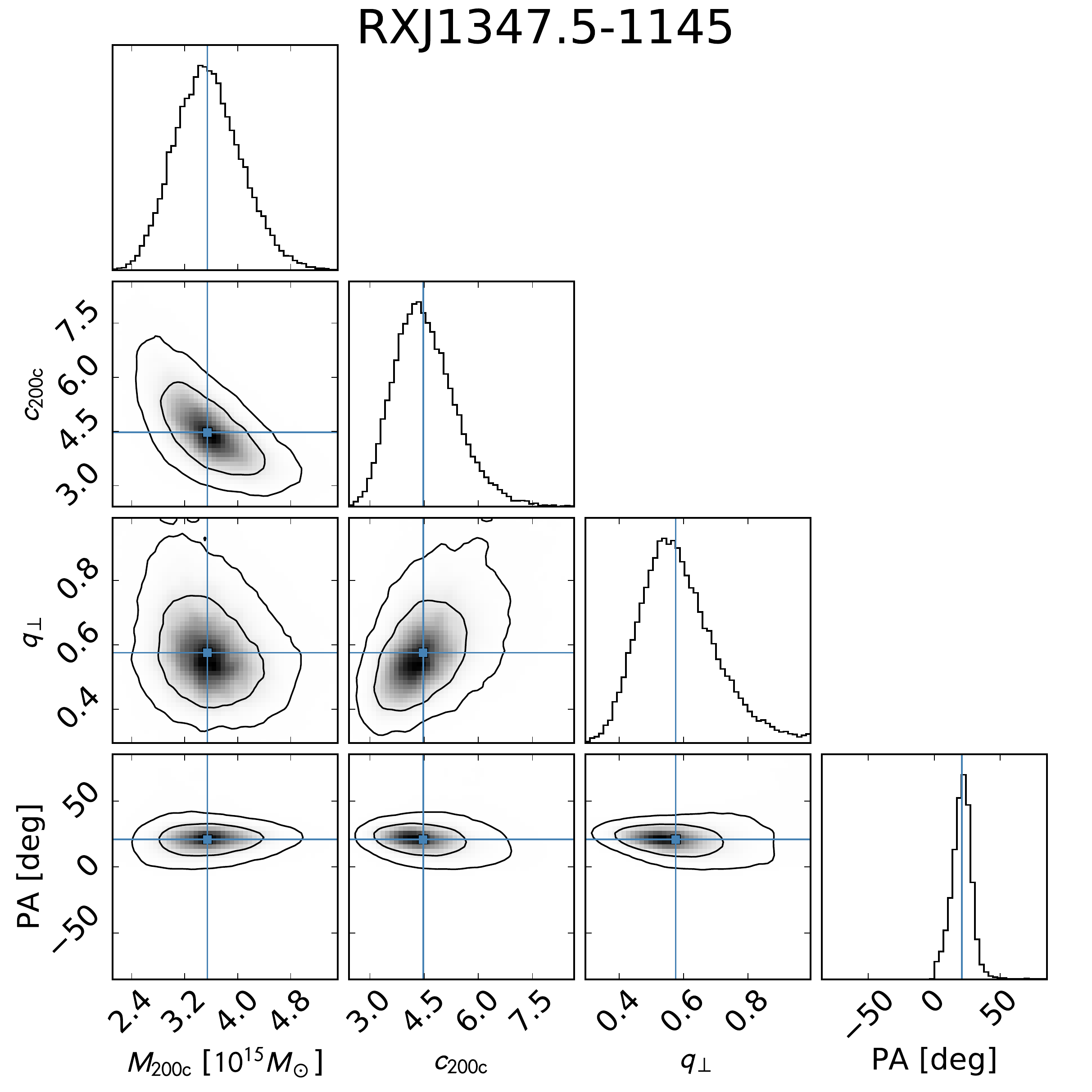}& 
   \includegraphics[width=0.4\textwidth,angle=0,clip]{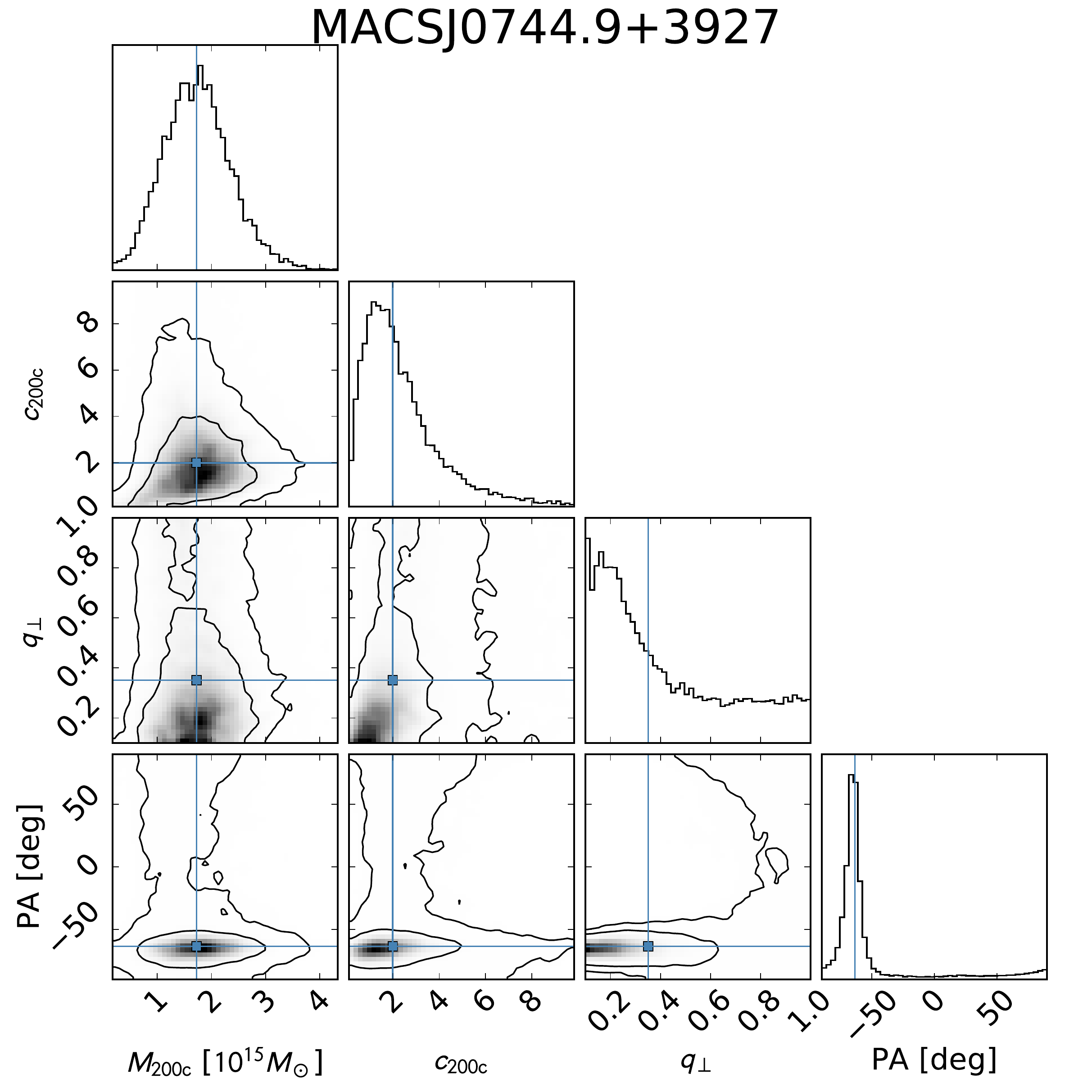}\\
 \end{array}
 $
 $
 \begin{array}
  {c@{\hspace{1.in}}c@{\hspace{1.in}}c}
   \includegraphics[width=0.4\textwidth,angle=0,clip]{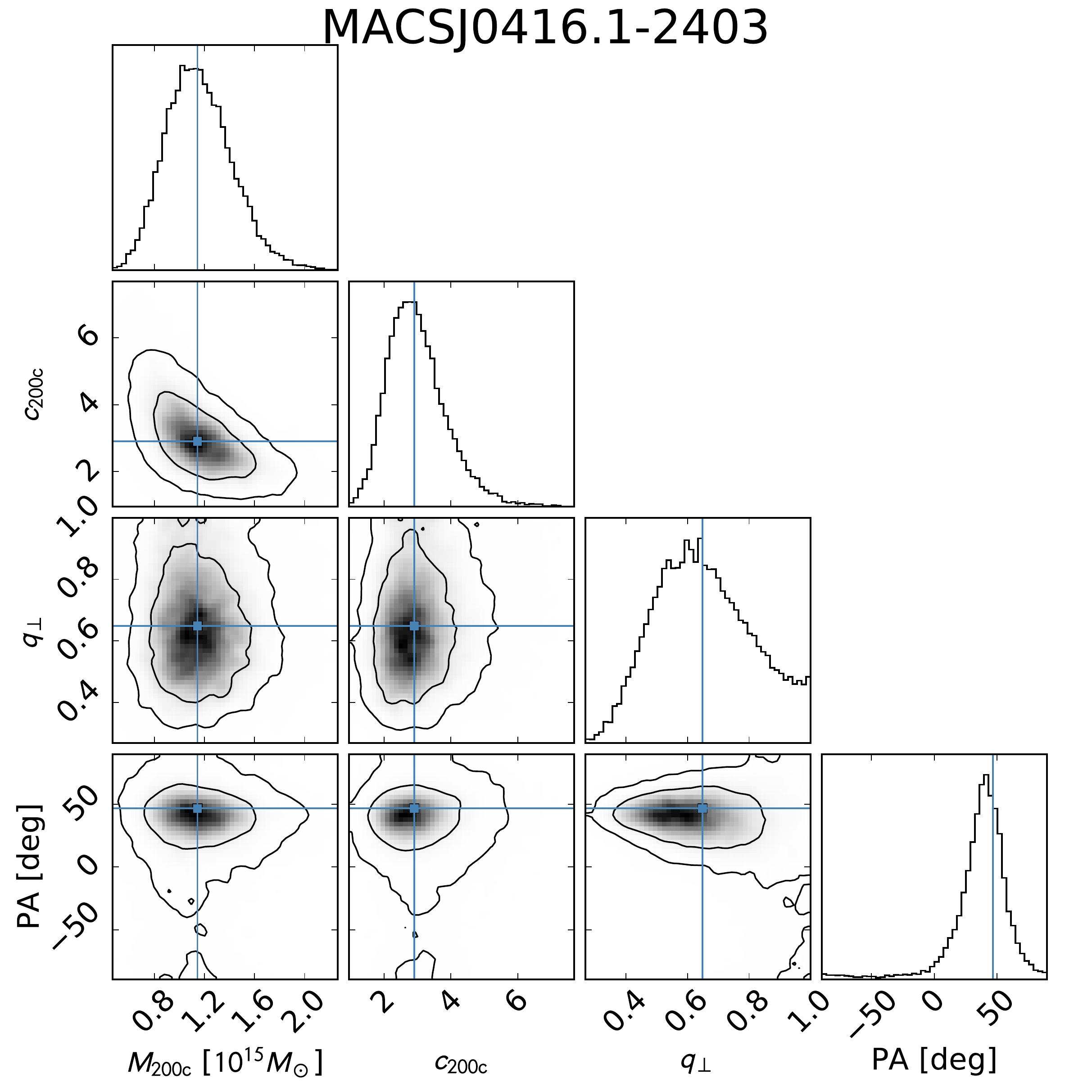}&  
   \includegraphics[width=0.4\textwidth,angle=0,clip]{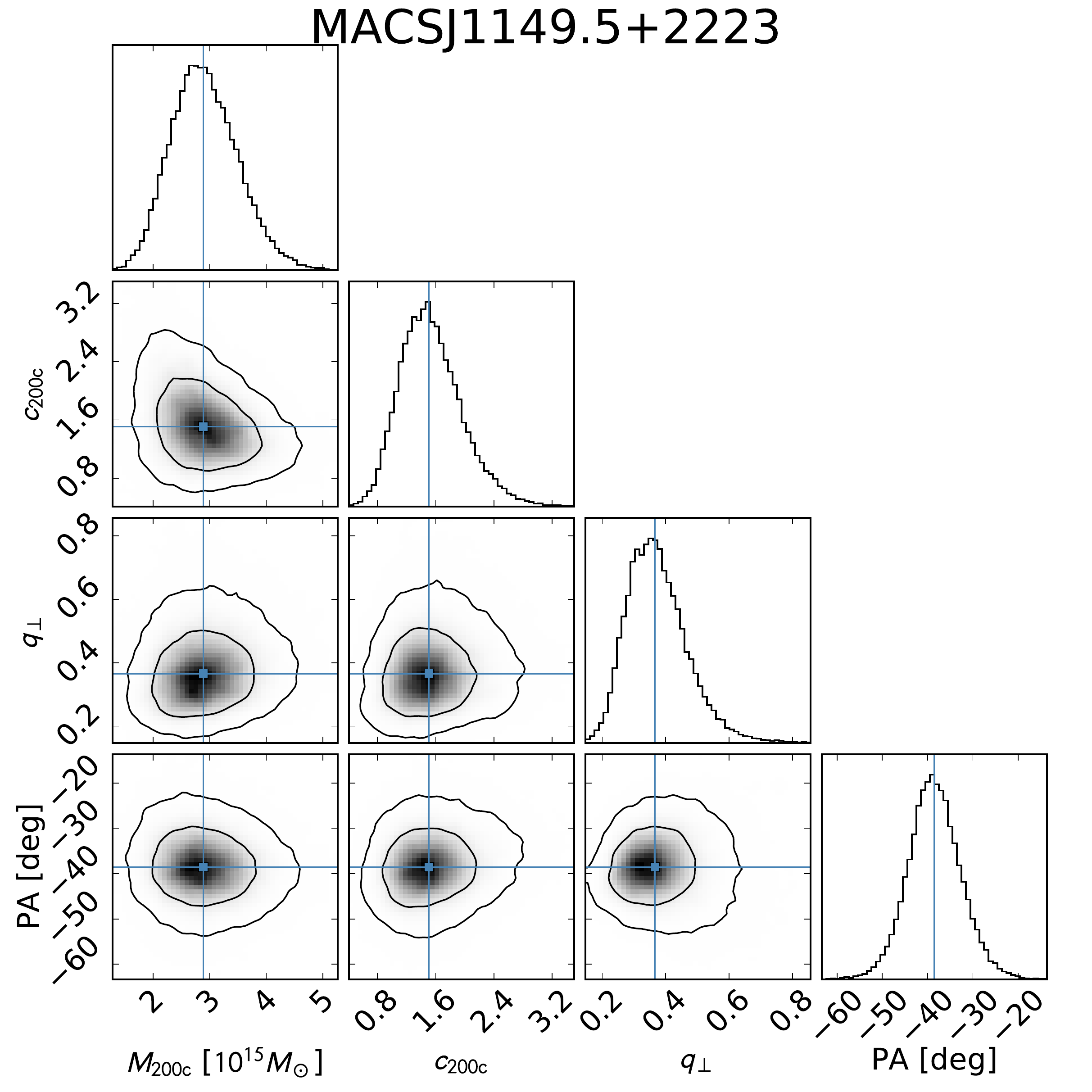}
 \end{array}
 $
 \end{center}
\caption{Continued: Posterior constraints on eNFW model parameters of
 CLASH clusters.}
 \end{figure*}

\begin{figure*}[!htb] 
\addtocounter{figure}{-1}
 \begin{center}
 $
 \begin{array}
  {c@{\hspace{1.in}}c@{\hspace{1.in}}c}
  \includegraphics[width=0.4\textwidth,angle=0,clip]{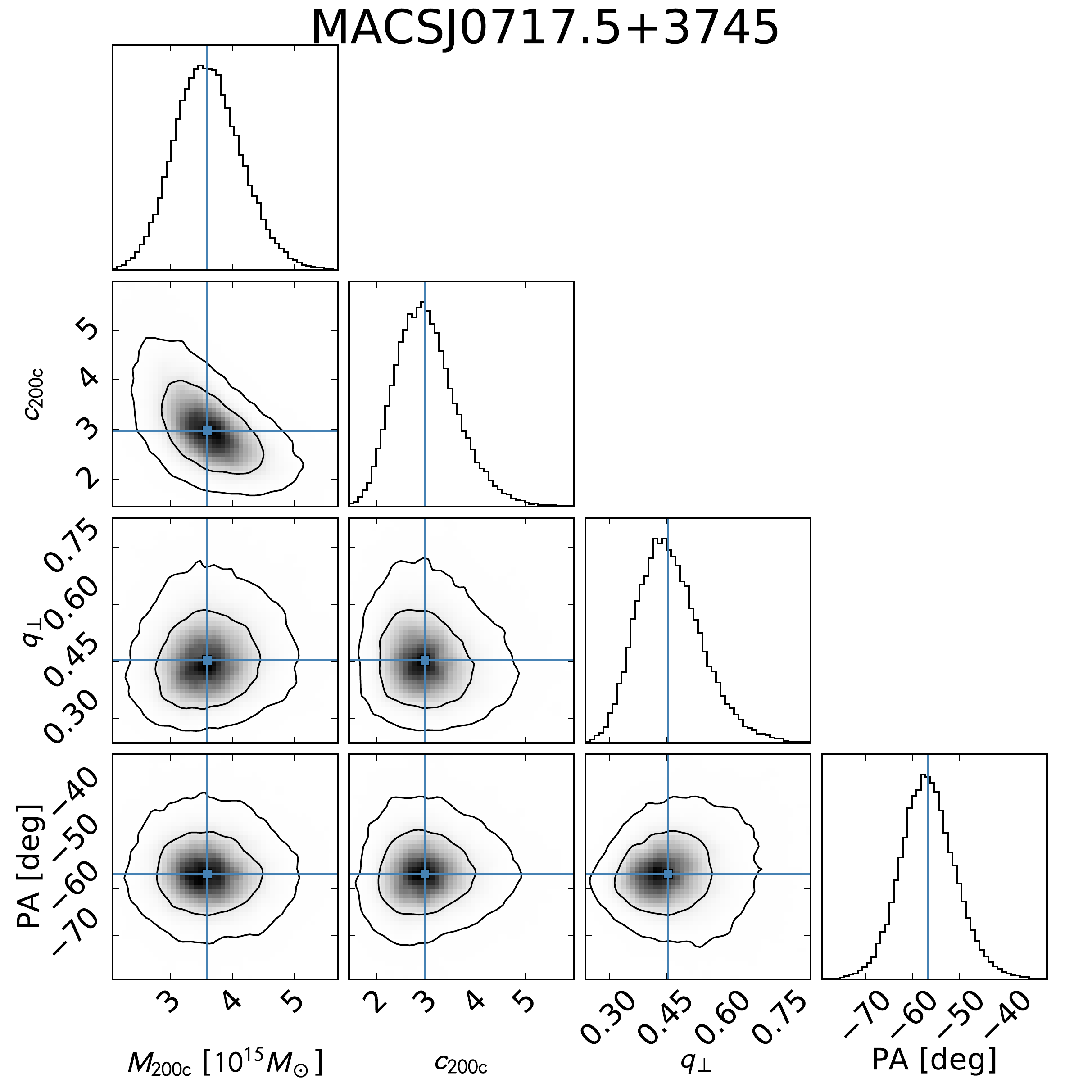}&
  \includegraphics[width=0.4\textwidth,angle=0,clip]{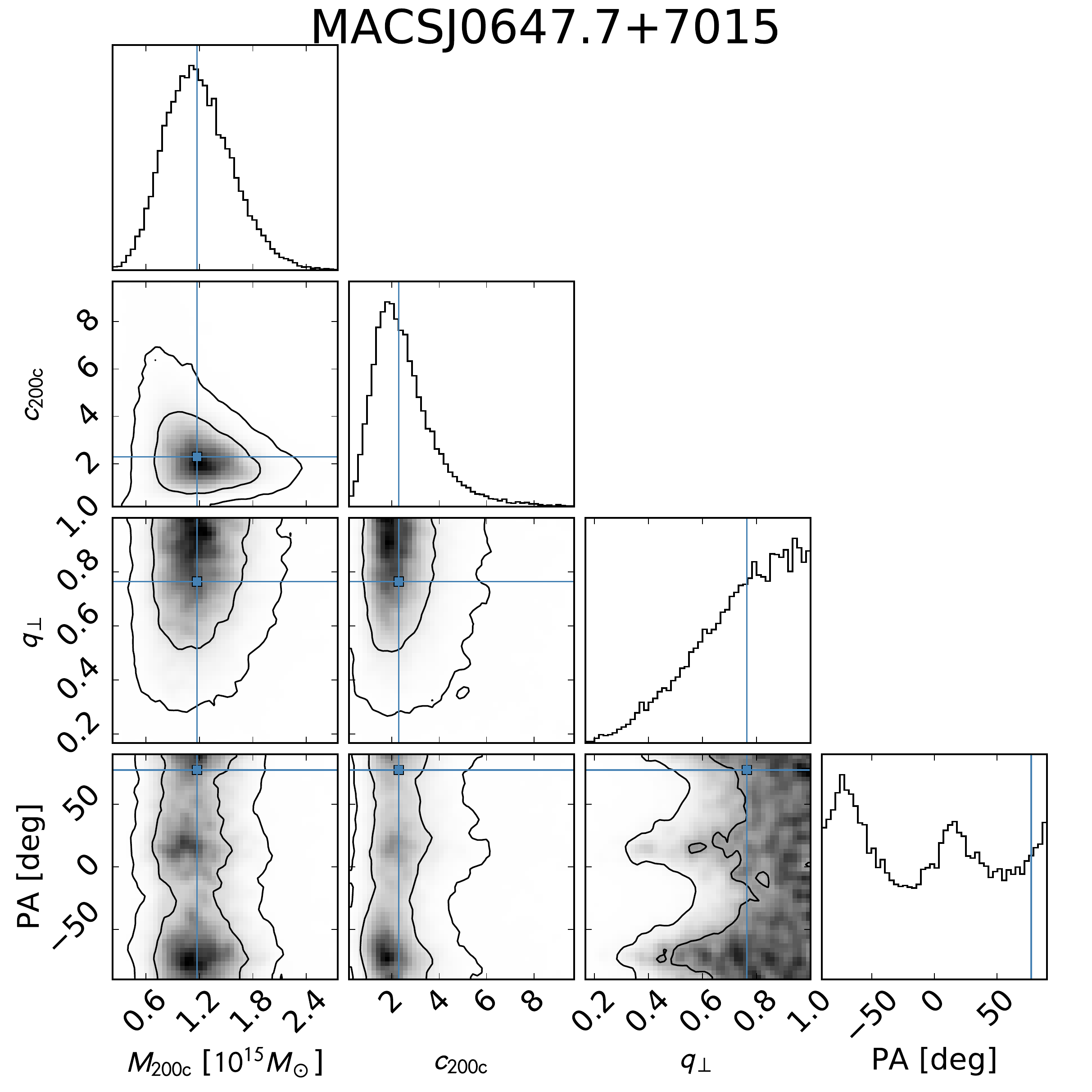}
 \end{array}
 $
 \end{center}
\caption{Continued: Posterior constraints on eNFW model parameters of
 CLASH clusters. For MACS~J0647.7$+$7015, the marginalized 1D
 distribution of $\mathrm{PA}$ is bimodal, so that the biweight estimate
 of the center location ($C_\mathrm{BI}$) lies between the two
 probability peaks.}  
\end{figure*}

\end{appendix}



\end{document}

%% file: citation_fix.tex
\usepackage{etoolbox}
\makeatletter
\patchcmd{\NAT@citex}
  {\@citea\NAT@hyper@{\NAT@nmfmt{\NAT@nm}\NAT@date}}
  {\@citea\NAT@nmfmt{\NAT@nm}\NAT@hyper@{\NAT@date}}
  {}
  {}
\patchcmd{\NAT@citex}
  {\@citea\NAT@hyper@{%
     \NAT@nmfmt{\NAT@nm}%
     \hyper@natlinkbreak{\NAT@aysep\NAT@spacechar}{\@citeb\@extra@b@citeb}%
     \NAT@date}}
  {\@citea\NAT@nmfmt{\NAT@nm}%
   \NAT@aysep\NAT@spacechar%
   \NAT@hyper@{\NAT@date}}
  {}
  {}
\patchcmd{\NAT@citex}
  {\@citea\NAT@hyper@{%
     \NAT@nmfmt{\NAT@nm}%
     \hyper@natlinkbreak{\NAT@spacechar\NAT@@open\if*#1*\else#1\NAT@spacechar\fi}%
       {\@citeb\@extra@b@citeb}%
     \NAT@date}}
  {\@citea\NAT@nmfmt{\NAT@nm}%
   \NAT@spacechar\NAT@@open\if*#1*\else#1\NAT@spacechar\fi%
   \NAT@hyper@{\NAT@date}}
  {}
  {}
\makeatother

%% file: table1.tex
\begin{deluxetable*}{lcrrrrrrr}
\centering
\tabletypesize{\scriptsize}
\tablecolumns{11}
\tablecaption{
\label{tab:sample}
Cluster sample and elliptical Navarro--Frenk--White model parameters
}
\tablewidth{0pt}
\tablehead{
 \multicolumn{1}{c}{Cluster} &
 \multicolumn{1}{c}{Redshift\tablenotemark{a}} &
 \multicolumn{1}{c}{R.A.\tablenotemark{b}} &
 \multicolumn{1}{c}{Decl.\tablenotemark{b}} &
 \multicolumn{1}{c}{$M_\mathrm{200c}$} &
 \multicolumn{1}{c}{$c_\mathrm{200c}$} &
 \multicolumn{1}{c}{$q_\perp$\tablenotemark{c}} &
 \multicolumn{1}{c}{PA\tablenotemark{d}} &
 \multicolumn{1}{c}{$\chi^2$/dof\tablenotemark{e}} 
 \\ 
\colhead{} &
\colhead{} &
\multicolumn{1}{c}{(J2000.0)} &
\multicolumn{1}{c}{(J2000.0)} &
\multicolumn{1}{c}{($10^{14}M_{\odot}\,h_{70}^{-1}$)} &
\colhead{} &
\colhead{} &
\multicolumn{1}{c}{(degrees)} &
\colhead{}
}
\startdata
X-ray Selected:\\
         ~~Abell 383 & $0.187$ & 02:48:03.40 & -03:31:44.9 & $10.15 \pm 4.27$ & $ 2.5 \pm  1.6$ & $0.70 \pm 0.19$ & $82.1 \pm 30.5$ & 1537/2300
\\
         ~~Abell 209 & $0.206$ & 01:31:52.54 & -13:36:40.4 & $19.31 \pm 3.58$ & $ 3.4 \pm  0.7$ & $0.61 \pm 0.10$ & $-28.6 \pm  9.8$ & 1581/2300
\\
        ~~Abell 2261 & $0.224$ & 17:22:27.18 & +32:07:57.3 & $24.80 \pm 3.50$ & $ 3.7 \pm  0.6$ & $0.78 \pm 0.11$ & $20.9 \pm 14.0$ & 1529/2300
\\
 ~~RX~J2129.7$+$0005 & $0.234$ & 21:29:39.96 & +00:05:21.2 & $7.78 \pm 2.43$ & $ 2.9 \pm  1.2$ & $0.81 \pm 0.14$ & $85.0 \pm 43.8$ & 1477/2300
\\
         ~~Abell 611 & $0.288$ & 08:00:56.82 & +36:03:23.6 & $13.44 \pm 3.39$ & $ 4.2 \pm  1.3$ & $0.56 \pm 0.19$ & $46.8 \pm  9.3$ & 1586/1932
\\
     ~~MS2137$-$2353 & $0.313$ & 21:40:15.17 & -23:39:40.2 & $10.78 \pm 3.17$ & $ 2.4 \pm  1.0$ & $0.76 \pm 0.17$ & $-53.8 \pm 38.7$ & 1441/1760
\\
 ~~RX~J2248.7$-$4431 & $0.348$ & 22:48:43.96 & -44:31:51.3 & $19.81 \pm 5.97$ & $ 1.6 \pm  0.7$ & $0.51 \pm 0.19$ & $55.5 \pm 12.2$ &  866/1440
\\
~~MACS~J1115.9$+$0129 & $0.352$ & 11:15:51.90 & +01:29:55.1 & $17.91 \pm 3.81$ & $ 2.5 \pm  0.7$ & $0.53 \pm 0.14$ & $-32.3 \pm  8.1$ & 1155/1440
\\
~~MACS~J1931.8$-$2635 & $0.352$ & 19:31:49.62 & -26:34:32.9 & $11.62 \pm 2.84$ & $ 7.8 \pm  1.7$ & $0.77 \pm 0.18$ & $38.9 \pm 44.9$ & 1818/1440
\\
 ~~RX~J1532.9$+$3021 & $0.363$ & 15:32:53.78 & +30:20:59.4 & $7.01 \pm 1.49$ & $ 7.1 \pm  1.7$ & $0.82 \pm 0.14$ & $52.2 \pm 44.7$ & 1120/1440
\\
   ~~MACS~J1720.3$+$ & $0.391$ & 17:20:16.78 & +35:36:26.5 & $11.18 \pm 2.38$ & $ 5.9 \pm  1.7$ & $0.73 \pm 0.15$ & $24.8 \pm 25.8$ & 1053/1292
\\
~~MACS~J0429.6$-$0253 & $0.399$ & 04:29:36.05 & -02:53:06.1 & $8.88 \pm 1.70$ & $ 7.7 \pm  1.6$ & $0.84 \pm 0.12$ & $79.4 \pm 50.4$ & 1081/1292
\\
~~MACS~J1206.2$-$0847 & $0.440$ & 12:06:12.15 & -08:48:03.4 & $15.05 \pm 3.20$ & $ 5.8 \pm  1.7$ & $0.78 \pm 0.14$ & $-80.6 \pm 29.9$ &  973/1152
\\
~~MACS~J0329.7$-$0211 & $0.450$ & 03:29:41.56 & -02:11:46.1 & $12.70 \pm 2.19$ & $ 5.4 \pm  1.3$ & $0.49 \pm 0.09$ & $-51.7 \pm  8.1$ &  563/1020
\\
 ~~RX~J1347.5$-$1145 & $0.451$ & 13:47:31.05 & -11:45:12.6 & $35.40 \pm 5.05$ & $ 4.5 \pm  0.9$ & $0.58 \pm 0.12$ & $20.9 \pm  7.4$ & 1349/1020
\\
~~MACS~J0744.9$+$3927 & $0.686$ & 07:44:52.82 & +39:27:26.9 & $17.23 \pm 6.16$ & $ 2.0 \pm  1.4$ & $0.35 \pm 0.27$ & $-63.6 \pm  6.4$ &  274/672
\\
\hline High Magnification:\\
~~MACS~J0416.1$-$2403 & $0.396$ & 04:16:08.38 & -24:04:20.8 & $11.43 \pm 2.66$ & $ 2.9 \pm  0.9$ & $0.65 \pm 0.16$ & $45.1 \pm 13.9$ &  867/1292
\\
~~MACS~J1149.5$+$2223 & $0.544$ & 11:49:35.69 & +22:23:54.6 & $28.86 \pm 5.92$ & $ 1.5 \pm  0.4$ & $0.37 \pm 0.09$ & $-38.5 \pm  5.8$ &  704/896
\\
~~MACS~J0717.5$+$3745 & $0.548$ & 07:17:32.63 & +37:44:59.7 & $35.96 \pm 5.43$ & $ 3.0 \pm  0.6$ & $0.45 \pm 0.09$ & $-56.8 \pm  6.0$ &  729/896
\\
~~MACS~J0647.7$+$7015 & $0.584$ & 06:47:50.27 & +70:14:55.0 & $11.73 \pm 3.79$ & $ 2.3 \pm  1.2$ & $0.76 \pm 0.18$ & $77.2 \pm 52.3$ &  574/780

\enddata
\tablenotetext{a}{Cluster redshifts were taken from \citet{Umetsu2014clash}.}
\tablenotetext{b}{The cluster center represents the location of the brightest cluster galaxy when a single dominant central galaxy is found. Otherwise, for MACS~J0717.5$+$3745 and MACS~J0416.1$-$2403, it is the center of the brightest red-sequence-selected cluster galaxies.}
\tablenotetext{c}{Projected minor-to-major halo axis ratio $q_\perp\le 1$.}
\tablenotetext{d}{Position angle (PA) of the projected halo major axis measured east of north, defined in the range $[-90^\circ, 90^\circ)$.}
\tablenotetext{e}{Minimum $\chi^2$ per degrees of freedom (dof).}\end{deluxetable*}

%% file: table2.tex
\begin{deluxetable}{lccccc}
\tablecolumns{8}
\tablewidth{0pt}
\tabletypesize{\scriptsize}
\tablecaption{\label{tab:q}
Weak-lensing halo ellipticity measurements
}
\tablehead{
 \multicolumn{1}{c}{Sample} &
 \multicolumn{1}{c}{$N$} &
 \multicolumn{1}{c}{$\langle \epsilon\rangle$} &
 \multicolumn{1}{c}{$\langle e\rangle$} 
}
 \startdata
 Full sample        & 20 & $0.33\pm 0.07$ & $0.38\pm 0.08$\\
 X-ray selected     & 16 & $0.28\pm 0.07$ & $0.33\pm 0.06$\\
 High-magnification & 4  & $0.45\pm 0.11$ & $0.53\pm 0.14$
 \enddata
\tablecomments{Median values and $1\sigma$ errors of weak-lensing
 cluster shape measurements derived for the full sample, the
 X-ray-selected subsample, and the high-magnification subsample. We
 adopt the halo ellipticity defined in two ways:
 $\epsilon=1-q_\perp$ and $e=(1-q_\perp^2)/(1+q_\perp^2)$ with
 $q_\perp\le 1$
 the projected halo axis ratio.}
\end{deluxetable}

%% file: table3.tex
\begin{deluxetable}{lccccc}
\tablecolumns{8}
\tablewidth{0pt}
\tabletypesize{\scriptsize}
\tablecaption{\label{tab:PA}
Weak-lensing halo misalignment statistics
}
\tablehead{
 \multicolumn{1}{c}{Data sets} &
 \multicolumn{1}{c}{$N$} &
 \multicolumn{1}{c}{$\langle\Delta\mathrm{PA}\rangle$} &
 \multicolumn{1}{c}{$\langle|\Delta\mathrm{PA}|\rangle$} &
 \multicolumn{1}{c}{$p$-value} &
 \multicolumn{1}{c}{$p_\mathrm{Bin}$} 
 \\ 
\multicolumn{1}{c}{} &
\multicolumn{1}{c}{} &
\multicolumn{1}{c}{(degrees)} &
\multicolumn{1}{c}{(degrees)} &
\multicolumn{1}{c}{} &
\multicolumn{1}{c}{} 
}
 \startdata
BCG/WL          & 16 & $-25\pm 14$ & $34\pm 13$ & $1.5\times 10^{-1}$ & $1.8\times 10^{-1}$\\
X-ray/WL        & 20 & $-3\pm 9$   & $21\pm  7$ & $4.1\times 10^{-3}$ & $4.6\times 10^{-3}$\\
SZE/WL          & 18 & $7\pm 17$   & $42\pm 10$ & $3.8\times 10^{-1}$ & $1.9\times 10^{-1}$\\ 
{\em HST}-GL/WL & 20 & $-12\pm 6$  & $16\pm  4$ & $3.1\times 10^{-4}$ & $1.1\times 10^{-3}$       
 \enddata
\tablecomments{The basic statistics of misalignment angles are
 listed. The results here are based on the point estimates of
 weak-lensing PAs from individual cluster posterior distributions
 (Figure \ref{fig:eNFWfit}). 
 Column 1: combination of data sets.
 Column 2: number of clusters in the overlapping sample.
 Column 3: median of the distribution of misalignment angles,
 $\Delta\mathrm{PA}\in [-90^\circ,90^\circ)$.
 Column 4: median of the distribution of absolute misalignment
 angles, $|\Delta\mathrm{PA}|\in[0^\circ,90^\circ]$, 
 {\em Column 5}: probability of finding the value 
 $\langle|\Delta\mathrm{PA}|\rangle$ smaller than or equal to the
 observed value when the null hypothesis of a uniform distribution in
 $[0^\circ,90^\circ]$ is true.
 {\em Column 6}: probability, obtained with the binomial test, that the
 observed distribution in $|\Delta\mathrm{PA}|$ has random  
 orientations \citep{West2017}.}
\end{deluxetable}